\documentclass[twocolumn,floatfix,aps,prd,showpacs]{revtex4}
\usepackage{graphicx}
\usepackage{bm}
\usepackage{epsf}

\def\agt{\mathrel{\raise.3ex\hbox{$>$}\mkern-14mu\lower0.6ex\hbox{$\sim$}}}
\def\alt{\mathrel{\raise.3ex\hbox{$<$}\mkern-14mu\lower0.6ex\hbox{$\sim$}}}

\newcommand{\beq}{\begin{equation}}
\newcommand{\eeq}{\end{equation}}
\newcommand{\beqn}{\begin{eqnarray}}
\newcommand{\eeqn}{\end{eqnarray}}
\newcommand{\pa}{\partial}

\newcommand{\varep}{\varepsilon}

\newcommand{\cB}{{\cal{B}}}
\newcommand{\cb}{{\cal{H}}}

\newcommand{\ve}[1]{\mbox{\boldmath $#1$}}
\def\bI{\hbox{$\,I\!\!\!$--}}

\begin{document}

\title{Magnetorotational collapse of 
massive stellar cores to neutron stars: \\
Simulations in full general relativity}

\author{Masaru Shibata$^1$}


\author{Yuk Tung Liu$^2$}

\author{Stuart L. Shapiro$^{2,3}$}

\author{Branson C. Stephens$^2$}

\affiliation{$^1$Graduate School of Arts and Sciences, 
University of Tokyo, Komaba, Meguro, Tokyo 153-8902, Japan\\
$^2$Department of Physics, University of Illinois at Urbana-Champaign,
Urbana, IL 61801-3080 \\
$^3$Department of Astronomy and NCSA, University
of Illinois at Urbana-Champaign, Urbana, IL 61801}

\begin{abstract}
We study magnetohydrodynamic (MHD) effects arising in the collapse of
magnetized, rotating, massive stellar cores to proto-neutron stars
(PNSs). We perform axisymmetric numerical simulations in full general
relativity with a hybrid equation of state. The formation and early
evolution of a PNS are followed with a grid of $2500 \times 2500$
zones, which provides better resolution than in previous (Newtonian)
studies.  We confirm that significant differential rotation results
even when the rotation of the progenitor is initially
uniform. Consequently, the magnetic field is amplified both by
magnetic winding and the magnetorotational instability (MRI).  Even if
the magnetic energy $E_{\rm EM}$ is much smaller than the rotational
kinetic energy $T_{\rm rot}$ at the time of PNS formation, the ratio
$E_{\rm EM}/T_{\rm rot}$ increases to 0.1--0.2 by the magnetic
winding. Following PNS formation, MHD outflows lead to losses of rest
mass, energy, and angular momentum from the system.  The earliest
outflow is produced primarily by the increasing magnetic stress caused
by magnetic winding.  The MRI amplifies the poloidal field and
increases the magnetic stress, causing further angular momentum
transport and helping to drive the outflow. After the magnetic field
saturates, a nearly stationary, collimated magnetic field forms near
the rotation axis and a Blandford-Payne type outflow develops along
the field lines. These outflows remove angular momentum from the PNS
at a rate given by $\dot{J} \sim \eta E_{\rm EM}C_B$, where $\eta$ is
a constant of order $\sim 0.1$ and $C_B$ is a typical ratio of
poloidal to toroidal field strength. As a result, the rotation period
quickly increases for a strongly magnetized PNS until the degree of
differential rotation decreases. Our simulations suggest that rapidly
rotating, magnetized PNSs may not give rise to rapidly rotating
neutron stars.
\end{abstract}

\maketitle

\section{Introduction}

The explosion mechanism in core-collapse supernovae has been 
pursued for several decades, but the problem has not yet been 
solved. Neutrino-driven convection has been suggested as a way 
of rejuvenating the stalled shock, while rotation and
magnetic fields have also been proposed as key driving 
mechanisms (see e.g.,~\cite{WJ05} for a review). 
Recently, acoustic power generated by accretion-triggered g-mode 
oscillations of the proto-neutron star (PNS) has been offered 
as an explanation for the explosion~\cite{OBDE06}.

Even if rotation and magnetic fields are not the main mechanisms, they
may still have an important influence on a supernova explosion,
especially when a long-soft gamma-ray burst is produced.
In addition, several observations can be explained naturally by
magnetic fields.  For example, the rapid spindown of anomalous X-ray
pulsars may result from the amplification of the star's magnetic field
during core collapse~\cite{AXP,TD}.  Soft gamma-ray
repeaters~\cite{AXP} are likely to be neutron stars with very strong
magnetic fields (so-called ``magnetars'')~\cite{DT}.

It has been speculated that the magnetic field strength could be
amplified to $\sim 10^{16}$ G during stellar core collapse, and the
resulting strong magnetic field could play a crucial role in the
supernova explosion~\cite{WMW,AHM,TCQ,TQB}. The collapse is
generically nonhomologous, with the inner core collapsing faster than
the outer core.  Thus, even if the progenitor has a rigid rotation
profile at the onset of the collapse, differential rotation naturally
develops with angular velocity decreasing outward. In the presence of
such differential rotation, the magnetic field is amplified both by
magnetic winding~\cite{spitzer,spruit,BSS,Shapiro} and the
magnetorotational (magneto-shear) instability
(MRI)~\cite{MRI0,MRI,MRIrev}. The field strength may thus grow by many
orders of magnitude, even if it is initially small. The field
amplification is likely to continue until the kinetic energy
associated with the differential rotation $T_{\rm diff}$ is converted
to magnetic energy $E_{\rm EM}$~\cite{Shapiro}. Typically, $T_{\rm
diff}$ is an appreciable fraction of the total rotational kinetic
energy $T_{\rm rot}$.  The amplified magnetic field results in a
strong magnetic stress, which could blow off the matter in the
vicinity of the PNS, converting magnetic energy back into matter
kinetic energy and driving an outflow.  The typical rotational kinetic
energy of a PNS is approximately
\beqn
T_{\rm rot}&=&{1 \over 2} \kappa_I M R^2
\Omega^2 \nonumber \\ &=& 4 \times 10^{51} \biggl({\kappa_I \over
0.3}\biggr) \biggl({M\over 1.4M_{\odot}}\biggr) \nonumber \\
&&~~~~~~\times \biggl({R \over 10~{\rm km}}\biggr)^2 \biggl({P_{\rm
rot} \over 2~{\rm ms}}\biggr)^{-2} ~{\rm ergs},\label{eq1}
\eeqn
where $M$, $R$, and $\Omega(=2\pi/P_{\rm rot})$ are the mass, radius,
and angular velocity of the PNS, and $\kappa_I \sim$0.3--0.6 denotes
the ratio of the moment of inertia to $MR^2$ and depends on the
structure of the neutron star \cite{FIP,CST}. Thus, if the rotation
period of the PNS is shorter than $\sim 4$ ms, and if a large fraction
of the rotational kinetic energy is converted to the kinetic energy of
the outward matter flow via magnetohydrodynamic (MHD) processes, the
supernova explosion may be significantly powered by magnetorotational
effects.

Numerical simulations of magnetorotational core collapse were
pioneered by LeBlanc and Wilson~\cite{LW} and by Symbalisty~\cite{Sym}
in the 1970s and 1980s, respectively. In the past few years
(e.g.,~\cite{YS,SKY,TKNS,Kotake,OAM,OADM,ABM,MBA}), this has become an
active research topic in computational astrophysics. All of the
simulations to date have been performed by assuming Newtonian
gravitation or by including the general relativistic effects
approximately~\cite{OADM}. Yamada and his collaborators have performed
a variety of simulations in axial symmetry with simplified
microphysics for a variety of rotational profiles and magnetic field
strengths~\cite{YS,SKY}. They found that the initially poloidal
magnetic field is primarily amplified by compression during infall and
by magnetic winding following core collapse. They also found that the
magnetic pressure, which is amplified during collapse, could drive a
strong outflow along the rotation axis. Their simulations were done in
the nonuniform spherical polar coordinates $(r, \theta)$ with a grid
size of at most (300, 50). At this resolution, they were not able to
resolve the MRI.  Takiwaki et al.~\cite{TKNS} performed simulations
similar to those of Yamada et al., but with a realistic equation of
state (EOS), and obtained essentially the same results. Kotake et
al.~\cite{Kotake} performed a simulation adopting a strong toroidal
magnetic field and an extremely high degree of differential rotation
as the initial condition and found that the toroidal magnetic field
can power the explosion for this special initial
condition. Obergaulinger, Aloy, and M\"uller~\cite{OAM} performed
similar simulations to those by Yamada et al.\ but with a slightly
better grid resolution, (380, 60), in spherical polar
coordinates. They indicated that not only magnetic winding but also
the MRI could play an important role in the supernova explosion and
evolution of the PNS.  However, the grid resolution they adopted was
also not sufficient for resolving the regions in which the MRI occurs
most strongly~\cite{OAM0}. Ardeljan, Bisnovatyi-Kogan, and
Moiseenko~\cite{ABM,MBA} performed axisymmetric simulations with a
Lagrangian scheme. In their work, a purely hydrodynamic simulation was
performed up to the formation of a PNS.  Then they added a poloidal
magnetic field to the PNS to study the MHD effects. They found that
such effects, and, in particular the enhancement of the magnetic field
strength by magnetic winding, could result in buoyancy and trigger a
supernova explosion. They also reported that an MRI-like instability
played an important role. However, in their simulation, the
instability is observed only after $\sim 100 P_{\rm rot}$, long after
the toroidal field first becomes significant.  But the fastest-growing
mode for the {\em (shear-type)} MRI is amplified exponentially in a
few rotation periods, irrespective of the strength of the toroidal
fields~\cite{MRIrev}. Thus, the instability that they found is
unlikely to be a shear-type instability~\cite{spruit}; rather it is
likely to be the magneto-convective instability (see discussion in
Sec.~VII).

All of these previous simulations have provided a qualitative picture
of magnetorotational core collapse and supernova explosions. However,
not all of the important magnetorotational effects have been
incorporated in the simulations, mainly because of insufficient grid
resolution. Hence, some fundamental questions concerning the
magnetorotational explosion scenario have not been answered. For
example, it has been established that the magnetic field is amplified
during the collapse and during the subsequent evolution of the PNS,
but what determines the saturated field strength and how large is it? 
How important are the effects of the MRI? Rotational kinetic energy of
the PNS can be converted to outflow kinetic energy via MHD processes
such as magnetic winding and the MRI. How efficiently is the
rotational kinetic energy of the PNS converted to matter outflow
energy? What is the rate of decrease of the rotational kinetic energy
of the PNS and the corresponding rate of its period increase? After
the magnetic field saturates, the magnetic configuration in the PNS
will settle down to a stationary state. What is this final state? What
mechanisms drive outflow?

Previous work has been performed mainly in the framework of Newtonian
gravitation. However, general relativistic effects are not negligible
and may have significant influence in the evolution of the PNS
(particularly for a massive progenitor). Furthermore, after the
collapse, shocks and outflows are often produced at relativistic
speed. From a technical point of view, general relativistic
simulations have an additional advantage. In the Newtonian case, the
Alfv\'en velocity may exceed the speed of light (especially in the
low-density region), and hence, the Courant condition for the time
step can be quite severe.  On the other hand, the Alfv\'en velocity is
guaranteed to be smaller than the speed of light in general
relativity, and so we do not have to deal with this unphysical
complication.

In this paper, we summarize results from axisymmetric simulations in
full general relativity using {\it two} general relativistic
magnetohydrodynamical (GRMHD) codes independently developed
recently~\cite{DLSS,SS05,DLSSS,DLSSS2}. One purpose is to demonstrate
that multi-dimensional MHD simulations for core collapse are now
feasible in full general relativity, without approximation. The main
purpose is to address the questions raised above by performing
simulations with much higher grid resolution than those reported
previously. We adopt a uniform grid in cylindrical coordinates so that
we can achieve uniformly high resolution everywhere in our computation
domain, and the MRI can be resolved wherever it occurs. We have also
performed simulations using fisheye coordinates (see~\cite{CLZ06} and
Appendix~A), which provides high resolution in the central region
while using relatively few of grid points.  As in previous
works~\cite{YS,SKY,OAM}, we employ a simplified hybrid EOS since we
want to focus on studying the MHD effects on the collapse and on the
evolution of the magnetized PNS.

The latest study on supernova progenitors~\cite{HWS} suggests that
magnetized, massive stars of solar metallicity may not be rotating as
rapidly as the models we consider in this paper. It suggests that the
typical value of $P_{\rm rot}$ at formation of a PNS is $\sim 15$
ms. The reason is that the magnetic field grows due to a dynamo
mechanism during stellar evolution. As a result, angular momentum is
transported outwards by magnetic braking, which decreases the rotation
period in the central region of a massive star.  If this scenario
holds for all progenitors, the magnetorotational explosion scenario
would not be effective since a substantial amount of rotational
kinetic energy is required ($P_{\rm rot} \alt$ 5~ms), as shown in
Eq.~(\ref{eq1}).  However, the rotational kinetic energy of the
supernova progenitor depends strongly on model parameters of the
dynamo theory and there is still a possibility of forming a rapidly
rotating progenitor if the progenitor is massive~\cite{HWS}. The
magnetorotational scenario thus warrants detailed investigation.  In
this paper, we consider initial pre-collapse stars with substantial
rotational kinetic energies.

The remainder of the paper is organized as follows. In Sec.~II, we
briefly review our mathematical formalism and numerical methods for
our GRMHD simulations. In Sec.~III, we provide a qualitative summary
of the magnetorotational effects that could play important roles in
the evolution of a differentially rotating PNS. Sections~IV and~V
describe our initial models and computational setup, respectively. In
Sec.~VI, we present our numerical results, focusing on the evolution
of the magnetic fields and of the newly formed PNS. Finally, we
summarize our conclusions in Section~VII. Throughout this paper, we
adopt geometrical units in which $G=1=c$, where $G$ and $c$ denote the
gravitational constant and speed of light, respectively.  Cartesian
coordinates are denoted by $x^k=(x, y, z)$. The coordinates are
oriented so that the rotation axis is along the $z$-direction. We
define the coordinate radius $r=\sqrt{x^2+y^2+z^2}$, cylindrical
radius $\varpi=\sqrt{x^2+y^2}$, and azimuthal angle
$\varphi=\tan^{-1}(y/x)$. Coordinate time is denoted by $t$. Greek
indices $\mu, \nu, \cdots$ denote spacetime components ($x, y, z$, and
$t$), small Latin indices $i, j, \cdots$ denote spatial components
($x, y$, and $z$), and capital Latin indices $I, J, \cdots$ denote the
poloidal components ($\varpi$ and $z$).

\section{Formulation}

\subsection{Brief summary of methods}
\label{sec:formulation}

The formulation and numerical scheme for the present GRMHD simulations are
the same as those we reported in~\cite{DLSS,SS05}, to which the reader 
may refer for details.

The fundamental variables for the metric evolution are the
three-metric $\gamma_{ij}$ and extrinsic curvature $K_{ij}$. We adopt
the Baumgarte-Shapiro-Shibata-Nakamura (BSSN)
formalism~\cite{SN,gw3p2,BS,POT,bina2} to evolve the metric. In this
formalism, the evolution variables are the conformal factor $\psi
\equiv e^{\phi}\equiv \gamma^{1/12}$, the conformal 3-metric $\tilde
\gamma_{ij}=e^{-4\phi}\gamma_{ij}$, three auxiliary functions
$F_i\equiv \delta^{jk}\pa_{j} \tilde \gamma_{ik}$ (or
$\tilde{\Gamma}^i \equiv -\tilde \gamma^{ij}{}_{,j}$), the trace of
the extrinsic curvature $K$, and the tracefree part of the extrinsic
curvature $\tilde A_{ij} \equiv e^{-4\phi}(K_{ij}-\gamma_{ij} K/3)$.
Here, $\gamma={\rm det}(\gamma_{ij})$. The full metric $g_{\mu \nu}$
is related to the three-metric $\gamma_{\mu \nu}$ by $\gamma_{\mu \nu}
= g_{\mu \nu} + n_{\mu} n_{\nu}$, where the future-directed, timelike
unit vector $n^{\mu}$ normal to the time slice can be written in terms
of the lapse $\alpha$ and shift $\beta^i$ as $n^{\mu} = \alpha^{-1}
(1,-\beta^i)$.

The Einstein equations are solved in Cartesian coordinates. We employ
the Cartoon method~\cite{cartoon,shiba2d} to impose axisymmetry. In
addition, we also assume a reflection symmetry with respect to the
equatorial plane and only evolve the region with $x>0$ and $z>0$.  We
perform simulations using a fixed uniform grid with size $N \times 3
\times N$ in $x-y-z$ which, covering a computational domain $0 \leq x
\leq L$, $0 \leq z \leq L$, and $-\Delta \leq y \leq \Delta$. Here,
$N$ and $L$ are constants and $\Delta = L/N$. The variables in the
$y=\pm \Delta$ planes are computed from the quantities in the $y=0$
plane by imposing axisymmetry.

The lapse $\alpha$ and shift $\beta^i$ are gauge functions that have
to be specified in order to evolve the metric. In the code
of~\cite{SS05}, an approximate maximal slice (AMS) condition $K
\approx 0$ is adopted following previous papers
\cite{gw3p2,gr3d,bina2}. In this condition, $\alpha$ is determined by
solving approximately an elliptic-type equation.  The shift is
determined by the hyperbolic gauge condition as in~\cite{S03,STU}.  In
the code of~\cite{DLSS}, the lapse and shift are determined by
hyperbolic driver conditions as in~\cite{DSY04}.

The fundamental variables in ideal MHD are the rest-mass density 
$\rho$, specific internal energy $\varep$, pressure $P$, four-velocity 
$u^{\mu}$, and magnetic field $b^{\mu}$ measured by an 
observer comoving with the fluid. 
The ideal MHD condition is written as $u_{\mu} F^{\mu\nu}=0$, 
where $F^{\mu\nu}$ is the electromagnetic tensor. The tensor 
$F^{\mu\nu}$ and its dual in the ideal MHD approximation are 
given by 
\beqn
&&F^{\mu\nu}=\epsilon^{\mu\nu\alpha\beta}u_{\alpha}b_{\beta}, \label{eqFF}\\
&&F^*_{\mu\nu} \equiv {1 \over 2}\epsilon_{\mu\nu\alpha\beta} F^{\alpha\beta}
=b_{\mu} u_{\nu}- b_{\nu} u_{\mu}, 
\eeqn
where $\epsilon_{\mu\nu\alpha\beta}$ is the Levi-Civita tensor. 
The magnetic field measured by a normal observer $n^{\mu}$ is given by 
\beqn
B^{\mu} \equiv n_{\nu} F^{*\nu\mu}
=\alpha (u^t b^{\mu} - b^t u^{\mu}), \label{defB}
\eeqn 
From the definition of $B^{\mu}$ and the antisymmetry of 
$F^{\mu\nu}$ we have $n_{\mu} B^{\mu}= 0 = B^t$. 
The relations between $b^{\mu}$ and $B^i$ are 
\beqn
&& b^t = {B^j u_j \over \alpha},\\
&& b^i={1 \over \alpha u^t}\Big(B^i + B^j u_j u^i \Big).  
\eeqn

Using these variables, the energy-momentum tensor is written as
\beqn
T_{\mu\nu}=T_{\mu\nu}^{\rm Fluid} + T_{\mu\nu}^{\rm EM}, 
\eeqn
where $T_{\mu\nu}^{\rm Fluid}$ and $T_{\mu\nu}^{\rm EM}$ denote the
fluid and electromagnetic pieces of the stress-energy 
tensor. They are given by 
\beqn
&&T_{\mu\nu}^{\rm Fluid}=
\rho h u_{\mu} u_{\nu} + P g_{\mu\nu}, \\
&&T_{\mu\nu}^{\rm EM}=
F_{\mu\sigma} F^{~\sigma}_{\nu}-{1 \over 4}g_{\mu\nu}
F_{\alpha\beta} F^{\alpha\beta} \nonumber \\
&&~~~~~~=\biggl({1 \over 2}g_{\mu\nu}+u_{\mu}u_{\nu}\biggr)b^2
-b_{\mu}b_{\nu}. 
\eeqn
Here, $h\equiv 1+\varep+P/\rho$ is the specific enthalpy and
$b^2\equiv b^{\mu}b_{\mu}$. Hence, the total stress-energy tensor becomes
\beq
  T_{\mu\nu}= (\rho h + b^2) u_{\mu} u_{\nu} + \left( P + \frac{b^2}{2}
\right) g_{\mu\nu} - b_{\mu} b_{\nu} \ .
\eeq
The quantity $b^2$ is often referred to as 
the magnetic energy density, and $b^2/2=P_{\rm mag}$ as 
the magnetic pressure.

In our numerical implementation of the GRMHD and magnetic 
induction equations, 
we evolve the weighted density $\rho_*$, 
weighted momentum density $S_i$, 
weighted energy density $S_0$, 
and weighted magnetic field $\cB^i$. They are defined as 
\beqn
&&\rho_* \equiv - \sqrt{\gamma}\, \rho n_{\mu} u^{\mu}, 
\label{eq:rhos} \\
&& S_i \equiv -  \sqrt{\gamma}\, T_{\mu \nu}n^{\mu} \gamma^{\nu}_{~i}, \\
&& S_0 \equiv  \sqrt{\gamma}\, T_{\mu \nu}n^{\mu} n^{\nu}, 
\label{eq:S0} \\
&& \cB^i \equiv  \sqrt{\gamma}\, B^i. 
\eeqn 
During the evolution, we also need the three-velocity $v^i = u^i/u^t$. 

The GRMHD and induction equations are written in conservative form for
variables $\rho_*$, $S_i$, $S_0$, and $\cB^i$ and evolved using a
high-resolution shock-capturing scheme (HRSC). In the code
of~\cite{SS05}, the high-resolution central
(HRC)~\cite{kurganov-tadmor,SFont} scheme is employed. In this
approach, the transport terms such as $\pa_i (\cdots)$ are computed by
a Kurganov-Tadmor scheme with a third-order (piecewise parabolic)
spatial interpolation. It has been demonstrated~\cite{SFont} that the
results obtained using this scheme approximately agree with those
using an HRSC scheme based on Roe-type reconstruction for the
fluxes~\cite{Font,shiba2d}.  In the code of~\cite{DLSS}, the transport
terms are evaluated by the HLL (Harten, Lax and van-Leer) flux
formula~\cite{HLL}. The cell interface data reconstruction is done by
the monotonized central (MC) scheme~\cite{vL77}. It has also been
demonstrated that the performance of the HLL scheme is as good as the
Roe-type and HRC schemes in many MHD simulations~\cite{valencia}.  The
magnetic field $\cB^i$ has to satisfy the no monopole constraint
$\partial_i \cB^i=0$. In the code of~\cite{SS05}, this magnetic
constraint is imposed by using the constrained transport scheme first
developed by Evans and Hawley~\cite{EH,SS05}. In the code
of~\cite{DLSS}, the flux-interpolated constrained transport (flux-CT)
scheme~\cite{t00} is employed. Both constrained transport schemes
guarantee that no magnetic monopoles will be created in the
computation grid during the numerical evolution. At each timestep, the
primitive variables $(\rho,P,v^i)$ must be computed from the evolution
variables $(\rho_*,S_0,S_i)$. This is done by numerically solving the
algebraic equations~(\ref{eq:rhos})--(\ref{eq:S0}) together with the
EOS $P=P(\rho,\epsilon)$. As in many hydrodynamic simulations in
astrophysics, we add a tenuous ``atmosphere'' that covers the
computational grid outside the star. The atmospheric rest-mass density
is set to $\approx 10^{-10} \rho_c(0)$ ($\approx 1~{\rm g/cm^3}$),
where $\rho_c(0)$ is the initial central density of the star.

The codes used here have been tested in relativistic MHD simulations,
including MHD shocks, nonlinear MHD wave propagation, magnetized Bondi
accretion, MHD waves induced by linear gravitational waves, and
magnetized accretion onto a neutron star. Furthermore, we have used
both codes to perform simulations of the evolution of magnetized,
differentially rotating, relativistic, hypermassive neutron stars, and
obtain good agreement~\cite{DLSSS,DLSSS2}.

\subsection{Equations of state}

A parametric, hybrid EOS is adopted following M\"uller and
his collaborators \cite{Newton4,HD,OAM}. In this EOS,
the pressure consists of the sum of a cold part and a
thermal part:
\beq
P(\rho,\epsilon)=P_{\rm P}(\rho)+P_{\rm th}(\rho,\epsilon). \label{EOSII}
\eeq
The cold part of the pressure $P_{\rm P}$ depends only on the density.
In this paper, we choose the following form of $P_{\rm P}$:
\beqn
P_{\rm P}(\rho)=
\left\{
\begin{array}{ll}
K_1 \rho^{\Gamma_1}, & \rho \leq \rho_{\rm nuc}, \\
K_2 \rho^{\Gamma_2}, & \rho \geq \rho_{\rm nuc}, \\
\end{array}
\right.\label{P12EOS}
\eeqn
where $K_1$, $K_2$, $\Gamma_1$ and $\Gamma_2$ are constants, and 
$\rho_{\rm nuc} \approx 2\times 10^{14}~\rm{g}/\rm{cm}^3$ is 
nuclear density. 
We set $K_2=K_1\rho_{\rm nuc}^{\Gamma_1-\Gamma_2}$ to make $P_{\rm P}$ 
continuous at $\rho=\rho_{\rm nuc}$. The adiabatic indices are
chosen as $\Gamma_1$=1.3 or 1.32 and $\Gamma_2$=2.5 or 2.75. 
Following \cite{Newton4,HD}, the value of $K_1$ is chosen to
be $5\times 10^{14}$ in cgs units. With this choice, the cold 
part of the EOS for $\rho < \rho_{\rm nuc}$ is 
approximately given by relativistic degenerate electron 
pressure. This simplified cold EOS is designed to mimic a
more complicated cold nuclear EOS.
Using the first law of thermodynamics at zero temperature, 
we obtain the specific internal energy $\varepsilon_{\rm P}$ associated 
with the cold part of the pressure $P_{\rm P}$: 
\begin{eqnarray}
\varepsilon_{\rm P} (\rho) &=& 
-\int P_{\rm P}(\rho)\, d \left( \frac{1}{\rho}\right) \cr
&=& \left\{
\begin{array}{ll}
\displaystyle
{K_1 \over \Gamma_1-1} \rho^{\Gamma_1-1}, & \rho \leq \rho_{\rm nuc}, \\
\displaystyle 
{K_2 \over \Gamma_2-1} \rho^{\Gamma_2-1}
+{(\Gamma_2-\Gamma_1)K_1 \rho_{\rm nuc}^{\Gamma_1-1}
\over (\Gamma_1-1)(\Gamma_2-1)},  & \rho \geq \rho_{\rm nuc}. \\
\end{array}
\right.
\end{eqnarray}
The thermal part of the pressure $P_{\rm th}$ plays an important
role when shocks occur. We adopt a simple form for $P_{\rm th}$: 
\beq
P_{\rm th}=(\Gamma_{\rm th}-1)\rho \varepsilon_{\rm th}, 
\label{eq:Pth}
\eeq
where $\varepsilon_{\rm th}=\varepsilon - \varepsilon_P$ is the thermal 
specific internal energy.
The value of $\Gamma_{\rm th}$ determines the efficiency of converting 
kinetic energy to thermal energy at shocks. We set $\Gamma_{\rm th}=\Gamma_1$ 
to conservatively account for shock heating. 

\subsection{Diagnostics}

We monitor the total baryon rest mass $M_*$, ADM
(Arnowitt-Deser-Misner) mass $M$, and angular momentum $J$, which are
computed as in~\cite{shiba2d,DLSSS2}. 
We also compute the internal energy $E_{\rm int}$,
thermal internal energy $E_{\rm heat}$, rotational kinetic energy
$T_{\rm rot}$, gravitational potential energy $W$, and 
electromagnetic energy $E_{\rm EM}$ using the 
formulae given in~\cite{DLSSS2}.

In a stationary, axisymmetric spacetime, the energy $E$ and
angular momentum $J$ are conserved, where
\beqn
E &=&\int \alpha \sqrt{\gamma}\, T^t_{~t} d^3x,\\
J &=&\int \alpha \sqrt{\gamma}\, T^t_{~\varphi} d^3x.
\eeqn
For a nearly stationary spacetime, which is achieved after
the formation of the PNS, $E$ and $J$ are approximately conserved. 
We can then define the fluxes 
of rest mass, energy, and angular momentum across a sphere of coordinate 
radius $r$ by 
\beqn
&&F_M(r)= \oint_{r={\rm const}} dA \rho_* v^r,\\
&&F_E(r)=-\oint_{r={\rm const}} dA \alpha \sqrt{\gamma}\, T^r_{~t},\\
&&F_J(r)= \oint_{r={\rm const}} dA \alpha \sqrt{\gamma}\, T^r_{~\varphi},
\eeqn
where $dA=r^2 \sin\theta d\theta d\phi$. The energy and angular momentum 
fluxes associated with the electromagnetic fields are defined as 
\beqn
&&F_{E,{\rm EM}}(r)=-\oint_{r={\rm const}} dA \alpha \sqrt{\gamma}\,
T^{\rm EM}{}^r_{~t},\\
&&F_{J,{\rm EM}}(r)= \oint_{r={\rm const}} dA \alpha \sqrt{\gamma}\,
T^{\rm EM}{}^r_{~\varphi}. 
\eeqn
The total energy flux $F_E$ is very close to the rest-mass flux $F_M$
since $F_E$ is primarily composed of the rest-mass energy flow. Thus,
we define another energy flux by subtracting the rest-mass flow:
$F_e=F_E-F_M$. We note that $F_e$ contains kinetic, thermal,
electromagnetic, and gravitational potential energy fluxes. If $F_e >
0$ at sufficiently large radius, an unbound outflow (overcoming
gravitational binding energy) is present.

It should be noted that even in a stationary spacetime, one can choose
gauges such that $\partial /\partial t$ is not a Killing vector. In
this case, the energy $E$ defined above is not conserved. Thus, these
are physical meaningful fluxes only in certain gauges. In our
evolution, we find that after the formation of the PNS, the system
relaxes to a stationary state and the metric does not change
significantly with coordinate time $t$, suggesting that $\partial
/\partial t$ is an approximate Killing vector in our
simulations. Hence we use the above formulae to compute the fluxes.

\subsection{Gravitational waveforms in terms of quadrupole formula}
\label{GW}

We compute gravitational waves in terms of the quadrupole formula
given by~\cite{SS,SS04a}:
\beq
h_{ij}=P_{i}^{~k} P_{j}^{~l} \biggl(
{2 \over r}{d^2\bI_{kl} \over dt^2}\biggr),\label{quadf}
\eeq
where $\bI_{ij}$ and $P_i^{~j}=\delta_{ij}-\hat n_i \hat n_j$
($\hat n_i=x^i/r$)
denote the tracefree quadrupole moment and the TT projection tensor, 
respectively.
From this expression, the $+$-mode of quadrupole gravitational waves 
in an axisymmetric spacetime can be written as
\beq
h_+^{\rm quad} = {\ddot I_{xx}(t_{\rm ret}) - \ddot I_{zz}(t_{\rm ret})
\over r}\sin^2\theta, \label{quadr}
\eeq
where $I_{ij}$ denotes the quadrupole moment, $\ddot I_{ij}$ is
its second time derivative, $\theta$ denotes the angle between the
rotation axis and the direction of observation, and
$t_{\rm ret}$ is retarded time 
$t_{\rm ret}\approx t - r$. In this paper, we characterize the 
gravitational waves by the variable 
$A_2(t)\equiv \ddot I_{xx}(t_{\rm ret}) - \ddot I_{zz}(t_{\rm ret})$, 
which has dimensions of length. In terms of $A_2$, the observed 
gravitational-wave strain is given by
\beqn
h=1 \times 10^{-20} \biggl({A_2 \over 3.1~{\rm m}}\biggr)
\biggl({10~{\rm kpc} \over r}\biggr) \sin^2\theta. \label{hamplitude}
\eeqn

In spacetimes with strong gravitational fields, 
there is no unique definition for the quadrupole moment.
Here, we choose for simplicity 
\beq
I_{ij} = \int \rho_* x^i x^j d^3x. 
\eeq
Using the continuity equation, 
we compute the first time derivative as 
\beq
\dot I_{ij} = \int \rho_* (v^i x^j +x^i v^j)d^3x.
\eeq
We then compute $\ddot I_{ij}$ by finite differencing the 
numerical values of $\dot I_{ij}$ in time. 

Because of the ambiguity in the definition of the quadrupole moment in
a spacetime with strong gravitational fields, one may choose
alternative expressions for $I_{ij}$. Hence, the gravitational
waveforms determined from the quadrupole formula depend on the chosen
definition of $I_{ij}$. In addition, they depend on the gauge
conditions, since the coordinates $r$ and $t$ appeared in $I_{ij}$ and
$\dot I_{ij}$ are gauge dependent. We have calibrated our quadrupole
formula in~\cite{SS}, and found that the magnitude of the error in the
amplitude of gravitational waves is of order $M/R$, where $M$ and $R$
denote the typical mass and radius of the emitter. On the other hand,
the phase of the waveforms is quite accurate. Thus, the wave
amplitudes shown in this paper are not accurate to better than
$\sim$10\%, but the computed radiation possesses the correct 
qualitative features.

We also note that in our formula, the contribution from
$T_{\mu\nu}^{\rm EM}$ is neglected. This is justified in the present
treatment, since the contribution from the electromagnetic part is
about 10\% of the matter part, and hence, the error in neglecting the
former is as large as the error of our quadrupole formula.

\section{Magnetorotational effects}

In this section, we summarize processes that play an important role in
magnetorotational core collapse and the subsequent evolution of the
PNS.

\subsection{Compression and magnetic winding during collapse}
\label{sec:compress}

The magnetic field is amplified by compression and magnetic winding during
stellar collapse. This can be understood by considering the magnetic 
induction equation in a perfectly conducting (ideal MHD) plasma: 
\beq
\pa_t \cB^i + \partial_j (v^j \cB^i - v^i \cB^j)=0 . \label{induction}
\eeq
Combining Eq.~(\ref{induction}) with the continuity equation 
\beqn
{d \rho_* \over dt}+ {\rho_* \over \varpi} \pa_I (\varpi v^I)=0, 
\label{eq:continuity}
\eeqn
we rewrite the induction equation as 
\beqn
&& {d\cb^I \over dt} =\cb^J \pa_J v^I,\label{eqb1}\\
&& {d\cb^{\varphi} \over dt} =\cb^J \pa_J \Omega, \label{eqb2}
\eeqn
where $\cb^i \equiv \cB^i/\rho_*$, $v^{\varphi}=\Omega$, and
\beq
{d\over dt} \equiv {\pa \over \pa t}+v^I{\pa \over \pa x^I}. 
\eeq

For a pre-collapse star having moderate angular momentum, the 
core collapses in an approximately spherical and homologous manner.
Hence $v^r \propto r$, which implies $v^{\varpi} \propto \varpi$, $v^z
\propto z$, and $|v^{\varpi}| \approx |v^z|$. It follows from the 
continuity equation that 
\beqn
{d \rho_* \over dt} \approx -3\rho_* {v^{\varpi} \over \varpi}. 
\eeqn
Note that $v^{\varpi}$ and $v^z$ are negative. Combining the above 
equation with Eq.~(\ref{eqb1}), we obtain 
\beqn
&& {d\cb^{\varpi} \over dt} \approx \cb^{\varpi}
{v^{\varpi} \over \varpi}
\sim - {\cb^{\varpi} \over 3}{d \ln \rho_* \over dt},\\
&& {d\cb^{z} \over dt} \approx \cb^{z}
{v^{\varpi} \over \varpi}  
\sim - {\cb^{z} \over 3}{d \ln \rho_* \over dt}. 
\eeqn
Hence for an observer comoving with the fluid, 
\beqn
\cb^I \propto \rho_*^{-1/3} \sim \rho^{-1/3}, 
\label{eq:40}
\eeqn
which means that the poloidal field $B^I \propto \rho^{2/3}$ during 
core collapse.

While the collapse proceeds roughly homologously in the bulk
of the star, this is not the case in the outer layers.
Thus, significant differential rotation develops in the outer 
layers. We see from Eq.~(\ref{eqb2}) that toroidal magnetic 
fields are amplified when there is differential rotation along 
the poloidal field lines (magnetic winding).  The toroidal
field is thus expected to grow during the collapse.  However,
since the growth depends on the non-homologous nature of the 
collapse in the outer layers, a simple prediction for the 
dependence of $\cb^{\varphi}$ on $\rho$ [as in Eq.~(\ref{eq:40})] 
is not available.

To obtain a rough idea of how the toroidal field evolves 
during the collapse, we proceed as follows.  Assume that 
$\Omega$ in the differentially rotating outer regions is described 
by $\varpi^{-p}$, where $p$ is constant with time.  This will
not apply in regions where the homology is seriously violated. 
However, the true behavior of $\Omega$ in the region of interest
is not known analytically, and we are only seeking a rough estimate. 
Equation~(\ref{eqb2}) gives 
\beq
{d\cb^{\varphi} \over dt} \approx -p 
{\cb^{\varpi} \Omega \over \varpi}.
\eeq
The evolution of $\cb^{\varphi}$ measured by the
comoving observer is given by 
\beqn
\cb^{\varphi} \approx -\int dt\, p 
{\cb^{\varpi} \Omega \over \varpi}. \label{eqb3}
\eeqn
For a comoving observer, $\varpi^{-1} \propto \rho^{1/3}$ and hence
$\cb^{\varpi}/\varpi$ is approximately constant according 
to Eq.~(\ref{eq:40}). (This again assumes homologous collapse, which 
does not strictly hold in the regions of interest for the toroidal 
field growth.)  When the magnetic field is weak and the spacetime is 
axisymmetric, the specific angular momentum of a fluid particle
$j=hu_{\varphi} \approx \Omega \varpi^2$ is approximately conserved, 
which implies $\Omega \propto \varpi^{-2} \propto \rho^{2/3}$. 
We then have 
\beqn
\cb^{\varphi} \propto \int \Omega dt \propto \int \rho^{3/2} dt . 
\label{eqb4}
\eeqn
Next, we obtain an approximate relation between $dt$ and $d\rho$ 
from the following Newtonian analysis. Since the collapse is 
approximately spherical and homologous, conservation of energy 
implies 
\beq
 \frac{1}{2}v^2 - \frac{m(r)}{r} = -\frac{m(r_0)}{r_0}  ,
\eeq
where $r_0$ is the radius of the fluid particle at the onset of 
collapse ($t=0$). The mass interior to the radius $r$, $m(r)=m(r_0)=
(4\pi/3) r_0^3 \rho_0$ is independent of time during 
homologous collapse. Here $\rho_0$ is 
the density of the fluid particle at $t=0$. Setting $v=-dr/dt$, 
we have 
\beq
  \frac{dr}{dt} = -\sqrt{m(r_0) \left( \frac{1}{r}-\frac{1}{r_0} \right)} . 
\eeq
Combining the above equation with $\rho = \rho_0 (r_0/r)^3$, we obtain 
\beq
  dt \propto \rho_0^{-1/2} \frac{d\rho/\rho_0}{(\rho/\rho_0)^{4/3}
\sqrt{ (\rho/\rho_0)^{1/3} - 1} }. \label{eq:dt}
\eeq
Substituting Eq.~(\ref{eq:dt}) into Eq.~(\ref{eqb4}), we 
have $\cb^{\varphi} \propto \sqrt{(\rho/\rho_0)^{1/3}-1}$, and 
hence we provide the following estimate of the toroidal field
during the collapse:
\beq
  B^{\varphi} \propto \rho \  \sqrt{ \left( \frac{\rho}{\rho_0} 
\right)^{1/3} - 1} \ .
\eeq
When $\rho \gg \rho_0$, the above equation simplifies to 
\beqn
\cb^{\varphi} \propto \rho^{1/6}~~{\rm and}~~
B^{\varphi} \propto \rho^{7/6}. \label{eqbb}
\eeqn
In practice, we find that $B^{\varphi} \propto \rho^q$, 
where $q$ is slightly less than 1.  However, with the
assumptions that went into Eq.~(\ref{eqbb}), only rough
agreement is to be expected.  

\subsection{PNS magnetic winding}
\label{sec:winding}

When the central density of the collapsing star exceeds the nuclear
density, the core bounces due to the stiff nuclear EOS. The star
quickly settles down to a quasistationary state and a PNS is
formed. In this quasistationary state, the toroidal magnetic field
grows linearly with time. This can be seen from the induction
equation~(\ref{induction}). If the magnetic field is weak and has a
negligible back-reaction on the fluid, the velocities will remain
constant with time.  In cylindrical coordinates, we have (assuming
axial symmetry)
\beqn
&& \partial_t \cB^I \approx 0 ,\\
&& \partial_t \cB^{\varphi} 
\approx {1 \over \varpi} \partial_I(\varpi \Omega \cB^I) ,
\eeqn
where we have assumed that $|v^{\varpi}|\ll \varpi\Omega$ and 
$|v^z| \ll \varpi\Omega$ when the PNS is in quasiequilibrium. Then, 
\beqn
\pa_t \cB^{\varphi} \approx \cB^I \partial_I\Omega, \label{eqn:dtBphi_int}
\eeqn
where we used the no-monopole constraint [$\pa_I (\varpi \cB^I)=0$
in axisymmetry]. During the early phase of the PNS evolution,
Eq.~(\ref{eqn:dtBphi_int}) indicates that the toroidal component of
the field $\cB^T(=\varpi \cB^{\varphi})$ grows linearly according to
\beqn
&&|\cB^T(t;\varpi,z)| \nonumber \\
&&\approx |\cB^T(t_p;\varpi,z)| + 
t\varpi |\cB^I(t_p;\varpi,z)\partial_I\Omega(t_p;\varpi,z)| \cr
&& \approx |\cB^T(t_p;\varpi,z)| + {3\pi t \over P_{\rm rot}} 
|\cB^{\varpi}(t_p;\varpi,z)| \nonumber \\
&&= |\cB^T(t_p;\varpi,z)| \nonumber \\
&& + 10^{15} \biggl({t \over 100~{\rm ms}}\biggr)
\biggl({P_{\rm rot} \over 1~{\rm ms}}\biggr)^{-1} 
\biggl({|\cB^{\varpi}(t_p;\varpi,z)| \over 10^{12}~{\rm G}}\biggr)~{\rm G},
\label{eqn:dtBphi}
\eeqn
where $t_p$ is the time at which the PNS first settles down to a quasiequilibrium 
state, $P_{\rm rot}$ denotes the local rotation period, and we have 
assumed a Keplerian angular velocity profile
$\Omega \propto \varpi^{-3/2}$.
The growth of $B^T$ is expected to deviate from this linear relation 
when the magnetic tension is large enough to change the angular 
velocity profile of the fluid (magnetic braking).
Magnetic braking transports angular momentum 
on the Alfv\'en timescale~\cite{spitzer,spruit,BSS,Shapiro}:
\beqn
t_A  &\sim & \frac{R}{v_A} \approx 10~{\rm ms}
\left( \frac{|\cB^{\varpi}|}{10^{15}~{\rm G}} \right)^{-1}  \nonumber \\
&&~~~~~~~~~\times \left( \frac{R}{10~{\rm km}}\right)
\left( \frac{\rho}{10^{14}~{\rm g/cm^3}}\right)^{1/2} \ ,
\label{eq:alfventime}
\eeqn
where $R$ is the characteristic radius of the PNS
and $v_A \approx |B|/\rho^{1/2}$ is the Alfv\'en speed. 
So Eq.~(\ref{eqn:dtBphi}) holds for $t_p \lesssim t \lesssim t_p + t_A$.

\subsection{Magnetorotational (shear) instability}\label{sec:MRI}

The MRI is present in a weakly magnetized, rotating fluid wherever
$\partial_{\varpi} \Omega < 0$~\cite{MRI0,MRI,MRIrev}.  When the
instability reaches the nonlinear regime, the distortions in the
magnetic field lines and velocity field lead to turbulence.  To
estimate the growth time scale $t_{\rm MRI}$ and the wavelength of the
fastest-growing mode $\lambda_{\rm max}$, we can use a Newtonian local
linear analysis given in~\cite{MRIrev}.  Linearizing the
MHD equations for a local patch of a rotating fluid and imposing a
plane-wave dependence ($e^{i({\bf k}\cdot{\bf x} - \omega t)}$) on the
perturbations leads to the dispersion relation given in Eq.~(125)
of~\cite{MRIrev}. Specializing this equation for a constant-entropy
medium leads to the dispersion relation
\beqn
&&\omega^4 - [2({\bf k}\cdot {\bf v}_A)^2 + \kappa^2]\omega^2  \nonumber \\
&&~~~~~+({\bf k}\cdot {\bf v}_A)^2[({\bf k}\cdot {\bf v}_A)^2 + \kappa^2 - 
4\Omega^2] = 0 \ ,
\label{disp}
\eeqn
where ${\bf v}_A = {\bf B}/\sqrt{\rho}$ is the (Newtonian)
Alfv\'en velocity vector and 
$\kappa$ is the epicyclic frequency of Newtonian theory:
\beq
\kappa^2 \equiv \frac{1}{\varpi^3}\frac{\partial(\varpi^4\Omega^2)}
{\partial \varpi}.
\eeq
Equation~(\ref{disp}) is modified for a medium of inhomogeneous
entropy~\cite{MRIrev}.
In this section, we neglect any entropy gradients and focus on the 
effects of shear.
We further simplify the analysis by considering only the vertical modes 
(${\bf k} = k{\bf e}_z$)
since these are likely to be the dominant modes. 
The value of $\omega^2$ can be found by solving Eq.~(\ref{disp}) 
and then minimized to
obtain the frequency of the fastest-growing mode, $\omega_{\rm max}$: 
\beq
\omega^2_{\rm max} = 
-\frac{1}{4}\left(\frac{\partial \Omega}{\partial \ln \varpi}\right)^2 .  
\eeq 
This maximum growth rate corresponds to 
\beq
(k v_A^z)^2_{\max} = \Omega^2-{\kappa^4 \over 16\Omega^2}.
\eeq
The growth time ($e$-folding time) and wavelength of the fastest-growing
mode are then given by 
\beqn
t_{\rm MRI} & = & 1/(i\omega_{\rm max}) = 
2 \left|\partial \Omega/\partial \ln \varpi \right|^{-1} 
\ ,  \label{tmri} \\
\lambda_{\rm max} & = & \frac{2\pi}{k_{\rm max}} = {2\pi v_A^z \over \Omega}
\biggl[1-\biggr({\kappa \over 2\Omega}\biggr)^4\biggr]^{-1/2}. 
\eeqn
For a Keplerian angular velocity distribution $\Omega \propto \varpi^{-3/2}$, 
we have 
\beqn
t_{\rm MRI} &=& {4 \over 3\Omega} \approx 1.3~{\rm ms} \ 
\left(\frac{\Omega}{1000~{\rm rad~s}^{-1}}\right)^{-1} , \label{tmri2} \\
\lambda_{\rm max} &=&  {8\pi v_A^z \over \sqrt{15}\Omega}  
\nonumber \\
& \approx & 2.1~{\rm km} \ 
\left(\frac{\Omega}{1000~{\rm rad~s}^{-1}}\right)^{-1} \nonumber \\
&&~~~~~~~\times \left(\frac{B^z}{10^{14} \ {\rm G}}\right)
\left(\frac{\rho}{10^{13}~{\rm g/cm^3}}\right)^{-1/2} . \ \ 
\label{lambdamax}
\eeqn 

If $B^z$ is comparable to the field strength of a canonical pulsar
($B^z \sim 10^{12}$ G), 
$\lambda_{\rm max}$ is much smaller than the typical 
radius $R$ of a PNS. Since $\lambda_{\rm max} \propto v_A$, larger
magnetic fields will result in longer MRI wavelengths.  When
$\lambda_{\rm max} \agt 100$ km, the MRI will be suppressed since the
unstable perturbations will no longer fit inside the region with a high
degree of differential rotation. Hence the MRI is regarded as
a weak-field instability. In this paper, the initial magnetic
field strength is chosen so that the field strength in the resulting 
PNS is $\sim 10^{14}$~G, giving $\lambda_{\rm max} \sim $ a few km. 

Unlike $\lambda_{\rm max}$, $t_{\rm MRI}$ does not depend on the magnetic field
strength but on the angular velocity profile. The Newtonian
local analysis indicates that the MRI always grows on the timescale 
of a rotation period for a configuration with $\partial \ln \Omega / 
\partial \ln \varpi \sim -1$. Hence, the MRI is expected to play an 
important role for a PNS, especially near its surface, which is in 
general differentially
rotating.  The resulting strong magnetic fields and turbulence tend to
transport angular momentum from the rapidly rotating inner
region of the PNS to the more
slowly rotating outer layers.  This causes the inner part to contract
and the outer layers to expand. We note, however, that once the magnetic 
field strength is saturated, the resulting
angular momentum transport and matter outflow will be governed
by the turbulence and is thus expected to occur on a time scale
longer than $t_{\rm MRI}$ (e.g., a turbulent transport time scale).

\subsection{MHD outflow by the magneto-spring effect}\label{sec:US}

After the formation of the PNS, MHD outflows may 
transport angular momentum outward (causing a spindown of the 
PNS) and may power 
a supernova explosion~\cite{WMW,AHM}. In the early stage of the 
evolution of the PNS, the MHD outflow may be 
driven by the toroidal magnetic field, which grows linearly due 
to magnetic winding according to Eq.~(\ref{eqn:dtBphi}). 
A helical magnetic field forms as a result. The rate of change 
of the magnetic energy associated with the toroidal field is 
\beqn
\dot{E}_{\rm EM}&& \sim {d (B^T)^2
\over dt} =2 B^T \dot B^T \nonumber \\
&&\approx 2 B^T B^{\varpi} \varpi
\pa_{\varpi} \Omega \approx -3 B^T B^{\varpi} \Omega,
\eeqn
where we have assumed a Keplerian angular velocity profile 
$\Omega \propto \varpi^{-3/2}$. We note that
$B^T B^{\varpi}$ should be negative 
[see Eqs.~(\ref{eqb2}) and~(\ref{eqn:dtBphi_int})]. Because of the growing 
magnetic pressure and hoop stress, material is lifted from
the PNS along the rotation axis,
producing a tower-like MHD outflow 
(see e.g.,~\cite{US,KHM,RUKL}). 

If a significant amount of energy gained from magnetic winding 
is used to drive the MHD outflow, the rate of energy loss from 
the PNS is approximately 
\beqn
\dot{E} \sim E_{\rm EM} \Big|{B^{\varpi} \over B^T}\Big| \Omega,
\eeqn
where we have assumed that the magnetic energy is dominated by the 
toroidal field $B^T$. 
This process is likely to be important only in the early stage 
before the toroidal field saturates.

\subsection{MHD outflow by magneto-centrifugal effect}\label{sec:BP}

Another type of MHD outflow may occur in a 
differentially rotating object penetrated both by poloidal and
toroidal field lines. Blandford and Payne showed 
that when a stationary disk is penetrated by open magnetic fields 
and when a mechanism for matter ejection operates at the disk surface, 
a stationary magneto-centrifugal outflow is launched~\cite{BP,AHM} 
(see also relevant computational work, e.g., \cite{BP1,BP2}).

We first consider the magneto-centrifugal effect operating on an 
accretion disk. 
To estimate the order of magnitude of the energy flux in the outflow,
we evaluate the Poynting flux approximately. We assume that the
disk rotates with an angular velocity $\Omega_d$ and that the
poloidal velocity is smaller than the rotational velocity. 
The Poynting flux along the poloidal field line is
$\sim |\ve{E} \times \ve{B}| \sim R_d \Omega_d B^P B^T$, 
where $R_d$ is the characteristic radius of the disk, 
$B^P\sim B^{\varpi}$ is the magnitude of the poloidal magnetic field, and
the electric field $\ve{E}$ is calculated by the (Newtonian) ideal 
MHD condition $|\ve{E}|=|\ve{v} \times \ve{B}| \sim R_d \Omega_d B^P$.
The energy generation rate is then given by
\beqn
\dot E_{\rm Poyn}
&=&({\rm Poynting~flux})\times ({\rm area})\nonumber \\
&\sim & \eta_d B^P B^T R_d^3 \Omega_d \nonumber \\
&\propto & E_{\rm EM} C_B
\Big({R_d \over H_d}\Big) \Omega_d,
\label{dotE}
\eeqn
where $\eta_d$ is a constant of order unity, $H_d$ is the thickness
of the disk, and $C_B$ denotes a typical ratio of $|B^{P}|$ to
$|B^T|$. Thus, the rate of energy loss of the disk is similar to that of
the magneto-spring mechanism.

In the Blandford and Payne model~\cite{BP}, the Poynting flux is 
accompanied by a kinetic energy flux of comparable amplitude from 
matter outflow in the vicinity of the disk.
Hence, the total rate of energy loss 
from the disk can be written in the same form of the last line of
Eq.~(\ref{dotE}).

The same qualitative analysis may be applied to a PNS by 
setting $H_d \sim R_d$, giving the energy loss 
rate as 
\beqn
\dot E=\eta_* E_{\rm EM} C_B \Omega, \label{dotE2}
\eeqn
where $\Omega$ is the typical angular velocity of the PNS and 
$\eta_*$ is a constant of order unity. 

Assuming that the main source for the MHD outflow (either by
the magneto-spring or magneto-centrifugal mechanism) is the rotational
kinetic energy of the star, $\dot E$
is related to the angular momentum flux $\dot J$ approximately by
\beqn
\dot E \approx \Omega\dot J,
\eeqn
and hence,
\beqn
\dot J \approx \eta_* E_{\rm EM}C_B. \label{dotJE}
\eeqn

For an MHD outflow, $\dot E$ and $\dot J$ should be approximately 
equal to the fluxes integrated over a closed surface far away from  
the source. Thus, in this paper, we calculate $F_M$, $F_e$, and
$F_J$ to obtain the rate of loss of the energy and angular momentum
of the PNS.

In the model of Blandford and Payne~\cite{BP}, a priori mass ejection
from the central object at a launching velocity as large
as the rotational velocity is required to drive the MHD outflow.
Thompson et al.~\cite{TCQ} suggest that a neutrino wind could be the
engine of the initial velocity. Here we anticipate a purely MHD
source. In the present context, the magnetic field in the vicinity of
the PNS is amplified by magnetic winding and 
the MRI. The resulting Poynting flux propagates outward and 
injects the energy into the Blandford-Payne wind. 

\section{Pre-collapse stellar models}

\begin{table*}[tb]
\begin{center}
\caption{Central density $\rho_c$, baryon rest mass $M_*$, ADM mass $M$,
equatorial radius $R$, ratio of the rotational kinetic energy to the
potential energy $T_{\rm rot}/W$, non-dimensional angular momentum
parameter $J/M^2$, central
value of the lapse function $\alpha_c$, angular velocity at the rotation
axis $\Omega_c$,
ratio of the angular velocity at the rotation axis to that at the
equatorial surface $\Omega_c/\Omega_s$,
ratio of polar to equatorial radii $z_s/R$, and $\hat A=\varpi_d/R$ of
the pre-collapse stars.}
\begin{tabular}{cccccccccccc} \hline
Model & $\rho_c~({\rm g/cm^3})$ & $M_*(M_{\odot})$
& $M(M_{\odot})$ & $R$~(km) & $T_{\rm rot}/W$ & $J/M^2$ & $\alpha_c$
& $\Omega_c$ (rad/s) &$\Omega_c/\Omega_s$ & $z_s/R$ & $\hat A$ \\ \hline
A & $1.00 \times 10^{10}$  & 1.503 & 1.503 & 2267 &
$8.9 \times 10^{-3}$ & 1.235  & 0.994 & 4.11 & 1.0 & 0.667 & $\infty$ \\ 
B & $1.00 \times 10^{10}$  & 1.486 & 1.486 & 1445 &
$4.9 \times 10^{-3}$ & 0.909  & 0.994 & 3.12 & 1.0 & 0.883 & $\infty$ \\ 
C & $1.00 \times 10^{10}$  & 1.504 & 1.504 & 1628 &
$9.0 \times 10^{-3}$ & 1.187  & 0.994 & 6.31 & 5.0 & 0.917 & 0.5 \\ \hline
\end{tabular}
\end{center}

\vspace{-5mm}
\end{table*}

The pre-collapse stars are modeled as rotating $\Gamma=4/3$ polytropes. 
Following \cite{HD,SS04a}, the central density is chosen to be
$\rho_c = 10^{10}~{\rm g/cm^3}$. The EOS is given by 
\beq
P=K_0 \rho^{4/3},\label{EOS43}
\eeq
where $K_0$ is set to be $5 \times 10^{14}$ in cgs units. 
This EOS corresponds to the degenerate pressure of ultra-relativistic 
electrons~\cite{ST}. We adopt a commonly used angular velocity 
profile~\cite{KEH,Ster}:
\beq
u^t u_{\varphi} = \varpi_d^2( \Omega_c - \Omega ),
\eeq
where $\Omega_c$ denotes the angular velocity along the rotation axis,
and $\varpi_d$ is a constant. 
In the Newtonian limit, this rotation
law reduces to the so-called `$j$-constant' law:
\beq
\Omega = \Omega_c{\varpi_d^2 \over \varpi^2 + \varpi_d^2}.
\eeq
Hence the parameter $\varpi_d$ controls the degree of differential 
rotation of the star. In this paper, we choose rigidly rotating cases
($\varpi_d \rightarrow \infty$) and a moderately differentially rotating case
with $\hat A \equiv \varpi_d/R=0.5$, where $R$ is the
equatorial radius. In the rigidly rotating cases,
we choose the ratios of polar to equatorial radii, $z_s/R$,
to be $0.667$ and 0.883. With $z_s/R=0.667$,
the angular velocity at the equatorial surface
is approximately equal to the Keplerian velocity (i.e., 
a uniformly rotating star at the mass-shedding limit).
These two models with rapid and moderate rotation are referred to as
models A and B, respectively. For the differentially rotating
case, we choose a model with the ratio of the rotational kinetic energy
$T_{\rm rot}$ to the gravitational potential energy $W (>0)$ of 
$\approx 0.009$, which is approximately the same as that of model A.
In Table~I we summarize the parameters of our models.

To induce collapse, we fix the density distribution of the star 
but recalculate the pressure $P$ and specific internal energy 
$\varepsilon$ using the hybrid EOS [Eqs.~(\ref{EOSII})--(\ref{eq:Pth})]. 
We set 
$K_1=K_0\rho_0^{4/3-\Gamma_1}$, where $\rho_0=1~{\rm g/cm^3}$. 
In a previous study of core collapse~\cite{SS04a}, the pressure 
is computed by $P=K_1 \rho^{\Gamma_1}$ following~\cite{HD}. 
The specific internal energy is then fixed by the hybrid EOS to be 
(note that we choose $\Gamma_{\rm th}=\Gamma_1$)
\beqn
\varepsilon={K_1 \over \Gamma_1 -1} \rho^{\Gamma_1-1}.
\eeqn
However, this choice reduces $\varepsilon$ by $\agt 50\%$ from 
the equilibrium value
for $\Gamma_1=1.3$, and the collapse is strongly accelerated in the
early times. As a result, the collapse is strongly non-homologous and hence not
very realistic. Therefore, in this paper, we do not decrease $\varepsilon$,
i.e., we set it according to 
\beqn
\varepsilon=3 K_0 \rho^{1/3}. 
\eeqn
The pressure is then determined by the hybrid EOS to be 
\beqn
  P = 3 K_0 (\Gamma_1-1) \rho^{4/3},
\eeqn
which is reduced uniformly from its initial value [cf.\ Eq.~(\ref{EOS43})] 
by a factor of $1-3 (\Gamma_1-1)$ (i.e., 10\% for $\Gamma_1=1.3$).
With this choice, the collapse is more uniform, although the collapse
time is longer. 

Next, we add a magnetic field to the stars.
Since the initial profile of the magnetic field in the core collapse
progenitor is not known, we follow our previous 
papers~\cite{DLSSS,DLSSS2} and add a dipole-like poloidal magnetic
field to the pre-collapse model by introducing a vector potential of
the following form:
\beqn
A_{\varphi}=A_b \varpi^2 {\rm max}[\rho^{1/n_b}-\rho^{1/n_b}_{\rm cut}, 0].
\label{eqnb}
\eeqn
Here the cutoff density $\rho_{\rm cut}$ is chosen as 
$10^{-4} \rho_c=10^6~{\rm g/cm^3}$. (Simulations have been performed for 
a different cutoff density,
$10^8~{\rm g/cm^3}$, and we find that the results depend only weakly on
this value.) The parameter $A_b$ determines the initial strength
of the magnetic field (see below), whereas $n_b$ in Eq.~(\ref{eqnb}) 
determines the initial profile of the magnetic field
lines. We choose $n_b=1$ and 4. In Fig.~1, we show the
density contour curves and the magnetic field lines (which 
coincide with contours of $A_{\varphi}$ in axisymmetry) for model~A with
$n_b=1$ (left) and with $n_b=4$ (right). This figure shows that for
the larger value of $n_b$, the maximum of the magnetic field strength
$B_{\rm max}$ is located at a larger cylindrical radius. This
results in a larger magnetic energy and a larger value of $(P_{\rm
mag}/P)_{\rm max}$ for a given value of $B_{\rm max}$.  

We choose $A_b$ such that the $z$-component of the magnetic field is
$7\times 10^{12}$--$4 \times 10^{13}$ G. We summarize in 
Table~II the maximum
value of $|B^z|$, the ratio of the magnetic energy to the rotational
kinetic energy $E_{\rm EM}/T_{\rm rot}$, and the maximum value of
$P_{\rm mag}/P$ for the models we consider. With these
values of $A_b$, the strength of the
poloidal magnetic field of the resulting PNSs can be $\agt
10^{15}$ G and the value of $\lambda_{\rm max}$ for the
fastest-growing mode of the MRI becomes $\agt 1$ km, which can be
resolved with our computational resources.  Furthermore, the
Alfv\'en timescale in the PNSs is short enough ($\alt 30$ ms)
to see the whole magnetic braking process with a reasonable 
computational cost.

Although the initial value of the magnetic field strength is large
(much larger than that of a presupernova stellar model~\cite{HWS} in
which the magnetic field is assumed to be amplified primarily by a
dynamo process), the magnetic energy is only $\sim 0.1$--5\% of the
rotational kinetic energy, and is an even smaller fraction of the
gravitational potential energy and the internal energy.  Even if the
simulation starts with a smaller magnetic field strength, the fields
are amplified globally by magnetic winding and locally by the MRI
until they saturate (see Sec.~\ref{sec:results}). Thus, the
qualitative features of the evolution of the PNS may not depend
strongly on the initial magnetic field strength.

According to the observed number ratio of magnetars to canonical
pulsars, the lower limit of the Galactic magnetar birth rate exceeds
0.01/year, which is as large as that of the canonical
pulsars~\cite{AXP}. Namely, neutron stars with magnetic field strength
$\sim 10^{15}$~G are not rare in the Universe. The origin of magnetars
is not clear. A simple hypothesis is that the stellar progenitor of a
magnetar has a strong magnetic field. The simulations in this paper
provide such a model for the formation of magnetars.

\begin{table}[tb]
\begin{center}
\caption{Parameters of the initial magnetic fields.}
\begin{tabular}{ccccc} \hline
Model & $n_b$ & $|B^z|_{\rm max}$(G) & $E_{\rm EM}/T_{\rm rot}$ &
$(P_{\rm mag}/P)_{\rm max}$ \\ \hline
A0 &---& 0 & 0     & 0 \\ 
A1 &1& $7.2\times 10^{12}$ & $1.9\times 10^{-3}$ & $2.1\times 10^{-4}$\\
A2 &1& $1.8\times 10^{13}$ & $1.2\times 10^{-2}$ & $1.3\times 10^{-3}$\\
A3 &1& $3.6\times 10^{13}$ & $4.8\times 10^{-2}$ & $5.2\times 10^{-3}$\\
A4 &4& $7.2\times 10^{12}$ & $2.3\times 10^{-2}$ & $1.7$\\
B  &1& $1.8\times 10^{13}$ & $2.1\times 10^{-2}$ & $1.3\times 10^{-3}$\\
C  &1& $7.2\times 10^{12}$ & $1.9\times 10^{-3}$ & $2.1\times 10^{-4}$\\\hline
\end{tabular}
\end{center}
\vspace{-5mm}
\end{table}

\begin{figure*}[t]
\begin{center}
\epsfxsize=3.in
\leavevmode
\epsffile{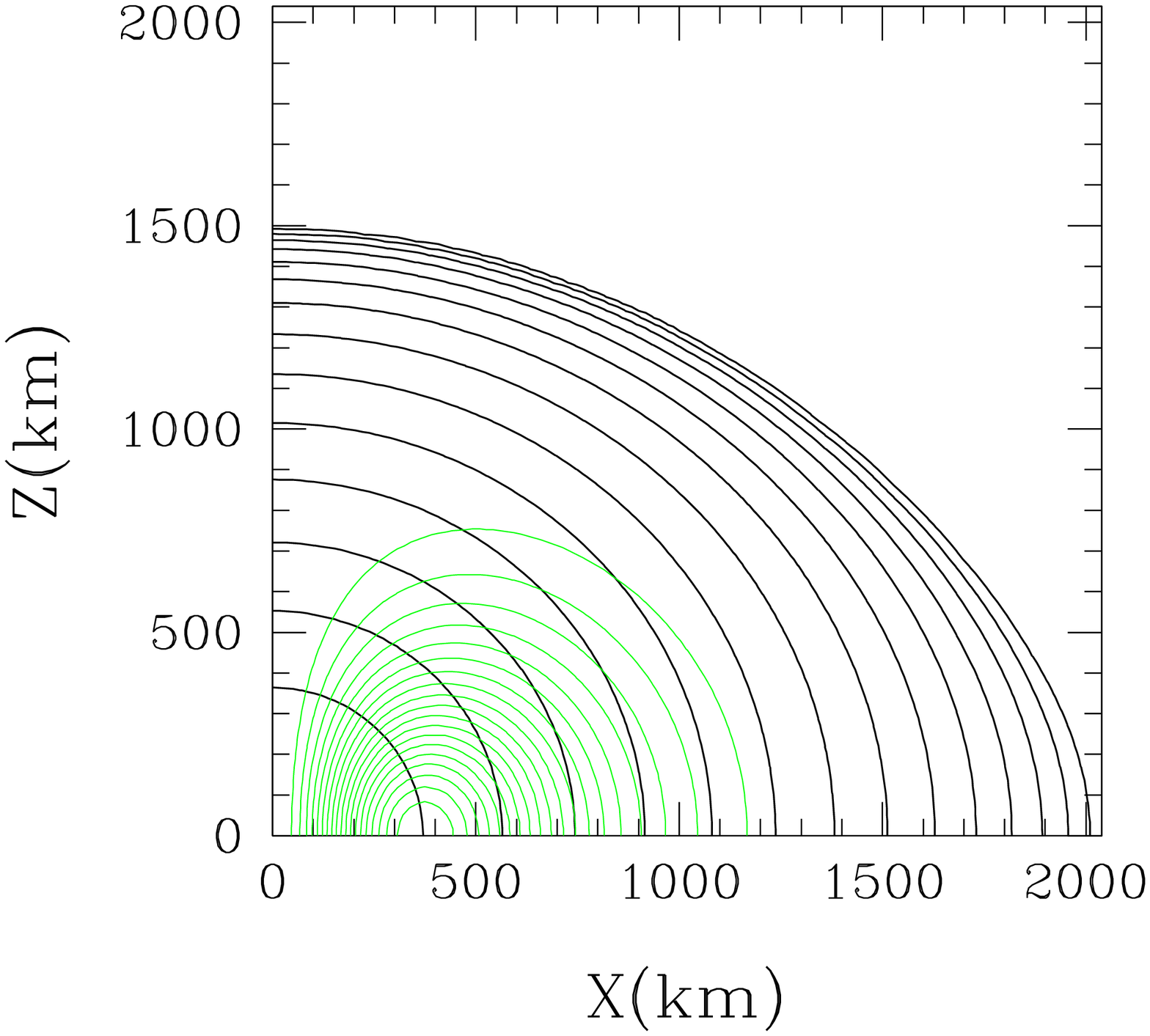}
\epsfxsize=3.in
\leavevmode
~~~~~~~~\epsffile{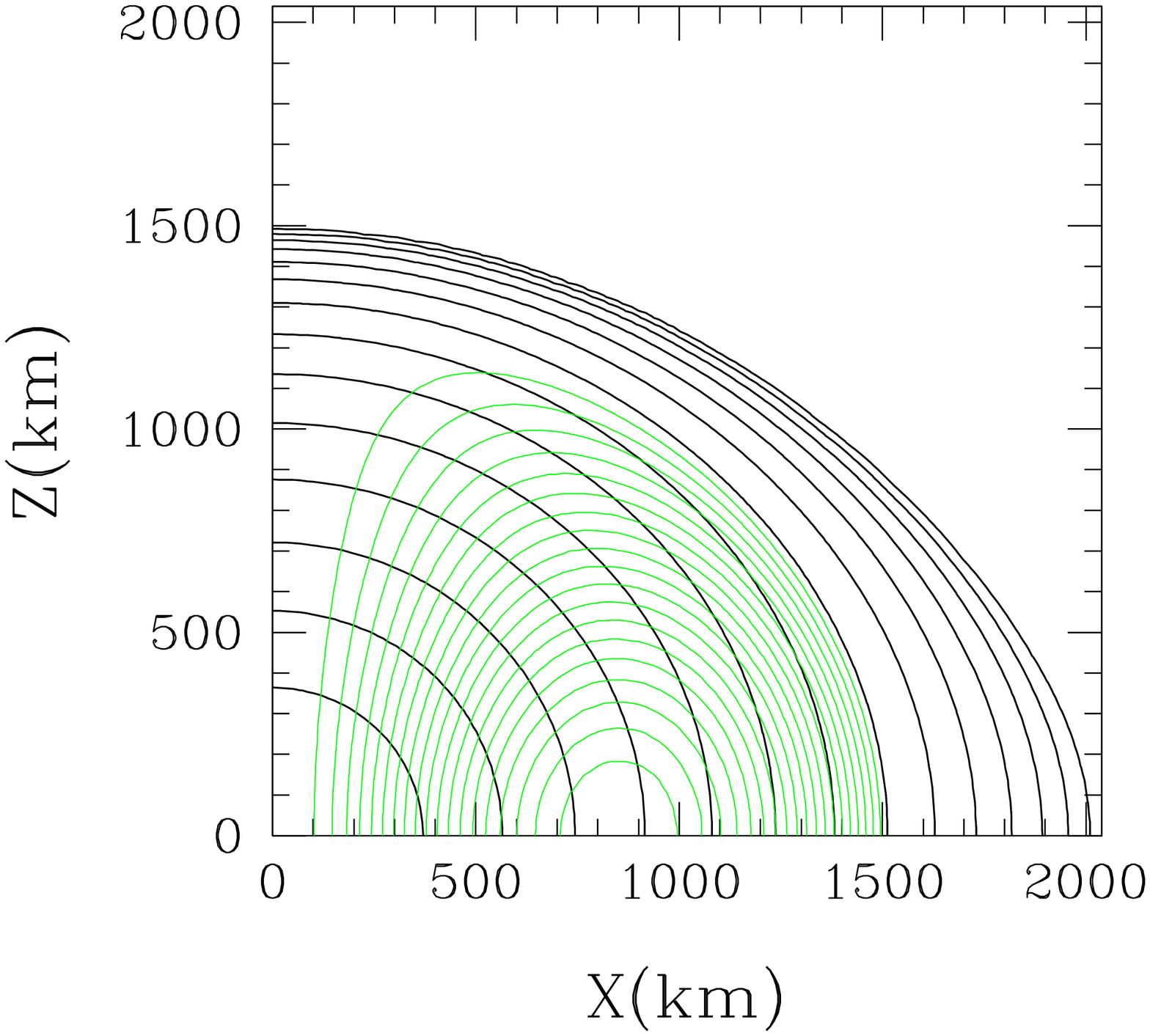}
\vspace{-4mm}
\caption{Initial density contour curves and magnetic field lines for
model A with $n_b=1$ (left) and with $n_b=4$ (right).  The density contour
curves (thick, black) are drawn for $\rho=10^{10-0.5j}~{\rm g/cm^3}$
with $j=1, 2, \cdots, 14$, and the magnetic field lines (thin, green)
are for $A_{\varphi}=A_{\varphi,{\rm max}}(1-0.05j)$ with $j=1, 2,
\cdots, 19$ where $A_{\varphi,{\rm max}}$ denotes the maximum of
$A_{\varphi}$.
\label{FIG1}
}
\end{center}
\end{figure*}

\section{Grid setup}
\label{sec:simulations}

During the collapse, the central density increases from $10^{10}~{\rm
g/cm^3}$ to $\sim 6 \times 10^{14}~{\rm g/cm^3}$.  This implies that
the characteristic length scale of the system varies by a factor of
$\sim 100$. One of the computational challenges in a stellar core
collapse simulation is maintaining numerical accuracy despite the
significant change in the characteristic length scale. In the early
phase of the collapse (infall phase; see Sec.~IV A), which proceeds in
a nearly homologous manner, we may follow the collapse with a
relatively small number of grid points, and successively move the
outer boundary inward (while keeping the same number of grid points)
to increase resolution.  As the collapse proceeds, the central region
shrinks more rapidly than the outer region, and hence a higher grid
resolution is necessary to accurately follow such rapid collapse in
the central region. On the other hand, the location of the outer
boundaries cannot be changed by a large factor because this would
discard matter in the outer envelope.

To follow the collapse accurately and save CPU time, we adopt the
regridding technique described in~\cite{SS02,SS04a}. Regridding is
carried out whenever the characteristic radius of the collapsing star
decreases by a factor of 2--3. At each regridding, the grid spacing is
decreased by a factor of 2. All the quantities in the new grid are
calculated using cubic interpolation. To avoid discarding the matter
in the outer region, we also increase the grid number at regridding,
ensuring that the discarded baryon rest mass is less than 3\% of the
total.

Specifically, the values of $N$ and $L$ (see
Sec.~\ref{sec:formulation}) in the present work are chosen in the
following manner. First, we define a relativistic gravitational
potential $\Phi_c \equiv 1 -\alpha_c~ (\Phi_c>0)$, which is $\approx
0.006$ at $t=0$ for all the models chosen in this work. Since $\Phi_c$
is approximately proportional to $M/R$, $\Phi_c^{-1}$ can be used as a
measure of the characteristic length scale. From $t=0$ to the time at
which $\Phi_c = 0.02$, we set $N=420$. The value of $L$ is chosen so
that the equatorial radius is initially covered by 400 grid points.
At $\Phi_c = 0.02$, the characteristic stellar radius becomes
approximately one third of the initial value. At this time, the first
regridding is performed. The grid spacing is cut in half and the grid
number is increased to $N=700$. Next, we set $N=1180$ for $0.04 \leq
\Phi_c \leq 0.08$, $N=1900$ for $0.08 \leq \Phi_c \leq 0.16$, and
$N=2500$ for $\Phi_c \geq 0.16 $ (the minimum value of $\Phi_c$ is ~
0.25 for all the models we consider).  In this treatment, the total
discarded fraction of the baryon rest mass is $\sim 1\%$ for model A,
$\sim 2\%$ for model B, and $\sim 2.5\%$ for model C with $(\Gamma_1$,
$\Gamma_2$, $\Gamma_{\rm th}$, $\rho_{\rm nuc})=$ (1.3, 2.5, 1.3,
$2\times 10^{14}~{\rm g/cm^3}$). In the following, we refer to this
resolution as the standard resolution.

To check the convergence of our numerical results, we also perform
simulations using higher and lower grid resolutions for model A2.  In
the higher-resolution case, the value of $N$ is changed as follows:
$N=500$ for $\Phi_c \leq 0.02$, $N=840$ for $0.02 \leq \Phi_c \leq 0.04$,
$N=1420$ for $0.04 \leq \Phi_c \leq 0.08$, $N=2268$ for $0.08 \leq
\Phi_c \leq 0.16$, and $N=2540$ for $\Phi_c \geq 0.16$.  In
this case, the equatorial radius is covered by 480 grid points initially. In
the lower-resolution simulations, the equatorial radius is 
initially covered by 300 and 240 grid points and the values of $N$ are 
changed as follows: 
$N=320$ and 260 for $\Phi_c \leq 0.02$, $N=540$ and 432 for 
$0.02 \leq \Phi_c \leq 0.04$,
$N=900$ and 720 for $0.04 \leq \Phi_c \leq 0.08$, $N=1440$ and 1200 
for $0.08 \leq \Phi_c \leq 0.16$, and $N=1900$ and 1520 for 
$\Phi_c \geq 0.16$. These
resolutions are referred to as the high, middle, and low resolutions,
respectively. 

We also performed several simulations using multiple transition fisheye 
coordinates~\cite{CLZ06}. This technique allows us to obtain high 
resolution in the central region with a relatively small grid number.
The details of our fisheye implementation are discussed 
in Appendix~A. We have confirmed that the results obtained using  
the fisheye coordinates agree with those obtained using the original 
coordinates. 

Simulations for each model with the standard resolution are performed
for about 150,000 time steps. This corresponds to the physical time 
of $\sim 150$~ms ($\sim 35$~ms after core bounce). At this time, 
most of the interesting MHD processes have occurred and the system 
is settling down to a stationary state. For the code of~\cite{SS05}, 
the required CPU time for one model is about 20 CPU hours
using 48 processors of FACOM VPP 5000 at the data processing center of
National Astronomical Observatory of Japan,
about 60 CPU hours using 8 processors of NEC SX8 at Yukawa Institute of
Theoretical Physics (YITP) in Kyoto University, and 
about 120 CPU hours using 8 processors of NEC SX6 at ISAS, JAXA. 
In order to check the validity of our results, several simulations were 
repeated using the code of~\cite{DLSS}. We have confirmed that the 
results obtained with these two codes agree. Thus, in the following 
section, we mainly present the numerical results from the code 
of~\cite{SS05}.

\section{Numerical Results}\label{sec:results}

\subsection{Dynamics}

\begin{figure}[th]
\vspace{-4mm}
\begin{center}
\epsfxsize=3.2in
\leavevmode
\epsffile{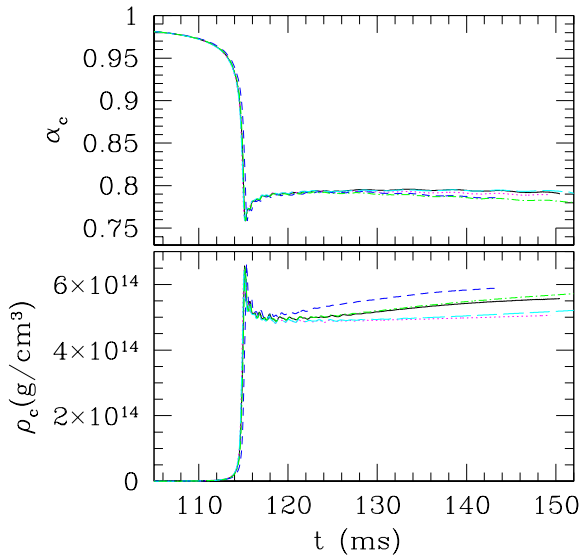}
\vspace{-6mm}
\caption{Evolution of the central lapse and central density 
for models A0 (dotted curves), A1 (long-dashed curves), A2 (solid
curves), A3 (dashed curves), and A4 (dot-dashed curves).
\label{FIG2}
}
\end{center}
\end{figure}

\begin{figure}[th]
\begin{center}
\epsfxsize=3.4in
\leavevmode
\epsffile{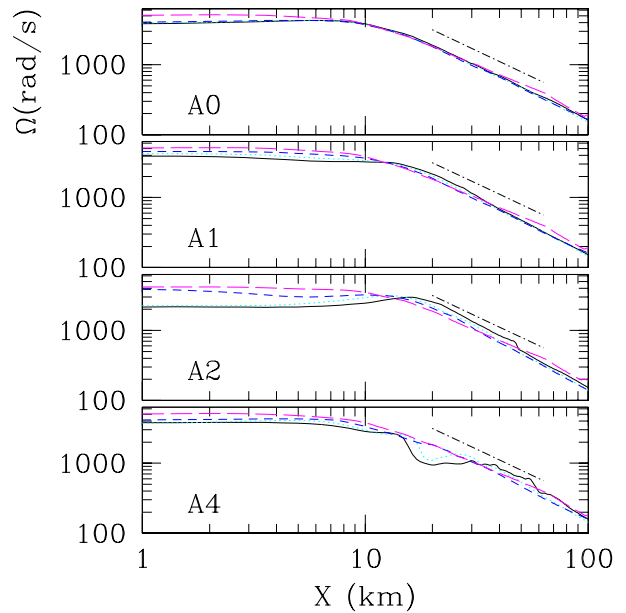}
\vspace{-4mm}
\caption{Evolution of the angular velocity of the resultant PNSs as 
a function of cylindrical radius in the equatorial plane for models
A0, A1, A2, and A4. The long-dashed, dashed, dotted, and solid
curves denote $\Omega$ at 118.4, 127.9, 136.7, and 149.1~ms
for model A0, at 118.5, 130.6, 140.3, and 150.6~ms
for model A1, and at 118.3, 131.3, 137.8, and 150.8~ms for
model A2, and at 118.6, 128.0, 137.2, and 149.5~ms for
model A4.  The dot-dashed line segments denote the slope of $\Omega
\propto \varpi^{-3/2}$.
\label{FIG3}
}
\end{center}
\end{figure}

As described in, e.g.,~\cite{Newton4,HD,SS04a}, the collapse of 
a rotating stellar core for which the initial $T/|W|$
is not too large (i.e., $\alt 0.01$)
and the degree of differential rotation is small ($\hat A \agt 0.5$), 
can be divided into the following three phases.  The first is the
infall phase, in which the collapse sets in due to the sudden softening 
of the EOS. During this phase, the central density monotonically increases
until it reaches nuclear density. The interior of the core that
collapses nearly homologously is referred to as the {\it inner core}. 
The duration of the infall phase (i.e., the interval between
the onset of the collapse and the time at which the central density
reaches a maximum) is of order $\rho_c^{-1/2}$. 

The second phase is the bounce, which sets in when the densities in
the central region exceed nuclear density $\rho_{\rm nuc}$. At this
phase, the infall of the inner core decelerates on a typical timescale
of a few ms ($\sim 10\rho_{\rm nuc}^{-1/2}$).  Because of its large
inertia and kinetic energy, the inner core does not settle down to a
stationary state immediately but overshoots and rebounds, driving
outgoing shocks at the outer edge of the inner core.

The third phase is the ring-down. The bounce occurs when the central 
density reaches $\sim 3\rho_{\rm nuc}$ due to a sudden stiffening of 
the EOS (if the centrifugal force is sufficiently small at the time 
that the density of the inner core exceeds nuclear density).
In this case, the inner core oscillates quasi-radially for about
10 ms, and then settles down to a stationary state. In the outer
region, on the other hand, shock waves propagate outward, sweeping
through infalling material from the outer envelope. With our simple
treatment of the microphysics, the shock propagates all the way
through to the surface of the outer core, which has a radius $\sim
1000$~km~\cite{note1}.

\begin{figure}[th]
\begin{center}
\epsfxsize=3.3in
\leavevmode
\epsffile{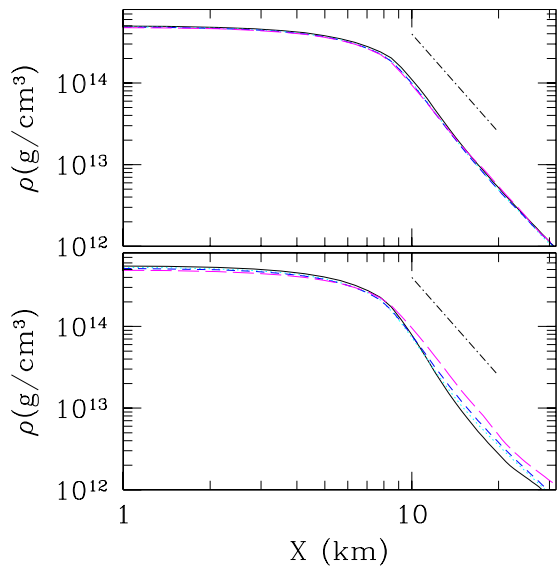}
\vspace{-6mm}
\caption{Evolution of the density profile of the PNSs as a function
of the cylindrical radius in the equatorial plane for models
A0 (upper panel) and A2 (lower panel). The profiles are given at 
times corresponding to those of Fig.~\ref{FIG3}.
The dot-dashed line segments denote the slope of
$\rho \propto \varpi^{-4}$.
\label{FIG4}
}
\end{center}
\end{figure}

Even in the presence of magnetic fields, these general features are
not significantly modified as long as the field is not extremely
strong (i.e., as long as the magnetic energy is smaller than the
rotational kinetic energy). In Fig.~\ref{FIG2}, we show the evolution
of the central lapse and central density for models A0--A4. The
results for models with non-zero magnetic fields (models A1--A4) are
qualitatively very similar to those for model A0, which has no
magnetic field.  However, the magnetic effects lead to some
quantitative changes.  Figure~\ref{FIG3} shows the evolution of the
angular velocity $\Omega$ for the newly-formed PNSs as a function of
the cylindrical radius in the equatorial plane for models A0, A1, A2,
and A4. For model A0 (see Fig.~\ref{FIG3}, upper panel), $\Omega$
relaxes to a stationary profile by $\sim 10$~ms after formation of the
PNS. The remaining panels show the evolution of $\Omega$ in the
presence of magnetic fields (up to $\sim 10^{16}$ G). In these cases,
the PNS slows down monotonically due to outward angular momentum
transport driven by magnetic braking, the MRI and the MHD outflow
(see Secs. \ref{sec:evo} and \ref{sec:BB} for details). In
particular, for model A2, in which the growth of the magnetic field by
winding saturates shortly after bounce (see Fig.~\ref{FIG7}), the
central spin period doubles in the first $\sim 20$~ms following the
bounce. For model A1, the
magnetic field is amplified more gradually (see Fig.~\ref{FIG7}). In
this case, $\Omega$ does not decrease as rapidly as for model A2.
For model A4, the PNS also spins down monotonically. This
model has a different initial magnetic field profile from those of
models A1--A3.  This difference is reflected in the evolution of the
angular velocity profile, which is not as smooth for A4 in the outer
region of the PNS at late times.  Since the magnetic field is stronger
in the outer layers for model A4, the MRI is more vigorous there and
the $\Omega$ profile is thus not as smooth (see also
Sec.~\ref{sec:evo}).

Figure~\ref{FIG4} shows the evolution of the density profiles in the
equatorial plane for models A0 and A2. For A0, the density
profile quickly relaxes to a stationary state, while it changes with
time for A2 due to the angular momentum transport. In accord 
with the spindown shown in Fig.~\ref{FIG3}, the central density
gradually increases (see Fig.~\ref{FIG2}). Since material is 
ejected in an MHD-driven outflow, the density in the surface region 
decreases. A more detailed discussion and a qualitative explanation of the
main mechanisms driving angular momentum transport and the outflow 
are given in Sec.~\ref{sec:BB}.


\begin{figure*}[t]
\begin{center}
\epsfxsize=3.2in
\leavevmode
(a)\epsffile{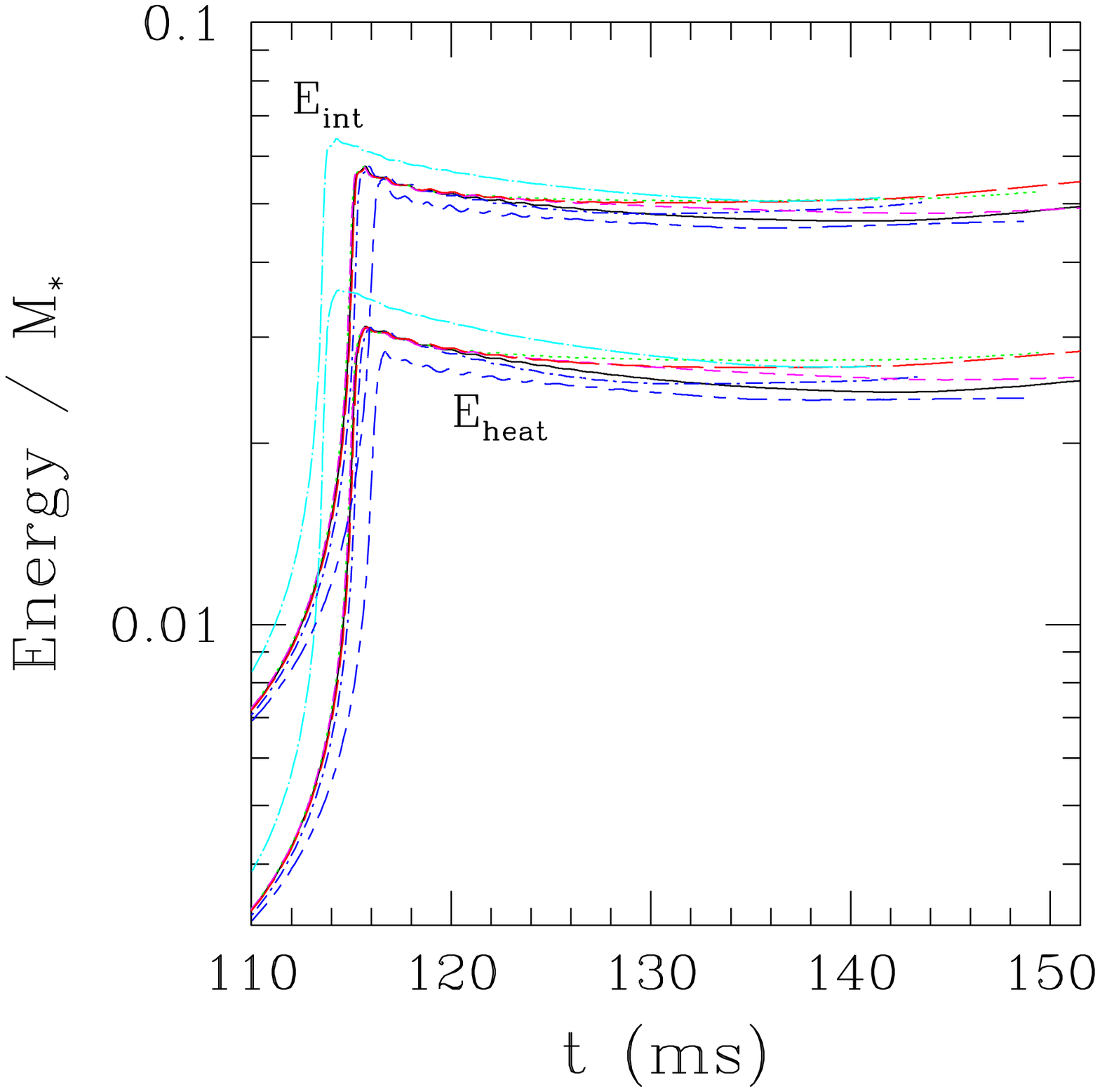}
\epsfxsize=3.2in
\leavevmode
~~~~(b)\epsffile{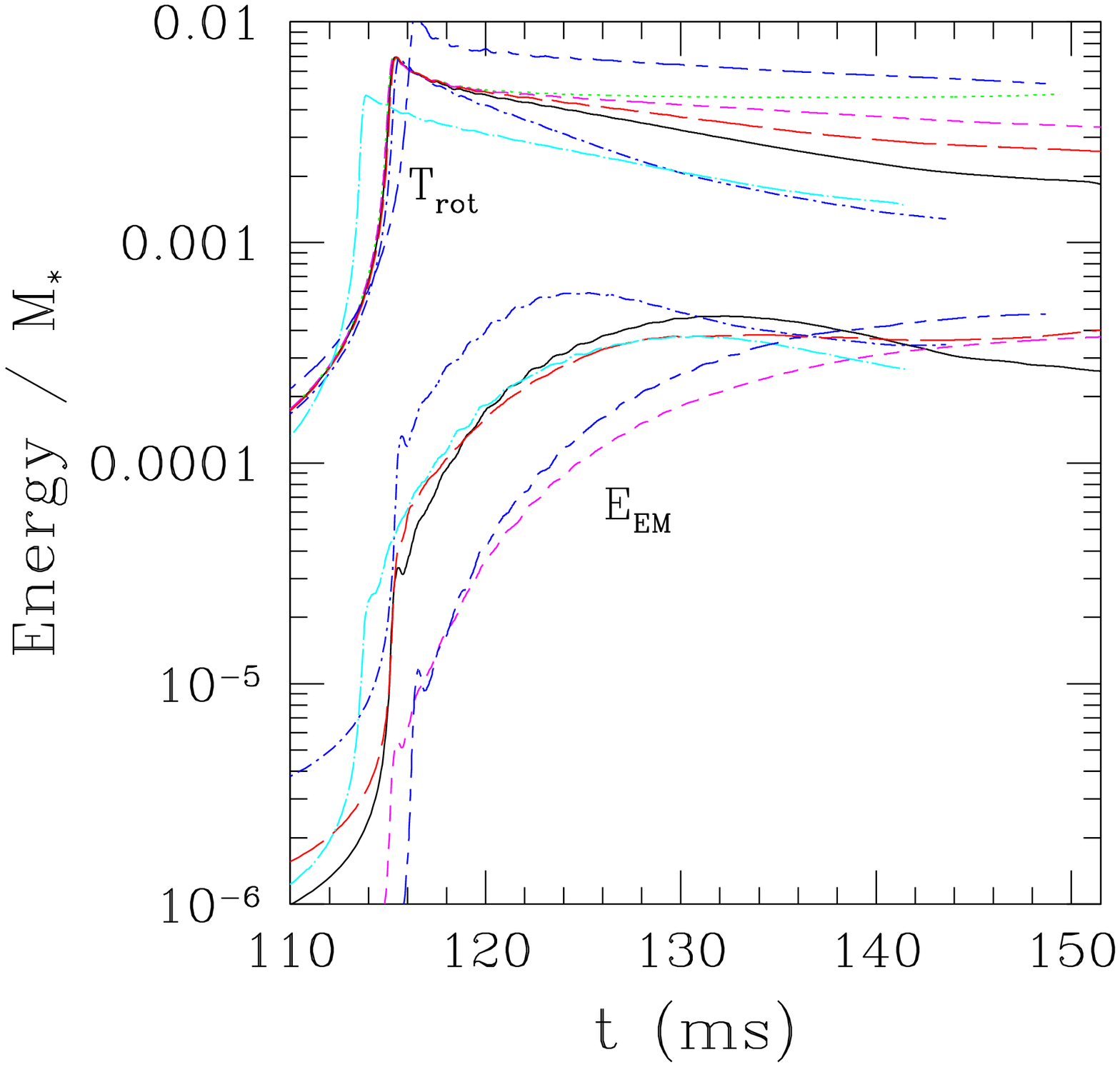}\\
\epsfxsize=3.2in
\leavevmode
(c)\epsffile{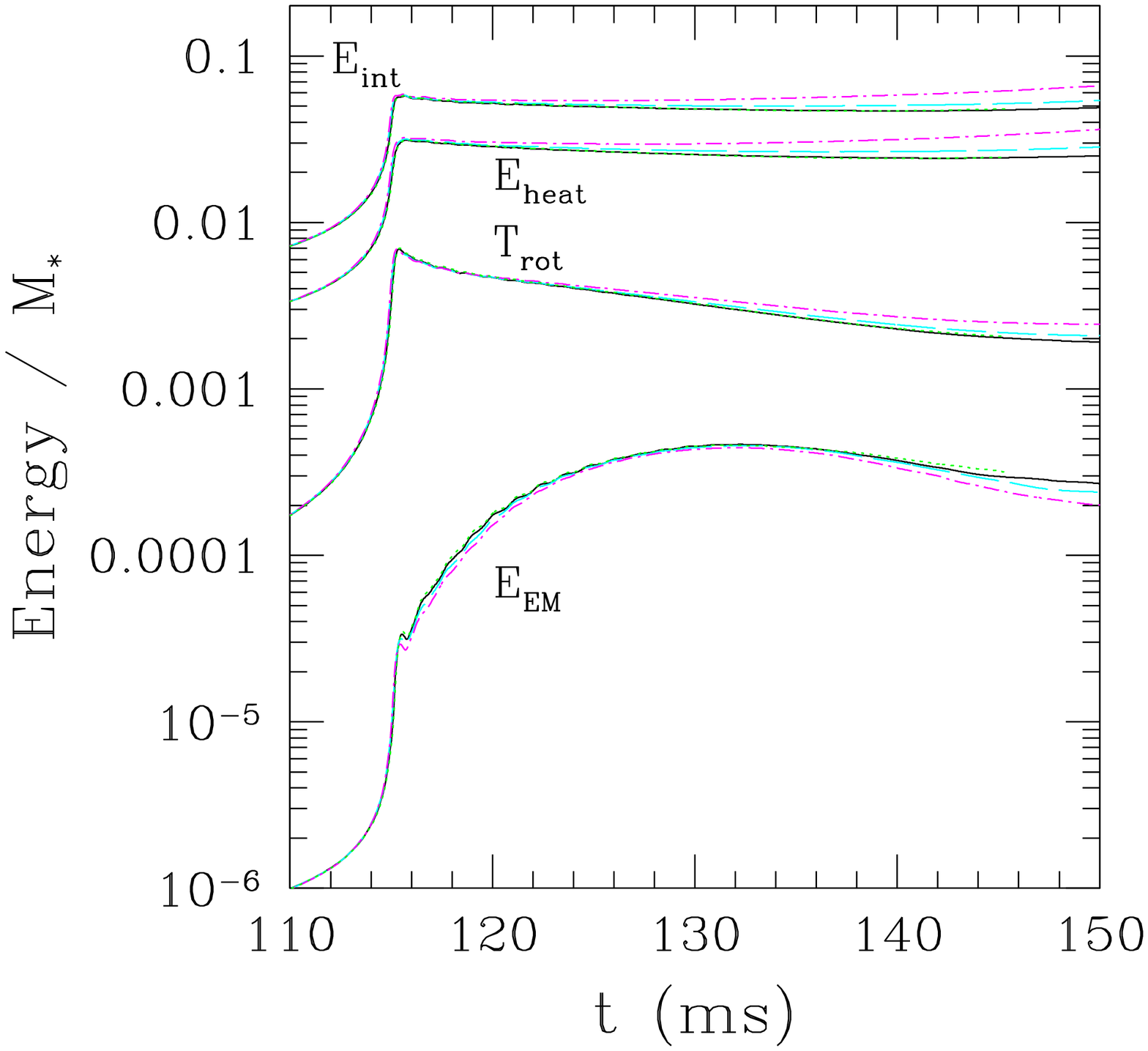}\\
\vspace{-4mm}
\caption{(a) 
Evolution of $E_{\rm int}$ and $E_{\rm heat}$
for models A0 (dotted curves), A1 (dashed curves),
A2 (solid curves), A3 (dot-dashed curves), A4 (long-dashed curves),
B (dot-long-dashed curves), and C (long-and-short dashed curves).
(b) The same as (a) but for $T_{\rm rot}$, and 
$E_{\rm EM}$. 
(c) Evolution of $E_{\rm int}$, $E_{\rm heat}$, $T_{\rm rot}$, and 
$E_{\rm EM}$ for model A2 with different
grid resolutions. The dotted, solid, dashed, and dot-dashed curves 
denote the results in the high-, standard-, middle-, and low-resolution runs.
\label{FIG6}
}
\end{center}
\end{figure*}


In Figs.~\ref{FIG6}(a) and (b), we show the various energies ($E_{\rm
int}$, $E_{\rm heat}$, $T_{\rm rot}$, and $E_{\rm EM}$) as functions
of time for models A0--A4, B, and C. All of the energies increase
during the infall phase and reach their maxima at the
bounce. Following the bounce, $E_{\rm int}$ and $E_{\rm heat}$ relax
to values of a quasistationary state which change very slowly (see
Fig.~\ref{FIG6}(a)).  On the other hand, $T_{\rm rot}$ rapidly
decreases in the presence of the magnetic field, reflecting the
spindown of PNSs by magnetic effects (see Fig.~\ref{FIG6}(b)). The
rate of this decrease is larger for larger values of $E_{\rm EM}$, as
anticipated in Sec.~\ref{sec:BP}.

Figure~\ref{FIG6}(c) shows the evolution of $E_{\rm int}$, $E_{\rm
heat}$, $T_{\rm rot}$, and $E_{\rm EM}$ for model A2 with different
grid resolutions. With the middle and low resolutions, convergence is
not well achieved. On the other hand, the results in the high and
standard resolutions agree well, indicating that the standard
resolution is an appropriate choice for this problem.

The simulations for models B and C are performed with the same initial
magnetic field profile and field strengths as models A2 and A1,
respectively (see Table~II), and the results for the various energies
are shown together in Fig.~\ref{FIG6}. It is found that the evolutions
for models B and C proceed in the same qualitative manner as for
models A2 and A1, respectively. Quantitatively, the initial rotational
kinetic energy is reflected in the evolution of $T_{\rm rot}$ and
$E_{\rm EM}$. For model B, the maximum rotational kinetic energy is
smaller than that for model A2. As a result, the growth rate of the
magnetic energy after bounce due to winding as well as the resulting
maximum value reached are smaller. Thus, the magnetic energy achieved
at saturation for a particular model depends on the rotational kinetic
energy.  This is expected since the magnetic energy is ultimately
drawn from the energy stored in differential rotation.  For model C,
$T_{\rm rot}$ at bounce is larger than that of model A1 because of the
differential rotation at the onset of collapse and resulting larger
rotation velocity of the PNS. As a result, the growth rate of $E_{\rm
EM}$ and the maximum value reached are larger than those for model A1.

\subsection{Evolution of the magnetic field}\label{sec:evo}

Figure~\ref{FIG3} shows that for $\varpi \alt 10$~km (i.e., in the
central region of the PNSs), the angular velocity has a
roughly flat profile. With so little differential rotation, the
magnetic field does not grow much, either by winding
or the MRI. MHD effects have little influence on the evolution
in this region. For $\varpi \agt 10$~km (i.e., near the surface 
and outside the PNSs), the
angular velocity is approximately Keplerian (i.e., $\Omega \propto
\varpi^{-3/2}$; see the dot-dashed line segments in
Fig.~\ref{FIG3}). Because of the strong differential 
rotation, the outer region with $\varpi \agt 10$ km is
subject to winding and the MRI (see Secs. \ref{sec:winding} and
\ref{sec:MRI}). In particular, the magnetic field is rapidly amplified
in the region with $\varpi \approx 10$--30 km, where both the 
angular velocity and the degree of differential rotation are large. 
We note that strong differential rotation results even for rigidly
rotating progenitors.

\begin{figure*}[t]
\begin{center}
\epsfxsize=3.3in
\leavevmode
\epsffile{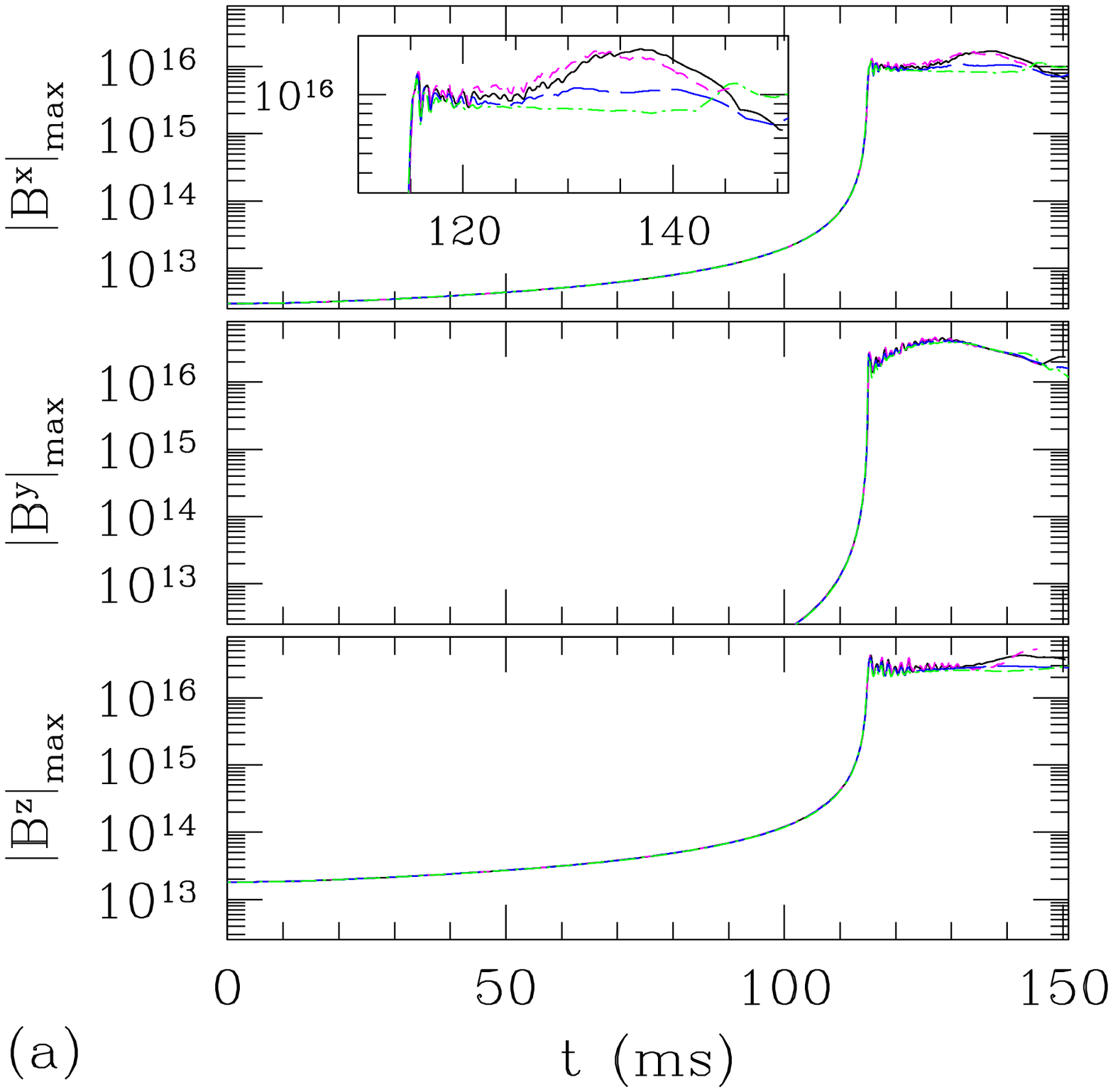}
\epsfxsize=3.3in
\leavevmode
~~~\epsffile{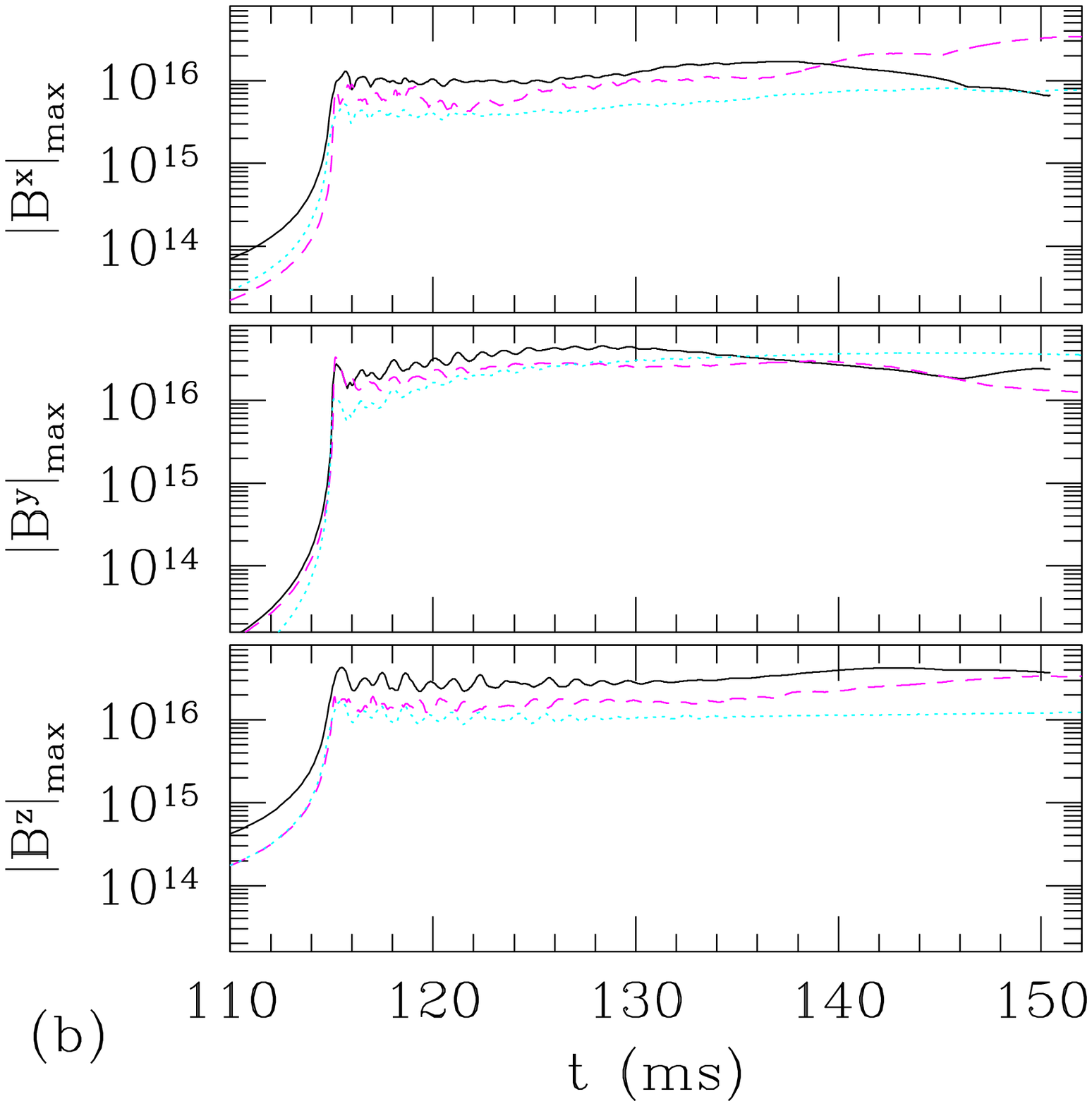}\\
\epsfxsize=3.3in
\leavevmode
~~~\epsffile{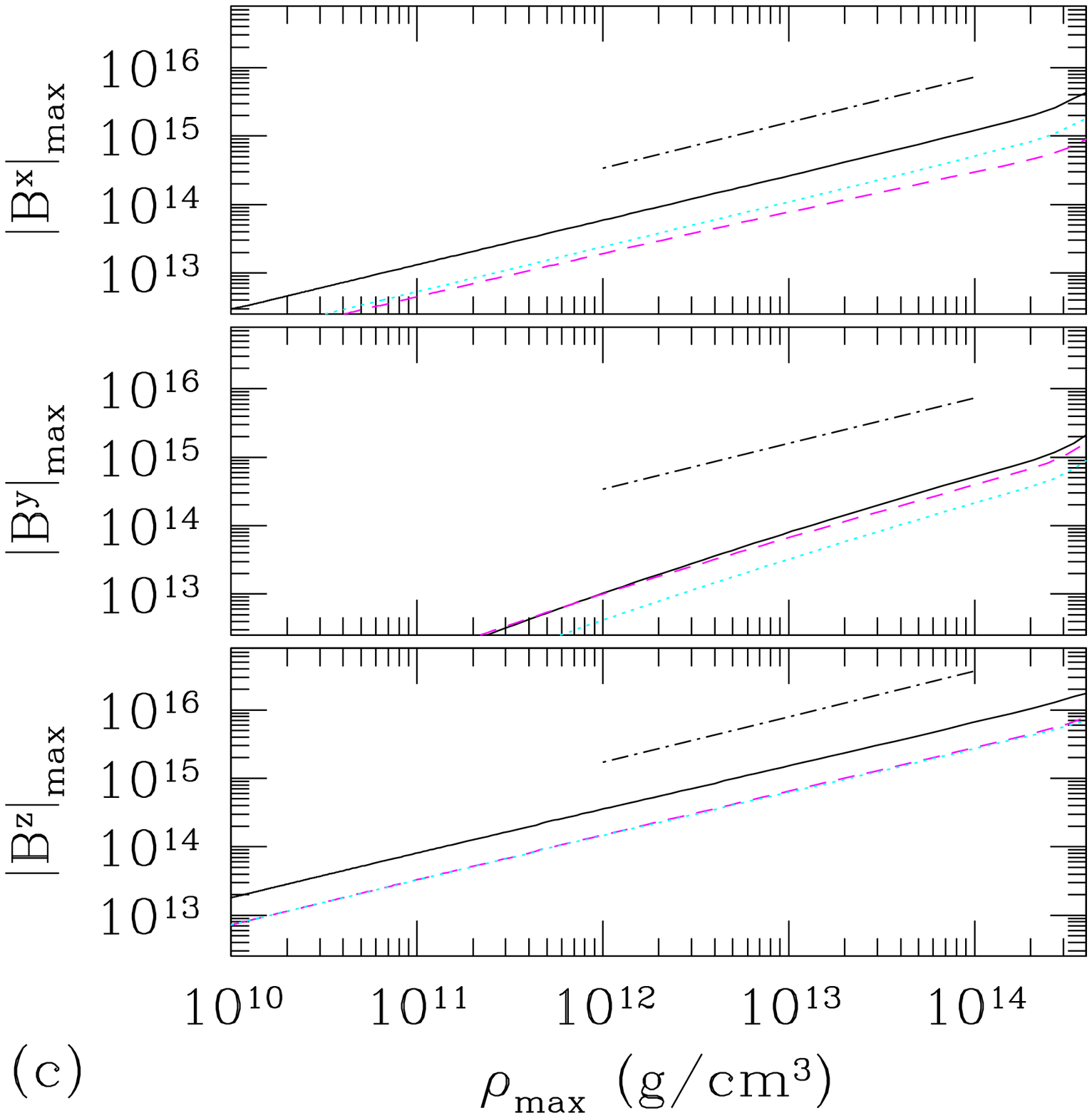}
\vspace{-6mm}
\caption{(a) Evolution of $|B^i|$ for model A2.
The solid, dashed, long-dashed, and dot-dashed curves denote the
results from the standard, high, middle, and low resolution runs for
model A2. The enlarged panel for the evolution of $|B^x|_{\rm max}$
illustrates that the growth of the field strength by the MRI is well
resolved only in the high and standard runs. (b) The same as (a) but
for models A1 (dotted curves), A2 (solid curves), and A4 (dashed
curves) with the standard resolution. (c) $|B^x|_{\rm max}$ as a
function of $\rho_{\rm max}$ during infall phase 
for models A1 (dotted curves), A2 (solid
curves), and A4 (dashed curves) with the standard resolution. The
dot-dashed line segments above the three curves denote the relation
$|B^i|_{\rm max}\propto \rho_{\rm max}^{2/3}$.
\label{FIG7}
}
\end{center}
\end{figure*}

In Fig.~\ref{FIG7}, we show the evolution of the maximum values of 
$|B^i|$ for models A1, A2, and A4. Note that, in our 
setup, $B^x$ and $B^z$ are the poloidal field components 
($B^x=B^{\varpi}$) and $B^y$ is the toroidal component 
($B^y=\varpi B^{\varphi}=B^T$) since we assume axisymmetry and solve 
the equations in $x$-$z$ plane. To demonstrate the effects of the grid 
resolution, the results for four different resolutions are
displayed together for model A2. 

During the infall phase, the magnetic field is amplified monotonically: 
The poloidal field grows due to the compression of matter, while the 
toroidal field grows primarily by winding (cf. Sec. \ref{sec:compress}).  
In the ideal MHD approximation, the magnetic field lines are frozen into 
the plasma, and hence, the poloidal field strength should grow approximately
as $\rho^{2/3}$ (see Sec. \ref{sec:compress}). As shown in Fig.~\ref{FIG7}(c),
this relationship holds for the maximum poloidal field components over
several decades of growth in the density.  Since $B^y$ grows by
winding, its growth accelerates rapidly just before the bounce due to 
the rapid growth of $|B^x|$.  As shown in 
Fig.~\ref{FIG7}, a significant amplification indeed occurs for about 2--3 ms 
before the bounce at $t \approx 115$ ms. Figure~\ref{FIG7}(c) also shows that
the maximum value of the toroidal magnetic field 
increases slightly faster than $\rho^{2/3}$ during the infall phase.
This qualitatively agrees with the estimate in Sec.~\ref{sec:compress}.

\begin{figure*}[t]
\vspace{-4mm}
\begin{center}
\epsfxsize=2.15in
\leavevmode
\epsffile{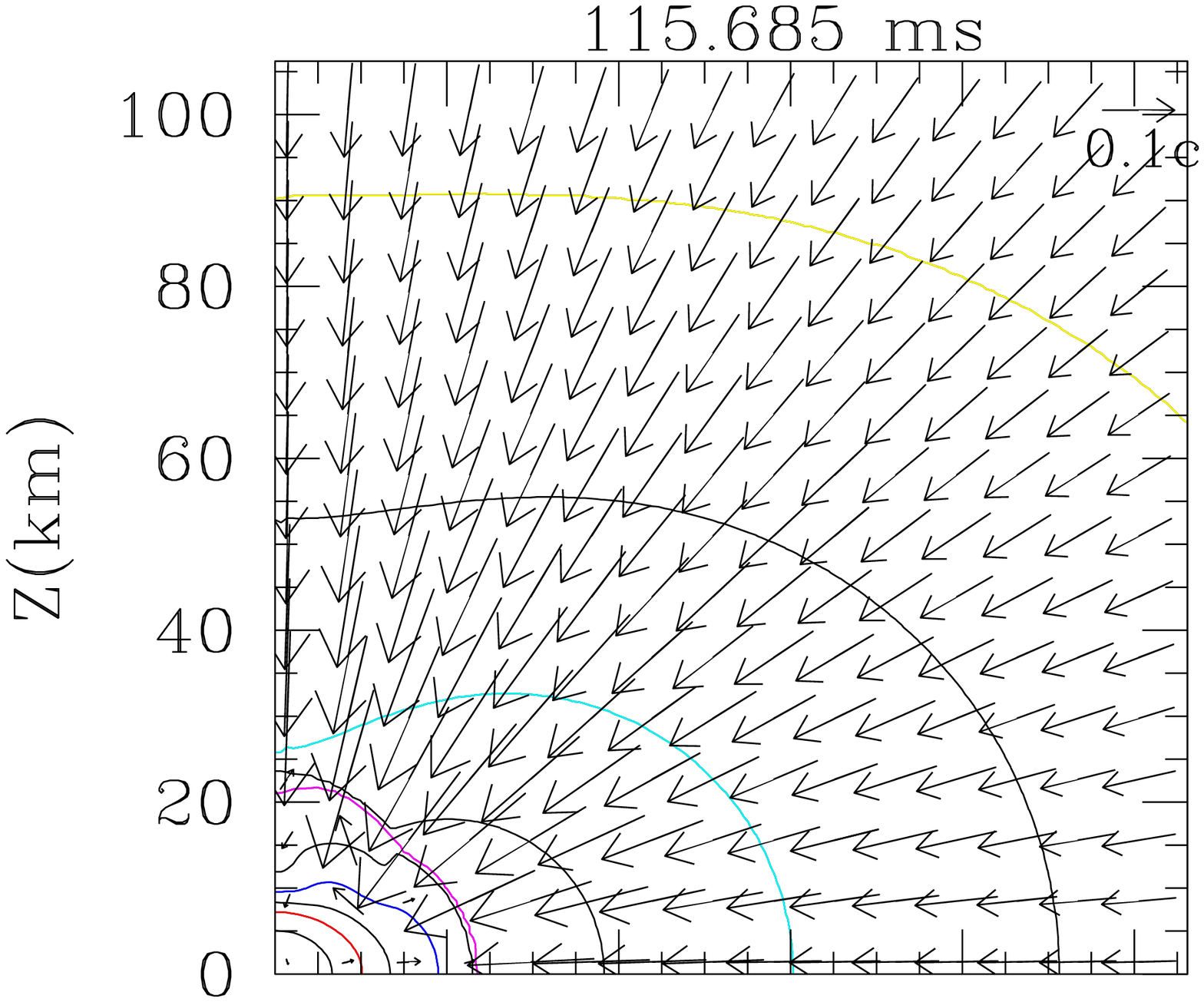}
\epsfxsize=2.15in
\leavevmode
\hspace{-1.45cm}\epsffile{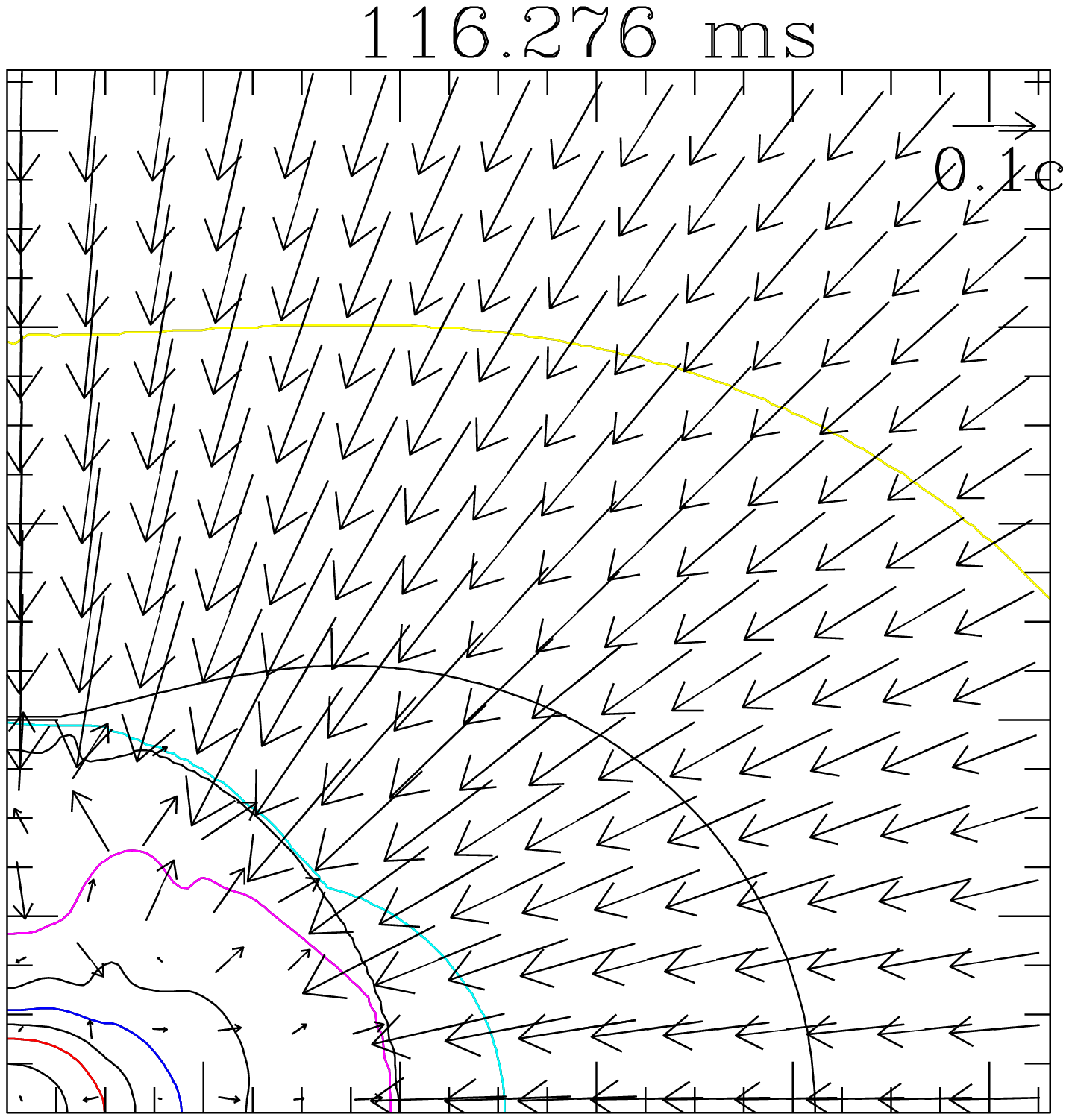}
\epsfxsize=2.15in
\leavevmode
\hspace{-1.45cm}\epsffile{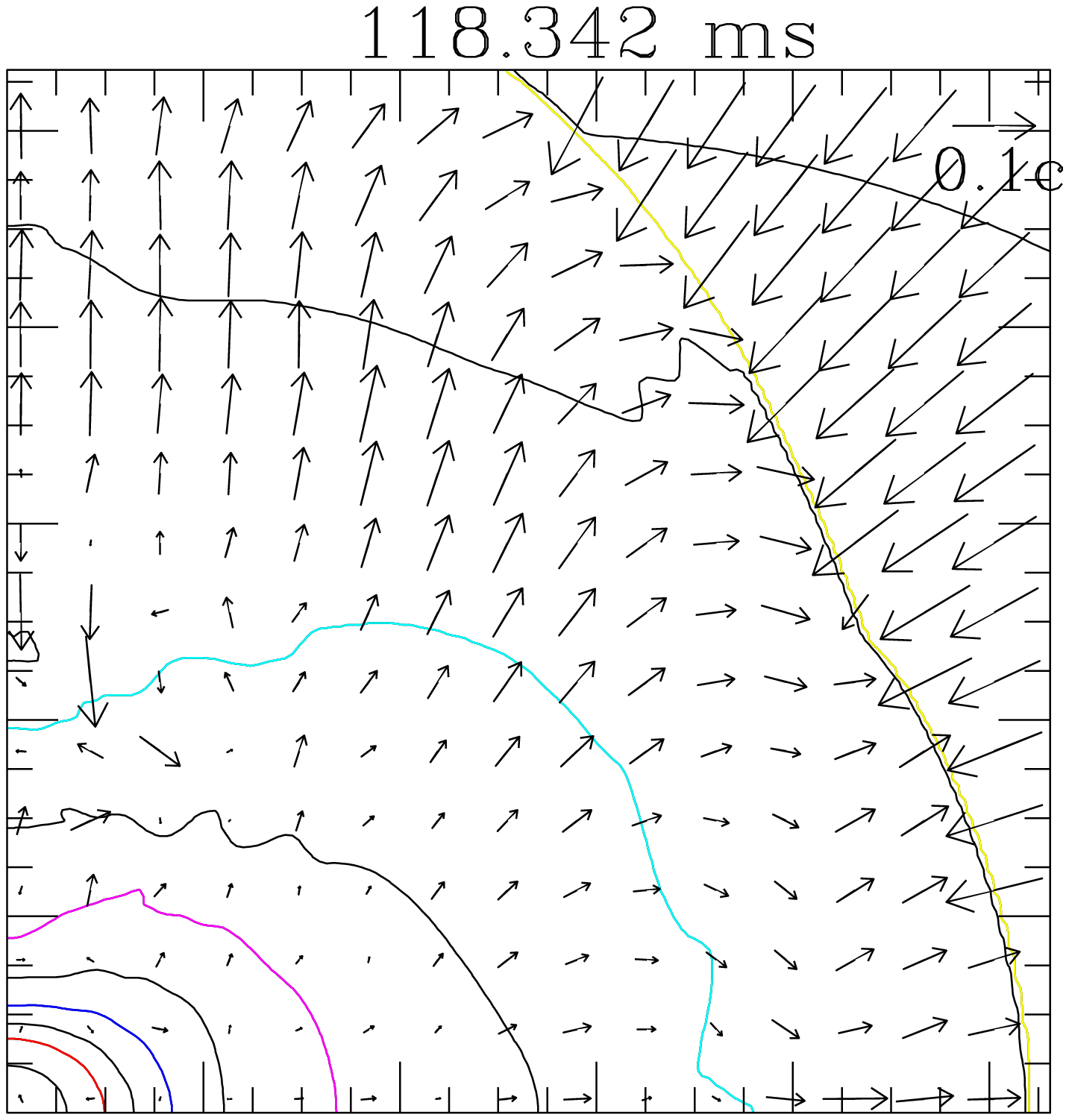}
\epsfxsize=2.15in
\leavevmode
\hspace{-1.45cm}\epsffile{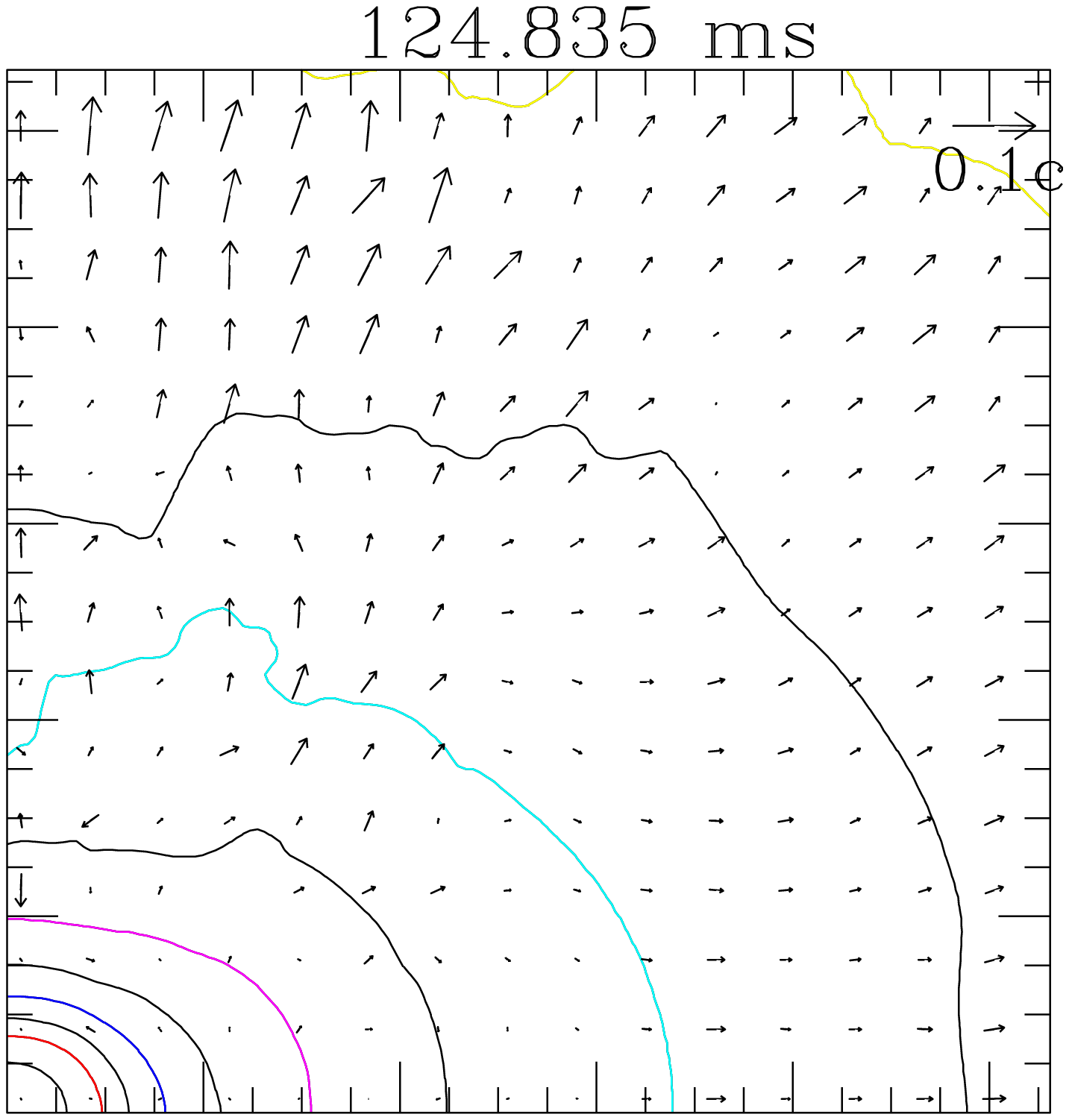}\\
\vspace{-1.6cm}
\epsfxsize=2.15in
\leavevmode
\epsffile{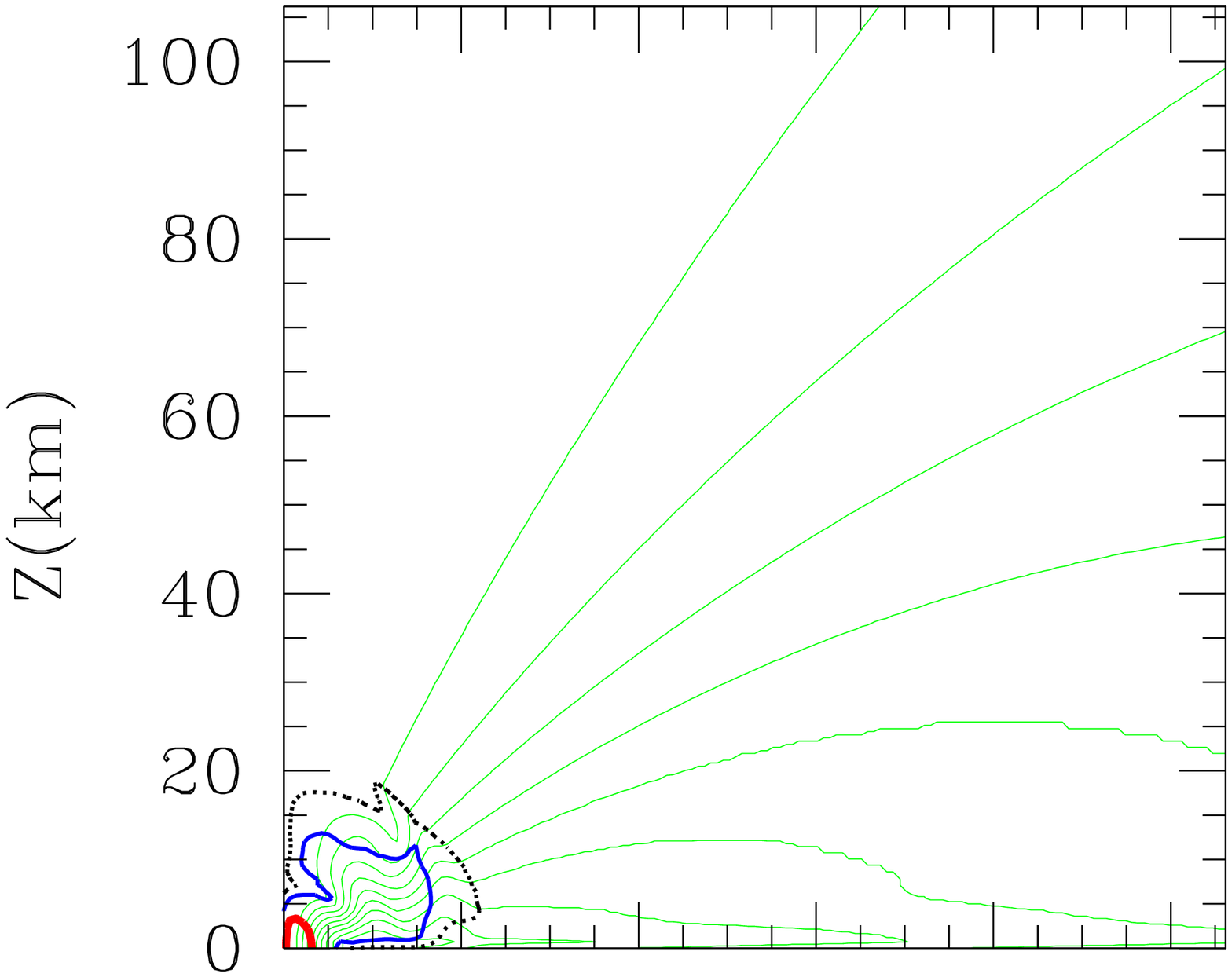}
\epsfxsize=2.15in
\leavevmode
\hspace{-1.45cm}\epsffile{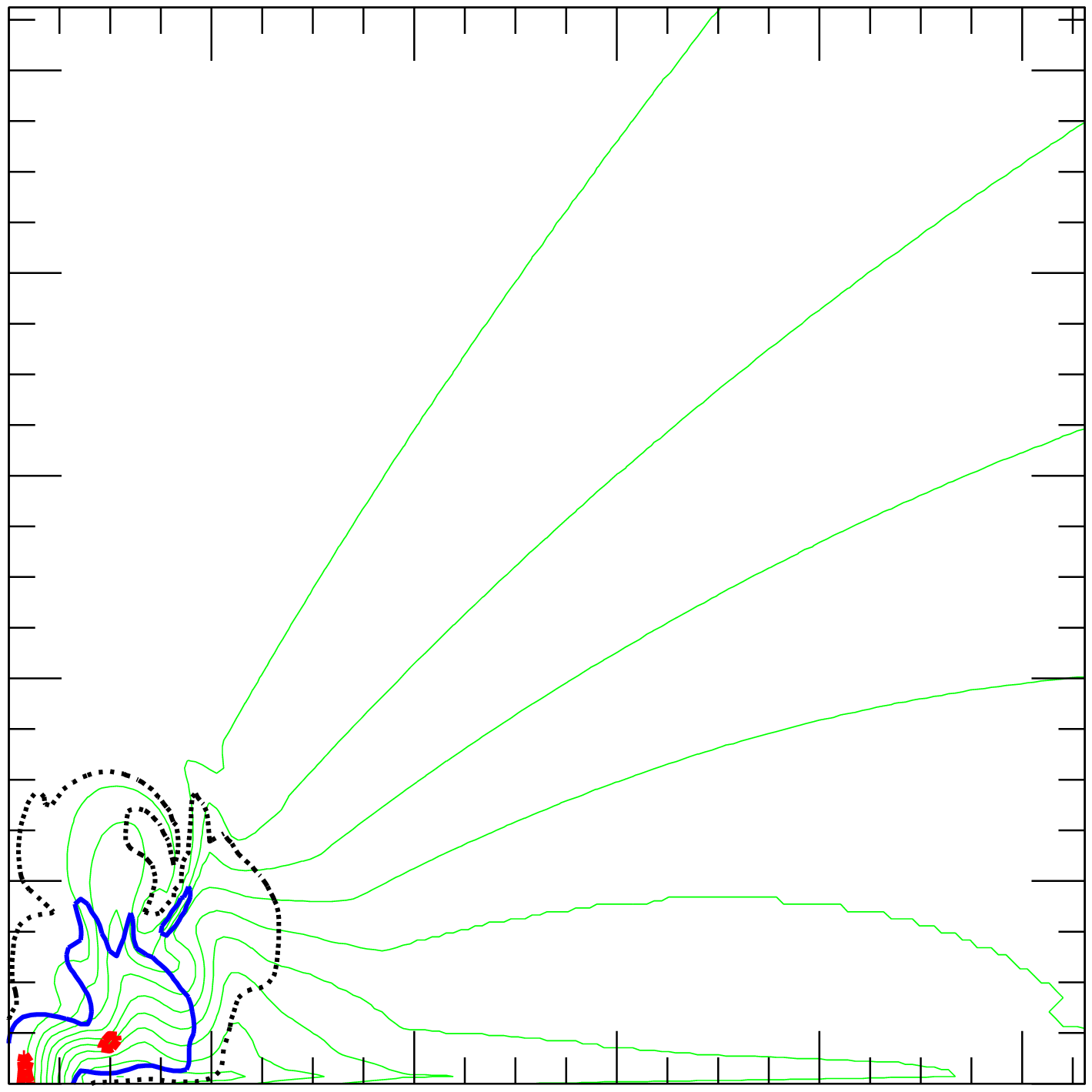}
\epsfxsize=2.15in
\leavevmode
\hspace{-1.45cm}\epsffile{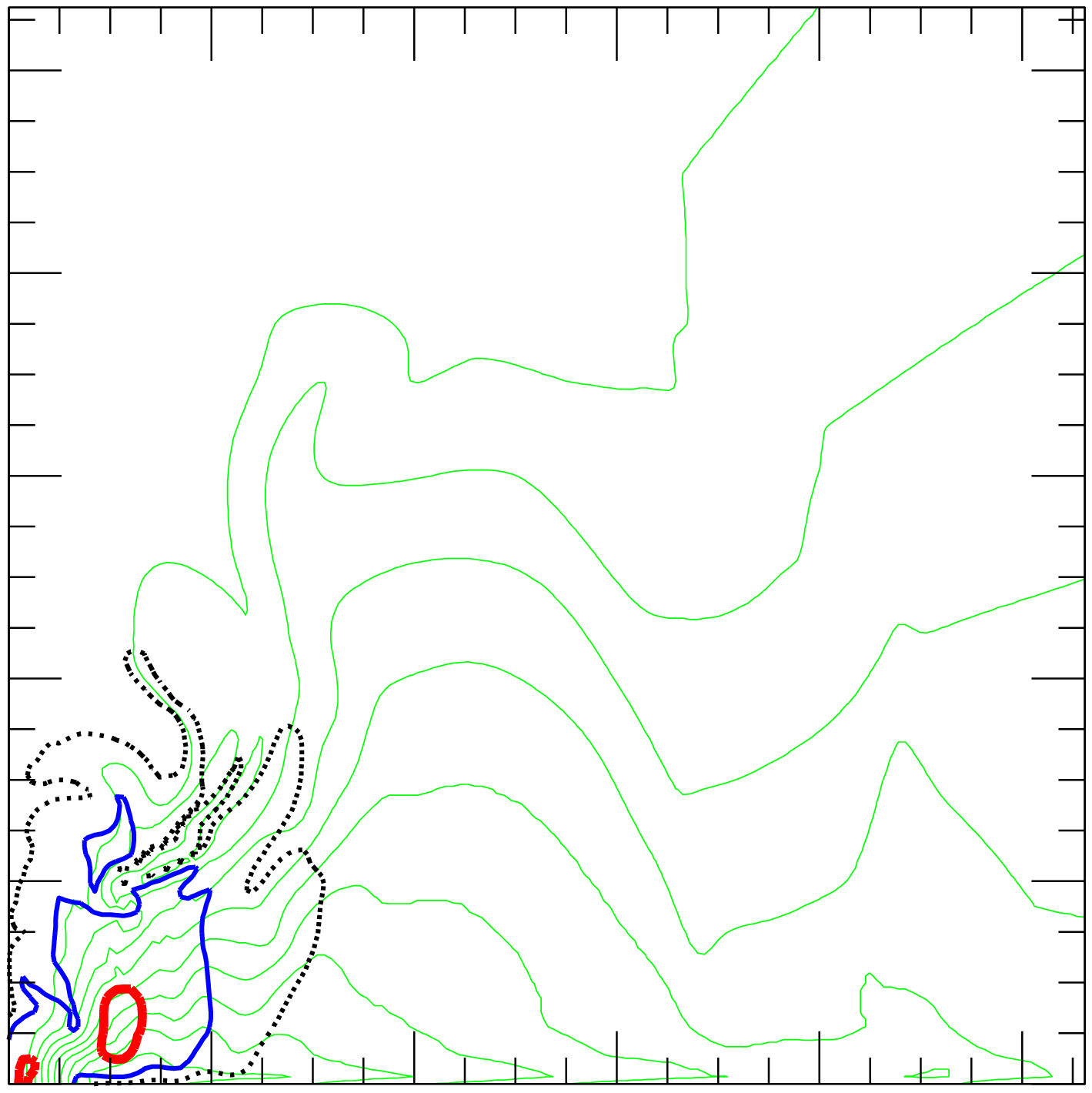}
\epsfxsize=2.15in
\leavevmode
\hspace{-1.45cm}\epsffile{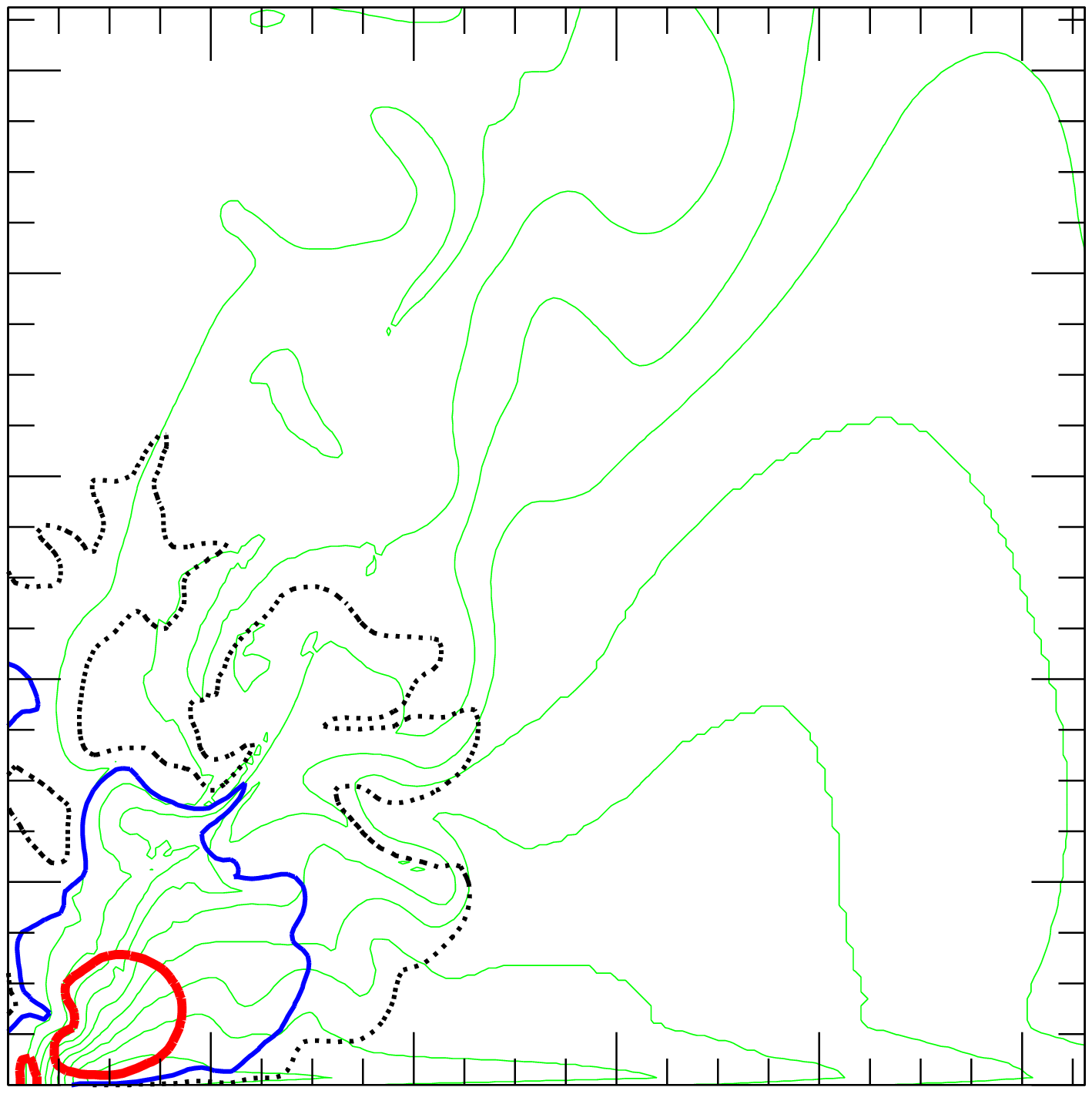}\\
\vspace{-0.8cm}
\epsfxsize=2.15in
\leavevmode
\epsffile{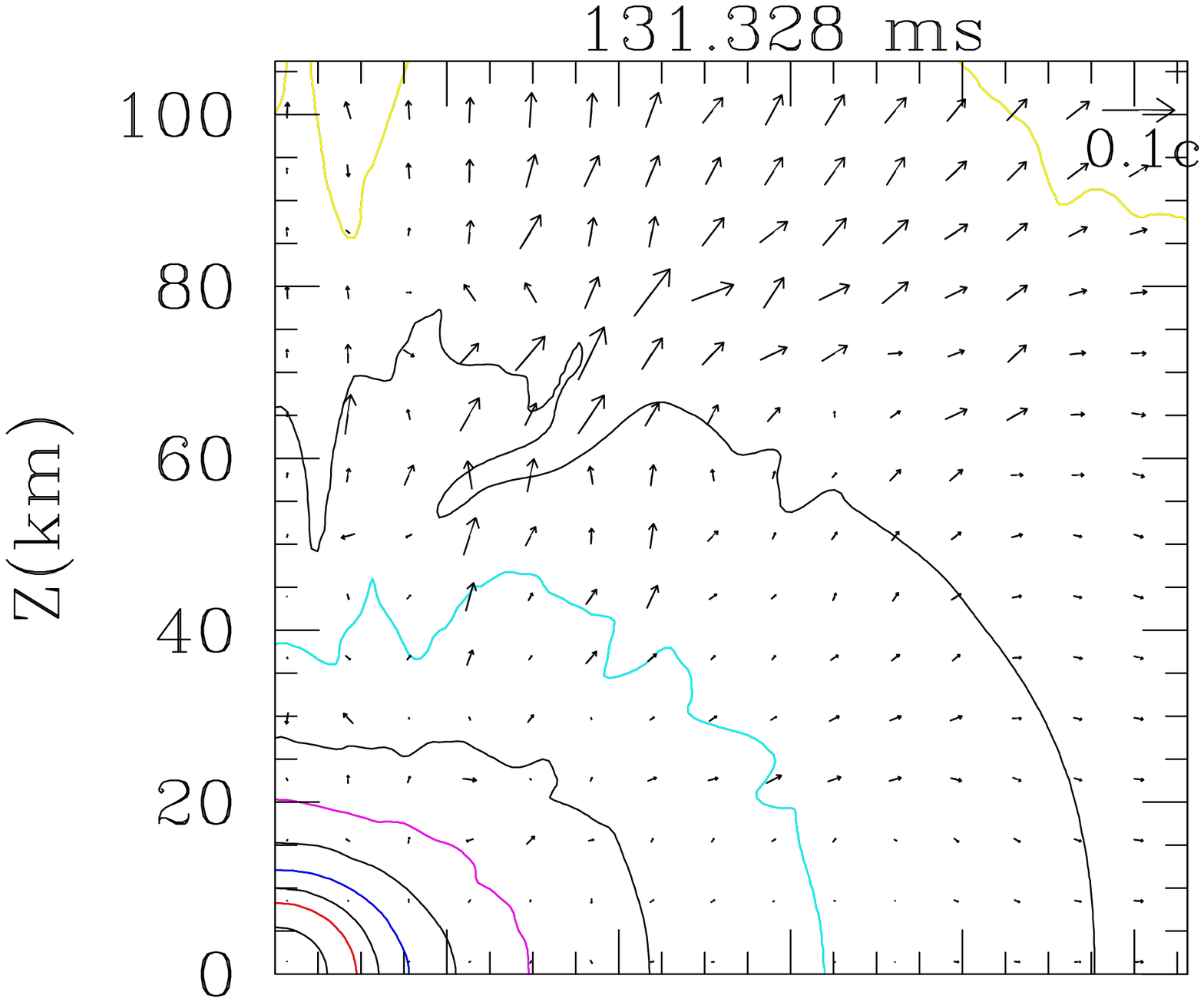}
\epsfxsize=2.15in
\leavevmode
\hspace{-1.45cm}\epsffile{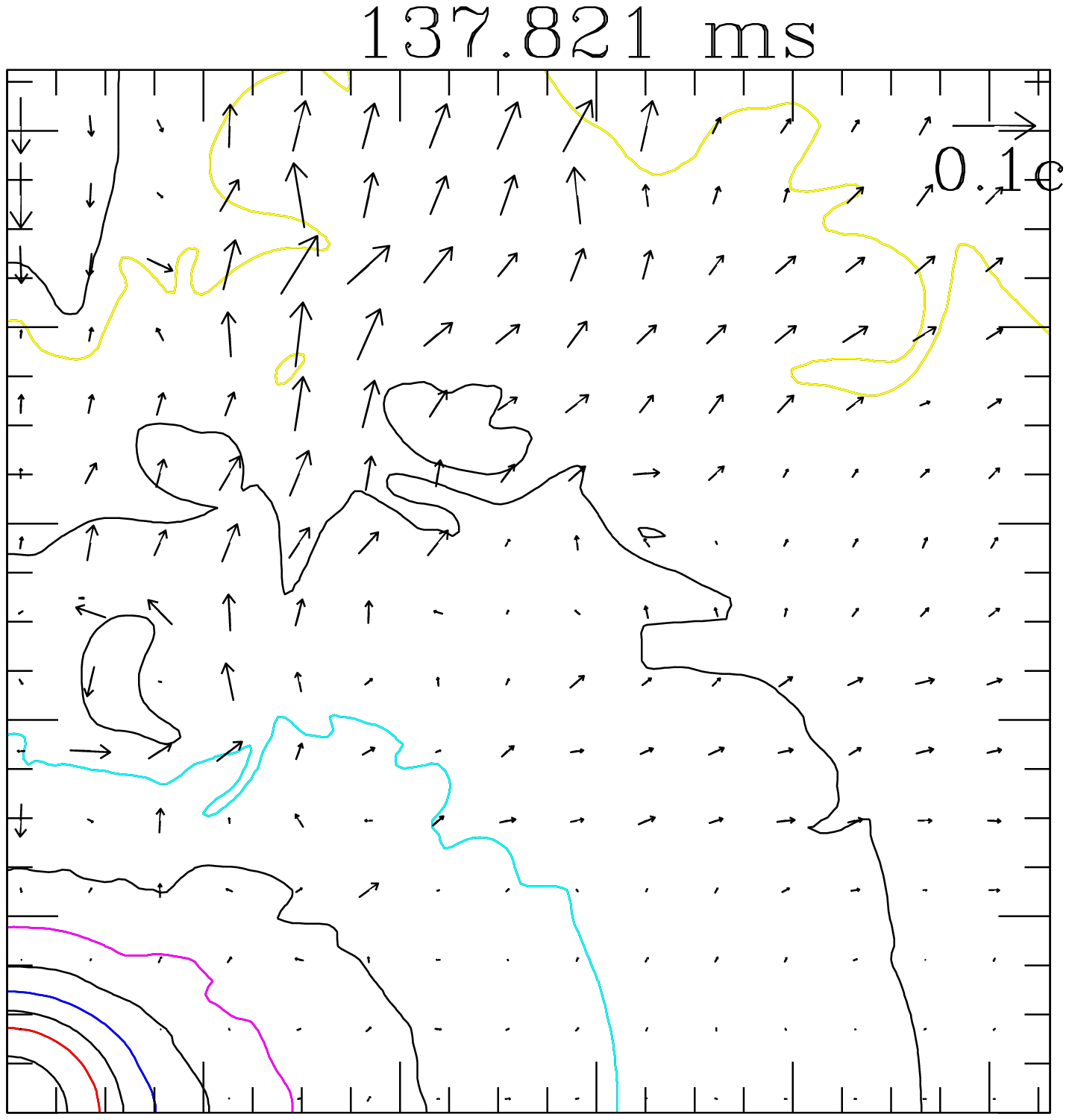}
\epsfxsize=2.15in
\leavevmode
\hspace{-1.45cm}\epsffile{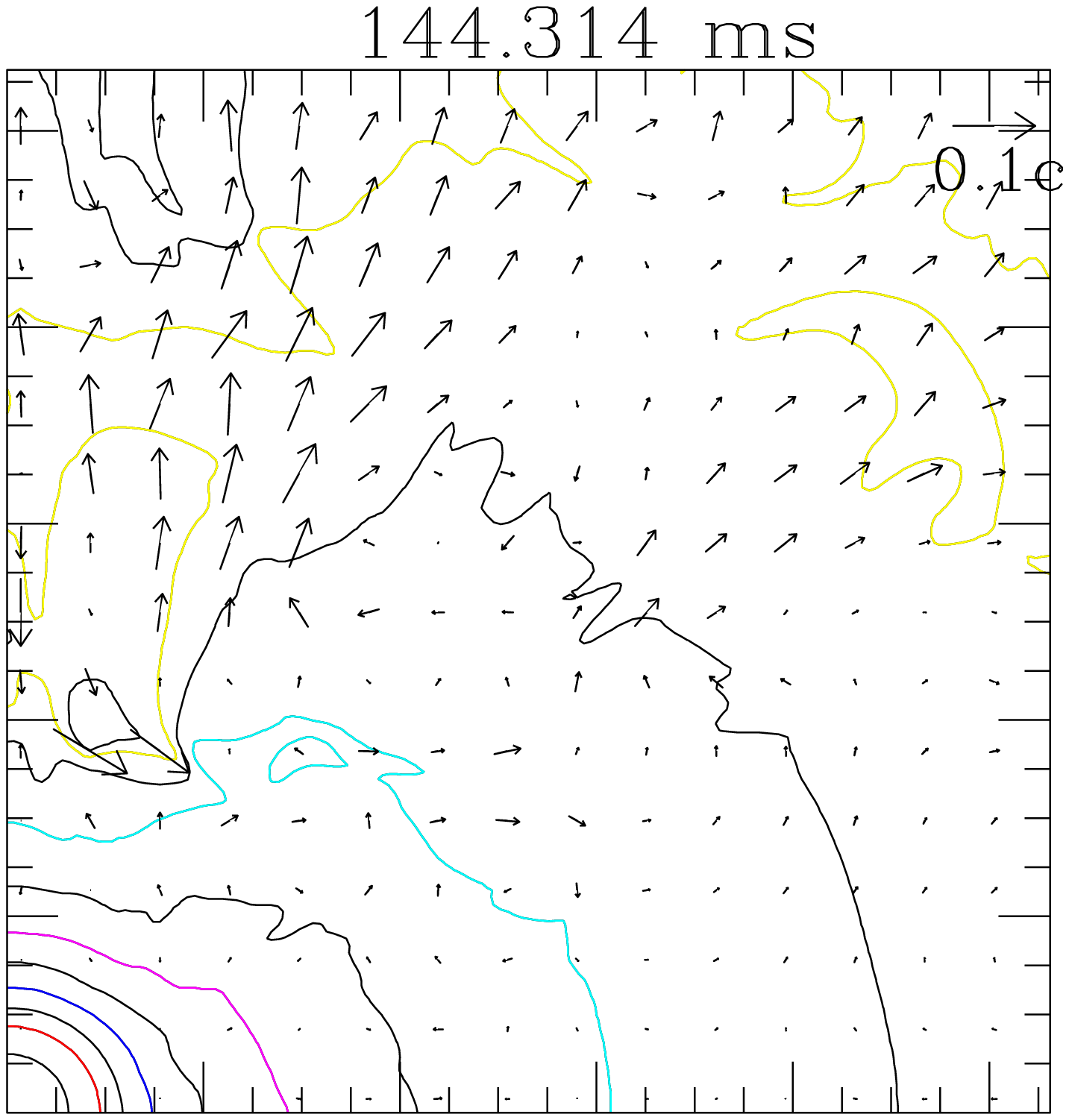}
\epsfxsize=2.15in
\leavevmode
\hspace{-1.45cm}\epsffile{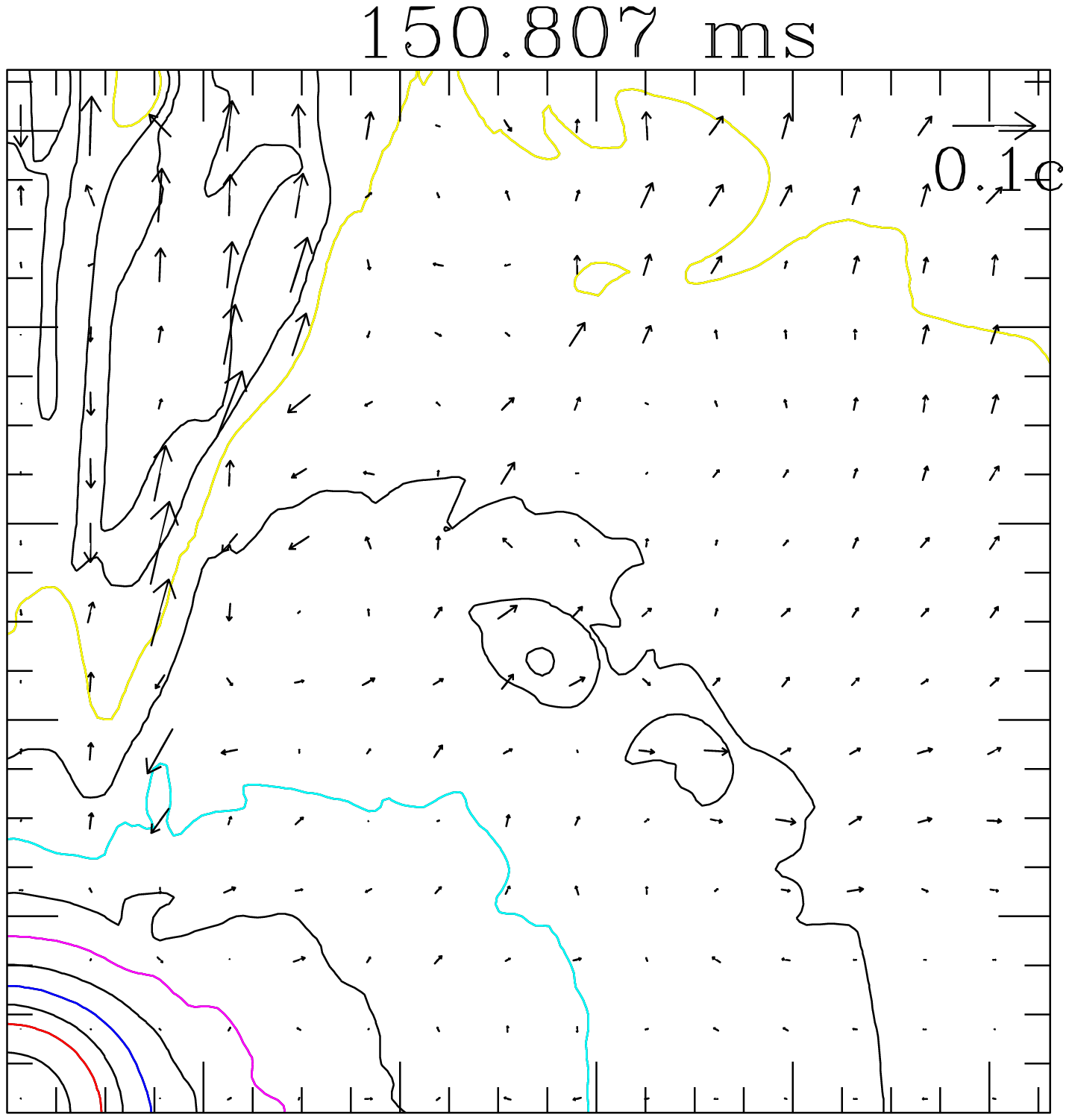}\\
\vspace{-1.6cm}
\epsfxsize=2.15in
\leavevmode
\epsffile{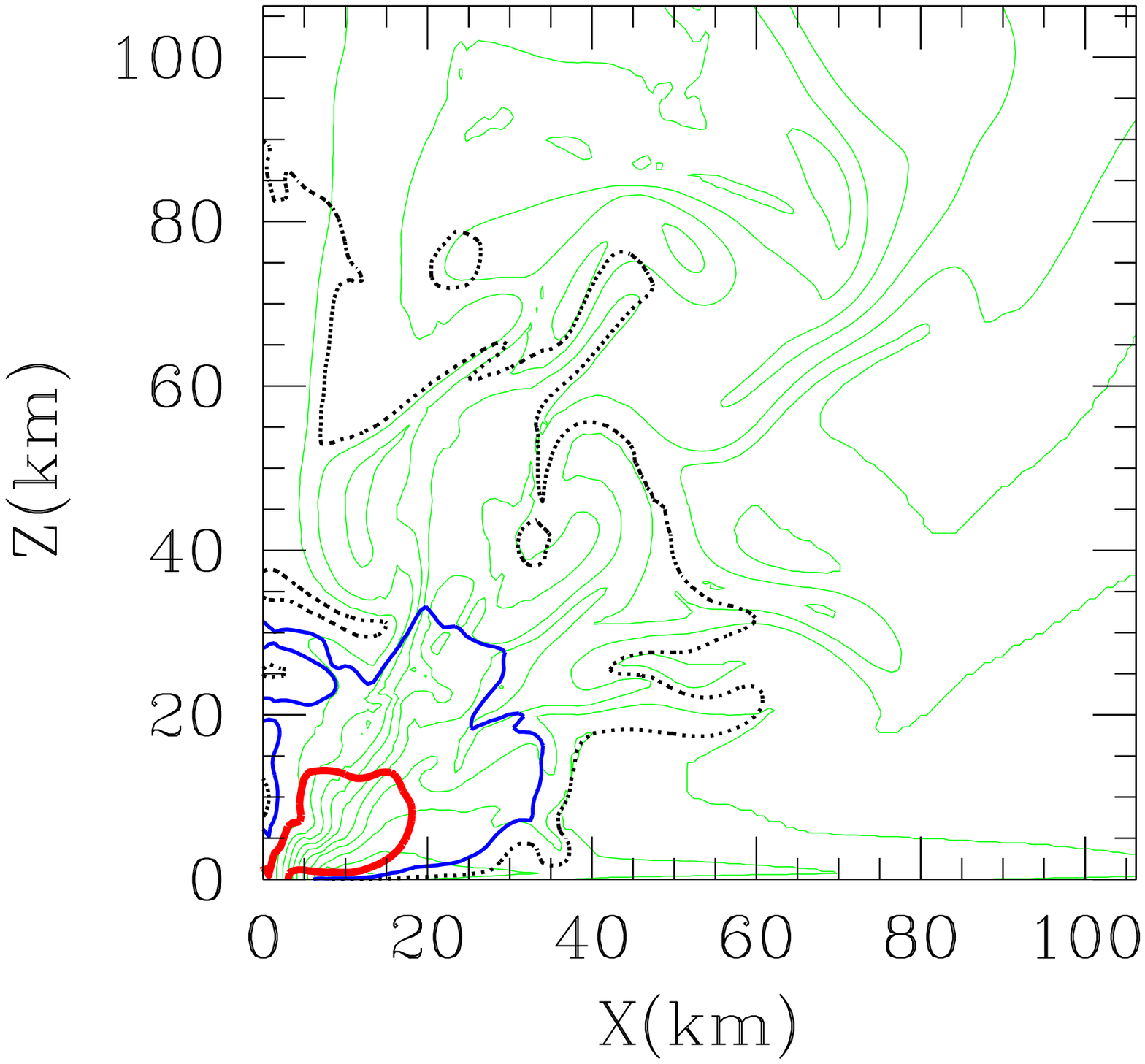}
\epsfxsize=2.15in
\leavevmode
\hspace{-1.45cm}\epsffile{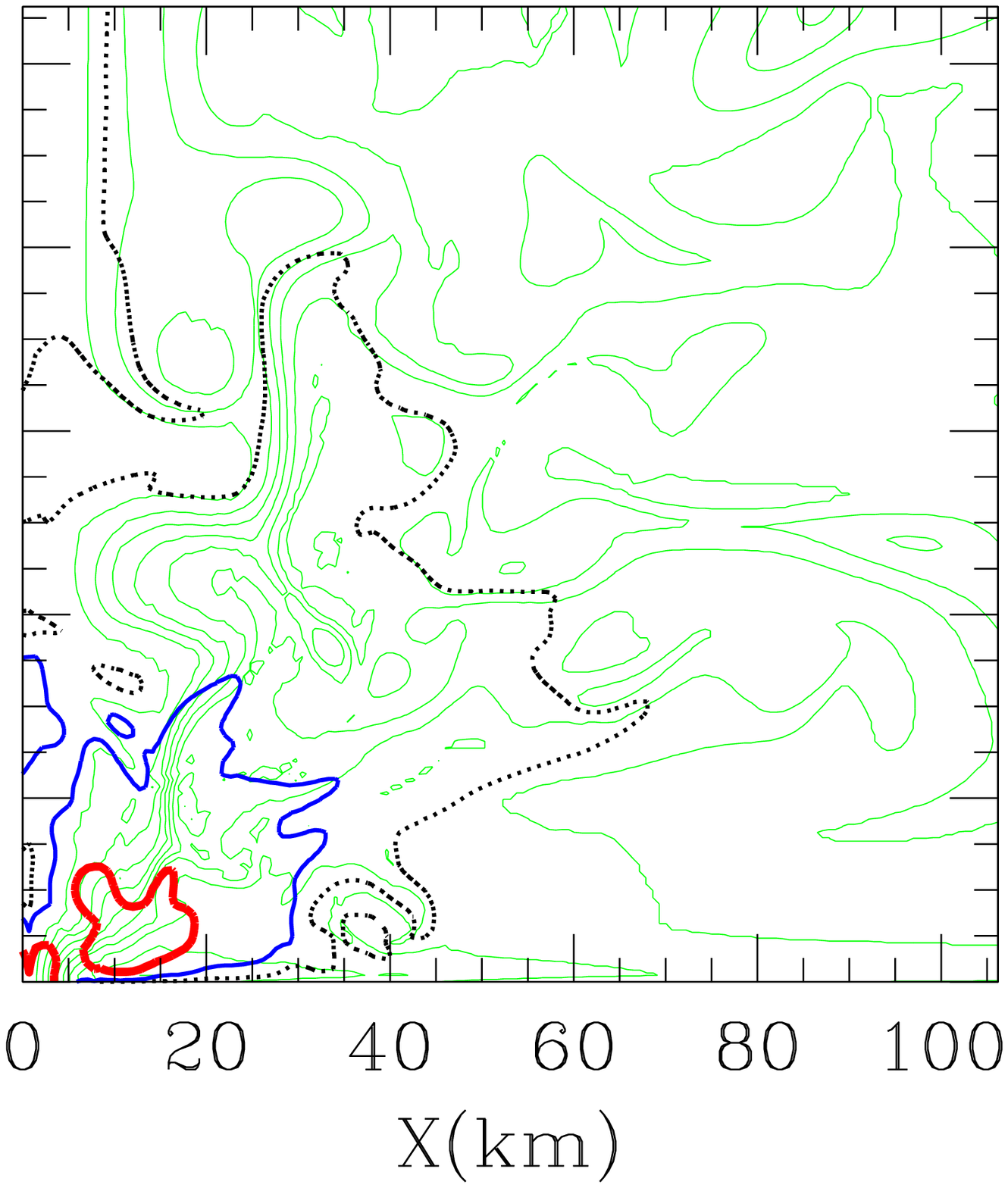}
\epsfxsize=2.15in
\leavevmode
\hspace{-1.45cm}\epsffile{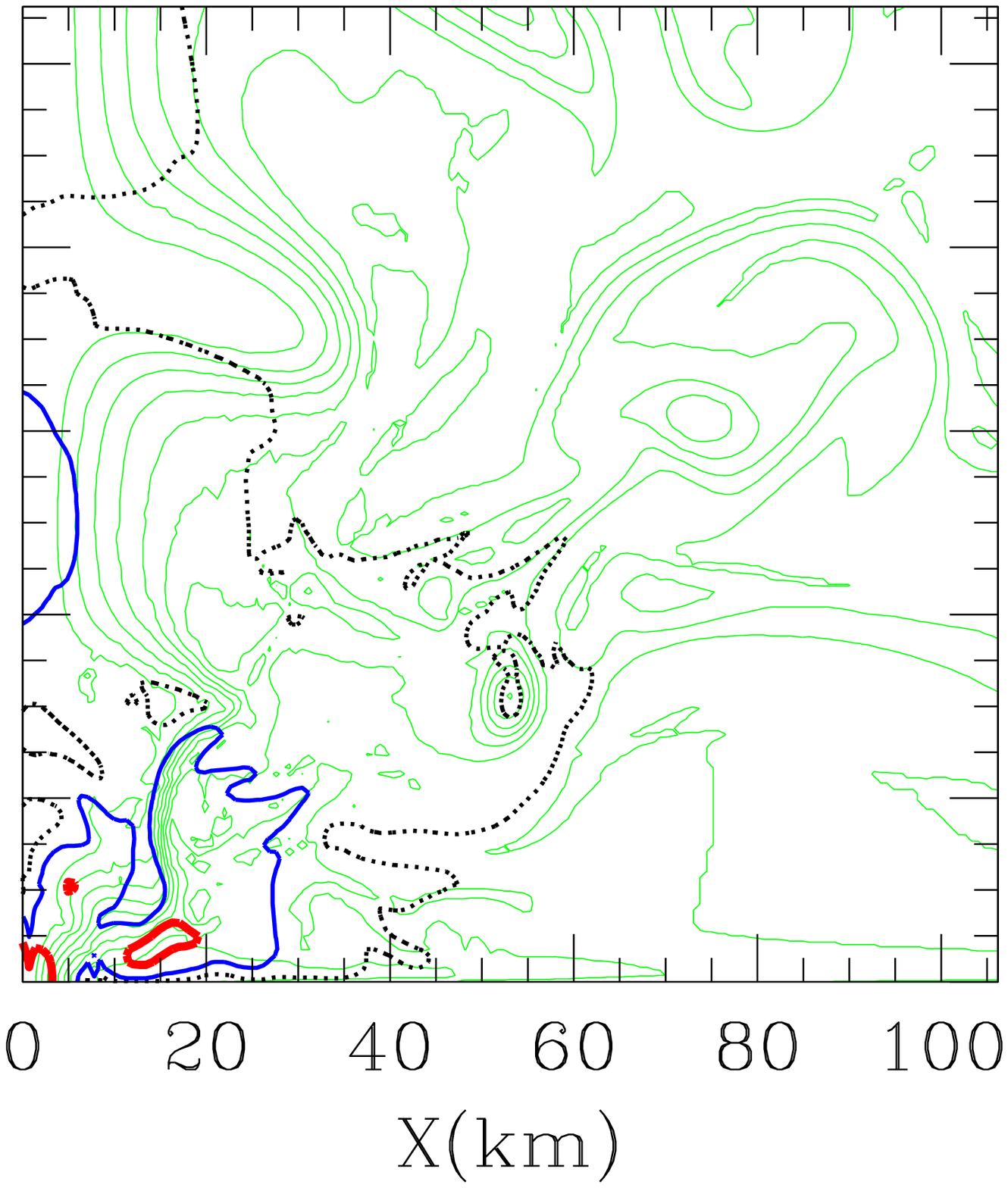}
\epsfxsize=2.15in
\leavevmode
\hspace{-1.45cm}\epsffile{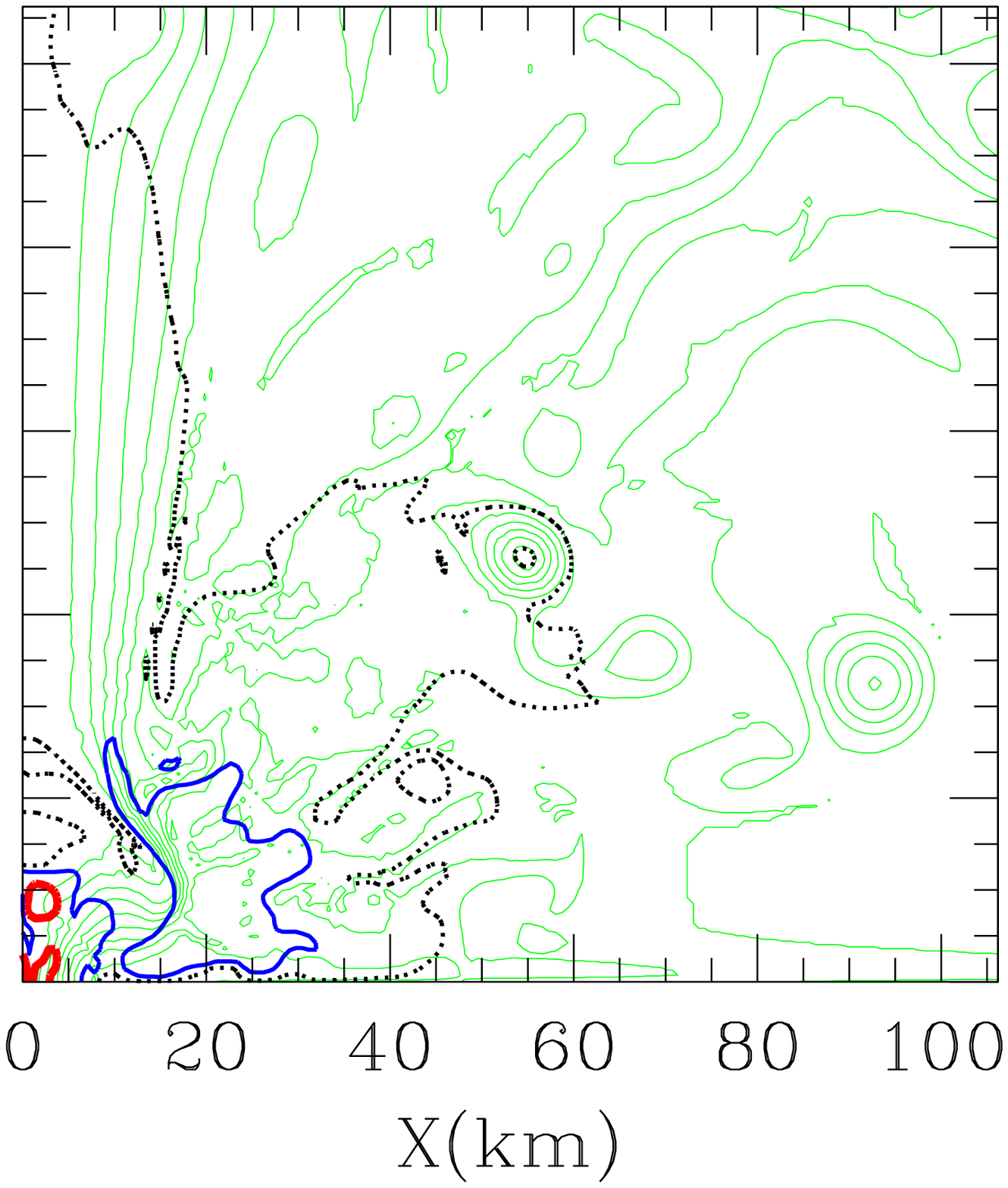}
\vspace{-5mm}
\caption{Density contour curves and velocity vectors for model 
A2 (first and third rows).   The contours are drawn for 
$\rho=10^{15-0.5i}~{\rm g/cm^3}~(i=1$--12).  (The red, blue, magenta, 
cyan, and yellow curves denote
$\rho=10^{14}$, $10^{13}$, $10^{12}$, $10^{11}$, and $10^{10}~{\rm g/cm^3}$,
respectively.)  The scale of the velocity is shown in the upper-right corner. 
The second and fourth rows show the poloidal magnetic field lines
(green) and contours of the magnetic pressure (thick red, blue, and 
normal black curves) at corresponding times.  The poloidal magnetic field 
lines are drawn as contours of $A_{\varphi}$, with levels given by 
$A_{\varphi}= (1-0.1i)A_{\varphi,{\rm max}}~(i=0$--9). 
Here, $A_{\varphi,{\rm max}}$ is the maximum value of $A_{\varphi}$ at
each time slice. The contour curves of the magnetic pressure are drawn 
for $10^{30}$ (very thick red curves), $10^{29}$ (thick blue curves),
and $10^{28}~{\rm dyn/cm^2}$ (thick black dotted curves). 
\label{FIG8}}
\end{center}
\end{figure*}

\begin{figure}[th]
\begin{center}
\epsfxsize=3.4in
\leavevmode
\epsffile{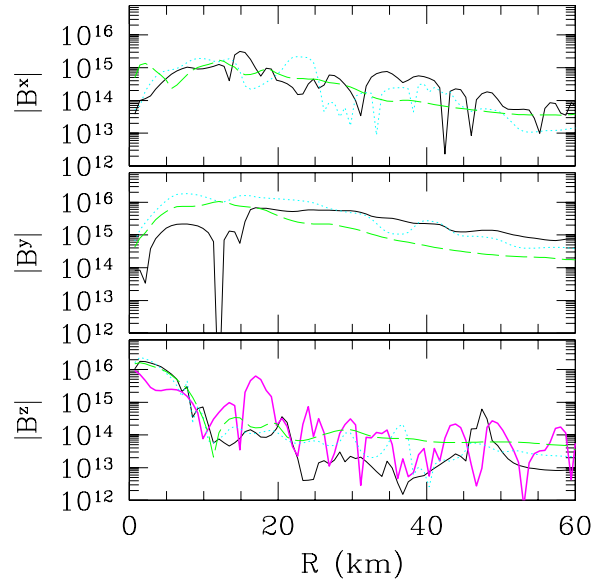}
\vspace{-4mm}
\caption{Evolution of $|B^x|$ and $|B^y|$ at $z \approx 14.2$ km and 
$|B^z|$ in the equatorial plane as a function of the cylindrical 
radius for model A2.  The solid, dotted, and dashed curves denote the 
data for $t=150.8$, 137.8, and 118.3 ms. The thick solid curve for 
$|B^z|$ is the result at $z\approx 10$ km at $t=150.8$ ms. 
\label{FIG9}
}
\end{center}
\end{figure}

To provide an overview of the evolution, we display snapshots 
of density contours, velocity vectors, poloidal magnetic field lines, 
and magnetic pressure contours for model A2 at selected time slices in 
Fig.~\ref{FIG8}.  In Fig.~\ref{FIG9}, the evolution of the magnetic 
field strength as a function of the cylindrical radius is shown for 
selected coordinate values of $z$.  (We do not plot $B^x$ and $B^y$ on
the equatorial plane since they vanish there by symmetry.)  We find 
that, after the bounce at $t \approx 115$~ms, shock waves propagate 
outward. Material behind the shocks moves outward in an anisotropic 
manner due to the rotation and to the anisotropic density profile of 
the collapsing star.  This matter outflow is strongest along the 
rotation axis. 

For $t \agt 120$ ms, differential rotation modifies the profile of the
magnetic field and amplifies the magnetic pressure (see Figs.~\ref{FIG7}, 
\ref{FIG8}, and Fig.~\ref{FIG9}). As Fig.~\ref{FIG7}
shows, the toroidal field continues to grow after bounce because of
the differential rotation near the surface and outside of the
PNS. The magnetic pressure monotonically
increases for 120~ms $\alt t \alt 130$~ms in the region 
$\varpi=10$--30 km and $z \alt 10$~km.  This is reflected in the 
third through fifth snapshots in Fig.~\ref{FIG8}, which show that the 
region with $P_{\rm mag} \geq 10^{30}~{\rm dyn/cm^2}$ expands. We note 
that the amplification at $z=0$ is very small since we impose the 
symmetry condition that $B^x$ and $B^y$ must vanish on the equatorial 
plane.  Comparison of Figs.~\ref{FIG6} and~\ref{FIG7} illustrates 
that the increase of the magnetic energy is 
primarily due to the growth of the toroidal field. This growth stops 
when the magnetic energy reaches about 10--20\% of the rotational 
kinetic energy irrespective of the initial field strength and profile 
(cf. Fig.~\ref{FIG6}). Figure~\ref{FIG7} shows that the toroidal field 
growth can be followed well even for the lower grid resolutions.


The growth of magnetic energy saturates when back-reaction from
magnetic braking becomes significant.  This occurs at a time
comparable to the Alfv\'en crossing time~\cite{Shapiro} $t_{\rm sat}
\approx t_{\rm A}$, where $t_A$ is given by Eq.~(\ref{eq:alfventime}).
For the models considered here, $t_{\rm A} \alt 30$~ms.  From the
numerical results for model A2, we see that $t_{\rm sat}\sim 10$~ms
(see Fig.~\ref{FIG7}).


The toroidal magnetic field would not achieve the saturation 
strength described above if significant magnetic reconnection 
were to operate before magnetic braking could take effect.  
However, the velocity of fluid entering the dissipation layer 
is expected to be a small fraction (probably $10^{-1}$ or $10^{-2}$)
of the Alfv\'en velocity (see, e.g.~\cite{parker,priest,Kulsrud}).  
Simple estimates show that this expectation is borne out in the solar 
magnetosphere and the geomagnetic tail of the earth~\cite{Kulsrud}.  
Thus, the reconnection timescale should, in general, be of order 10 
to 100 times longer than the Alfv\'en timescale which governs the 
growth of the toroidal field.  On the timescales of interest for
PNS evolution considered here, reconnection will thus not play an
important role.

Following saturation, the toroidal field strength begins to
decrease in the central region due to magnetic braking. This is
reflected in the decrease of the magnetic pressure for $\varpi\sim 10$
km and $z \sim 10$ km seen in the sixth and seventh snapshots of
Fig.~\ref{FIG8}. Magnetic braking also leads to 
outward transport of angular momentum by Alfv\'en waves. This is 
clearly seen in the middle panel of Fig.~\ref{FIG9}; $|B^y|$ decreases 
for $\varpi \alt 15$ km but increases for $\varpi \agt 15$ km.  The 
middle and lower panels of Fig.~\ref{FIG3} also show that the angular 
velocity in the central region decreases while it increases for 
$10~{\rm km} \alt \varpi \alt 50$~km for models A1 and A2. This 
angular momentum transport drives matter outflow at relatively low 
latitude (see for $t \agt 130$ ms in Fig.~\ref{FIG8}).  However 
this outflow is not as strong as that along the rotation axis~\cite{note2}.

At bounce, the compression stops and so does the rapid global
growth of the poloidal field.  However, the poloidal 
field in the outer region of the PNSs (see Fig.~\ref{FIG8}) 
continues to grow after the bounce by the MRI. 
Figure~\ref{FIG8} indicates that the poloidal field lines for 
$\varpi \approx 10$--30~km and $z=10$--50~km are highly distorted 
for $t \agt 130$~ms.  Evidence of the MRI is also seen 
in the first and third panels of Fig.~\ref{FIG9} which show that
initially smooth poloidal fields become highly inhomogeneous and 
the amplification is locally enhanced~\cite{note2} (the local 
amplification can be seen clearly in the region $R \sim 35$~km in 
the $B^x$ plot in Fig.~\ref{FIG9}).

The growth $|B^x|_{\rm max}$ shown in Fig.~\ref{FIG7}(a)
demonstrates the importance of resolution in capturing the MRI. 
In this figure, we plot results for four different resolutions. In 
the low and middle resolutions, the maximum value of $|B^x|$ increases 
until the bounce and holds steady thereafter. On the other hand, 
there is significant growth {\em after} the bounce in the standard and high 
resolution runs. Note that the wavelength of the fastest-growing mode 
of the MRI, $\lambda_{\rm max}$, is a few km for $B\sim 10^{14}$~G, 
$\rho \sim 10^{13}~{\rm g/cm^3}$, and $\Omega \sim 10^{3}~{\rm rad~s}^{-1}$
[cf. Eq. (\ref{lambdamax})] which are typical values in the outer
region of the PNS. In the standard resolution, the grid
spacing is $\sim 0.1\lambda_{\rm max}$, and the MRI is
marginally resolved, but it is unresolved for the low and
medium resolutions. This illustrates that, in MHD simulations, 
integrations with several grid resolutions are essential for correctly 
identifying the physical mechanisms at work.

Figures~\ref{FIG7}(a) and (b) show that $|B^z|_{\rm max}$ does not 
increase significantly. This does not imply that $|B^z|$ does not change 
due to the MRI. Indeed, we find that the $z$-component is amplified in 
the outer regions of the PNSs (see Fig.~\ref{FIG9}). However, 
these local fluctuations are not as large as $B^z$ in the central part of 
the PNSs (where the $z$-component dominates and is fairly constant
following the PNS formation).

The amplification of the magnetic field by the MRI is seen at $\sim\! 
10$~ms after the bounce.  The estimated growth time scale of the MRI
is $t_{\rm MRI} \approx 4/3\Omega$ [cf. Eq. (\ref{tmri2})] which is a
few milliseconds for $\varpi=10$--30 km. Our simulation indicates that
the actual $t_{\rm MRI}$ is somewhat longer than this estimate. The
tendency for the linear analysis to underestimate the MRI growth time
is also seen in our previous results for the evolution of magnetized
neutron stars \cite{DLSSS2}.  The linear analysis may be inaccurate by
a factor of several in the present context since the MRI is treated as
a purely local phenomenon, assuming a uniform background state over
length scales much longer than the wavelengths of the perturbations.
However, since the expected value of $\lambda_{\rm max}$ is only one
order of magnitude smaller than the characteristic radius, this
assumption may not hold. To understand the growth timescale of the MRI
in the PNS, a global analysis is necessary. Indeed, a global
stability analysis for magnetized accretion disks shows that the local
analysis overestimates the growth rate~\cite{CPS}. 

As in the case of the toroidal field, the growth of the poloidal field
saturates when $|B^x|_{\rm max}$ reaches $\sim\! 2$--$3 \times
10^{16}$ G.  The saturation may be partly due to the limited grid
resolution (i.e., due to numerical diffusion), but this effect does
not seem to be very large since the value of $|B^x|_{\rm max}$ is
approximately the same for the high and standard resolutions (see
Fig.~\ref{FIG7}).

The contribution to the growth of the magnetic field strength 
by winding is significantly larger than the contribution from the MRI.  
Indeed, we find that the average value of the poloidal field is smaller 
than that of the toroidal field. Because MRI is a local instability,  
it does not increase the global magnetic energy as significantly as winding. 

After the saturation of the field growth by winding and the MRI, a
collimated and nearly stationary magnetic field with $|B^z| \gg
|B^{\varpi}|$ is formed near the rotation axis (see the last snapshot
of Fig.~\ref{FIG8}). For the high latitude regions with $z \agt 50$ km
and with $\varpi \agt 10$ km, the absolute value of the toroidal
component $|B^y|$ is as large as $|B^z|$, indicating a helical
magnetic field. (On the other hand, very near the rotation axis,
$|B^z|$ is much larger than $|B^y|$). The field strength at $t \sim
150$ ms is $\agt 10^{15}$ G at $z=100$ km. Since the degree of
differential rotation is small near the rotation axis, this magnetic
field configuration is stable.  A nearly stationary outflow is driven
along the collimated field lines.  This is probably due to the
magneto-centrifugal mechanism (cf. Sec.~\ref{sec:BP}), and we discuss
this possibility in more detail in Sec.~\ref{sec:BB}.  In contrast to
the region near the axis, the toroidal field is much stronger than the
poloidal field at large cylindrical radius.

We find that the evolution path of the magnetic field depends
quantitatively (but not qualitatively) on the initial field strength
and profile.  In the right panel of Fig.~\ref{FIG7}, we show the
evolution of each component of the magnetic field for models A1, A2,
and A4 (see also Fig.~\ref{FIG6}).  For A1, the initial magnetic field
profile is the same as that for A2 while the initial strength is
weaker. The evolution of the magnetic field is similar for these two
models except that, for A1, it takes longer to amplify the magnetic
field because of the smaller initial value of $B^{\varpi}$.

\begin{figure*}[t]
\vspace{-4mm}
\begin{center}
\epsfxsize=2.15in
\leavevmode
\epsffile{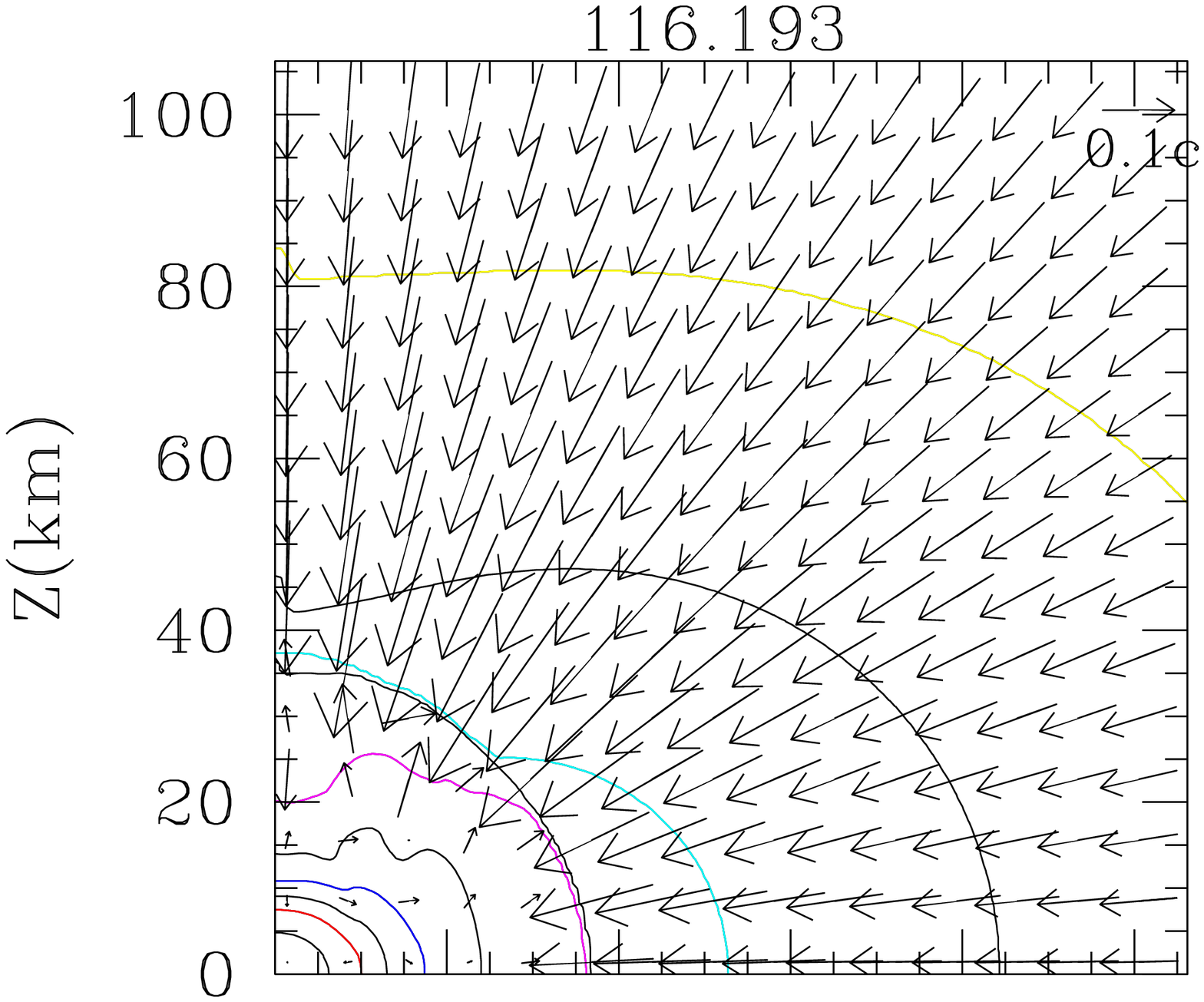}
\epsfxsize=2.15in
\leavevmode
\hspace{-1.45cm}\epsffile{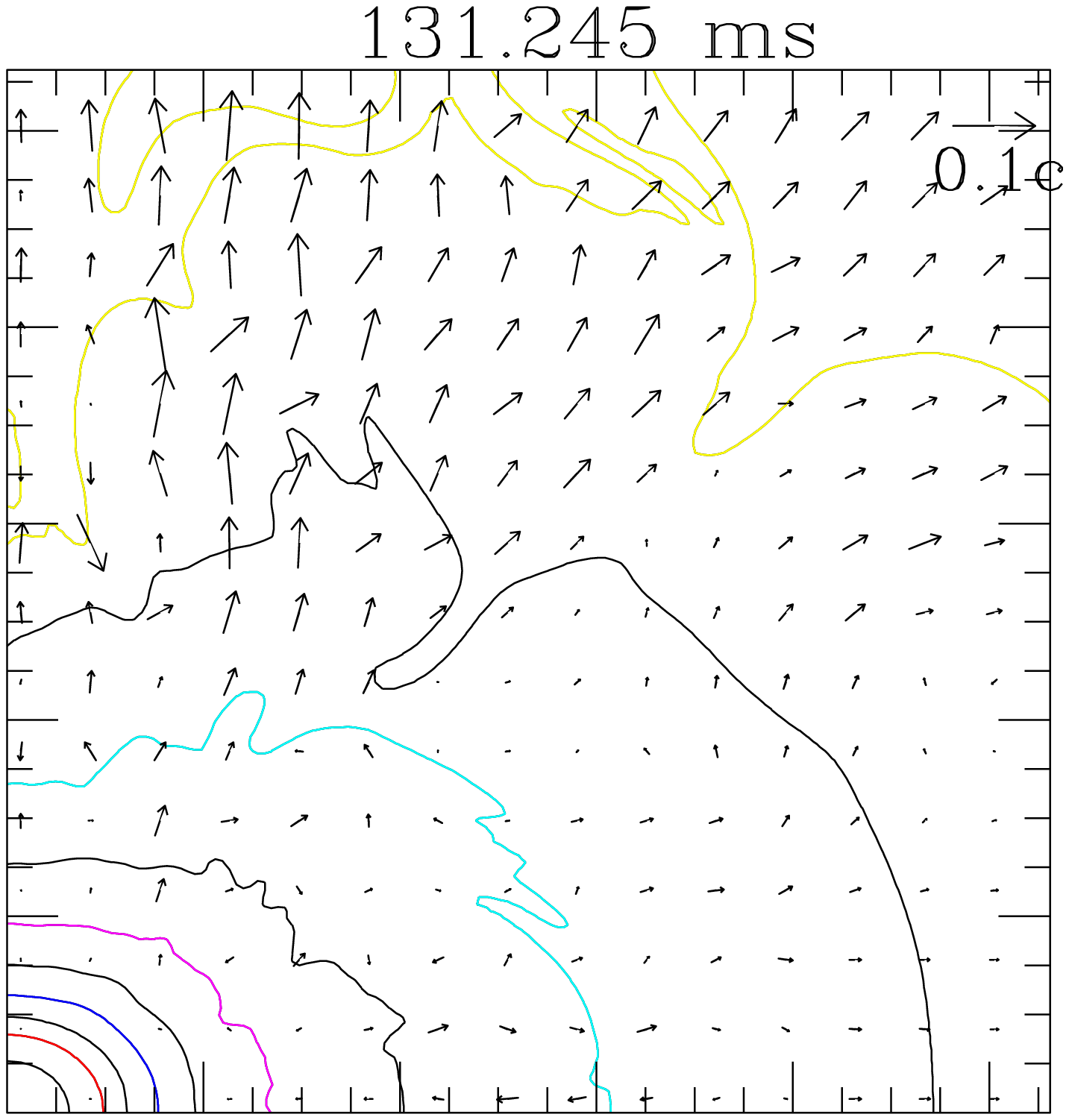}
\epsfxsize=2.15in
\leavevmode
\hspace{-1.45cm}\epsffile{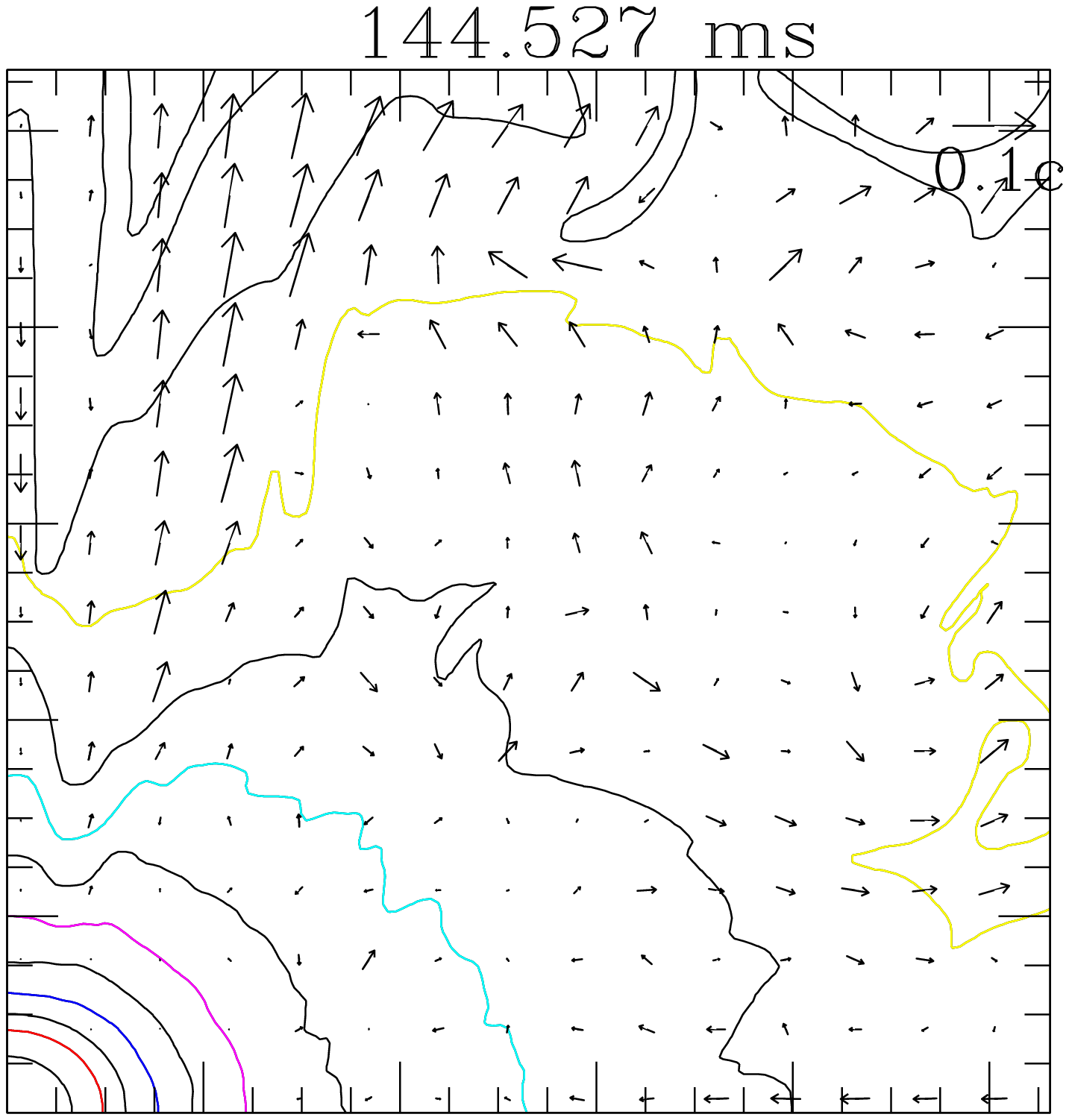}
\epsfxsize=2.15in
\leavevmode
\hspace{-1.45cm}\epsffile{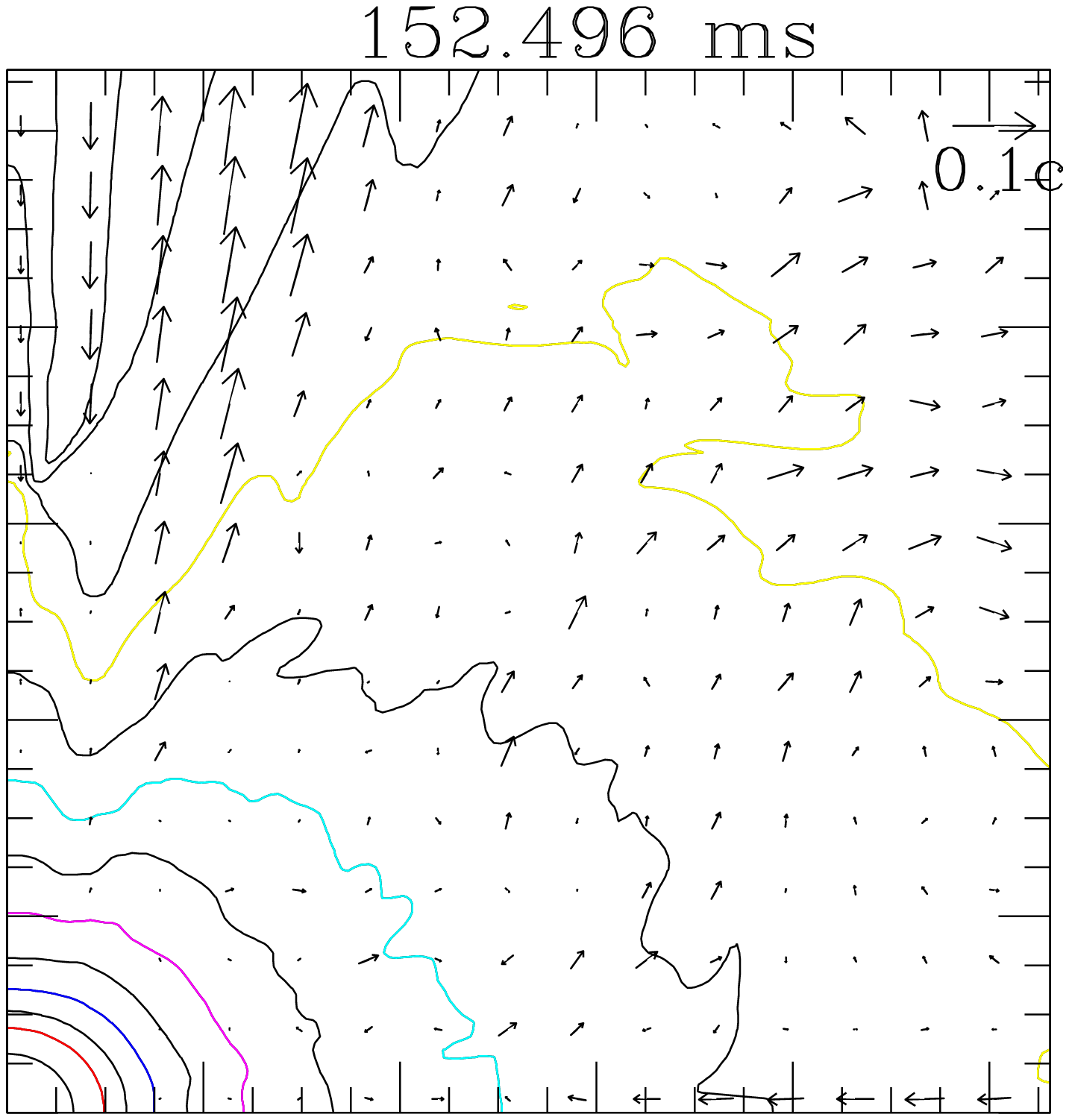}\\
\vspace{-1.6cm}
\epsfxsize=2.15in
\leavevmode
\epsffile{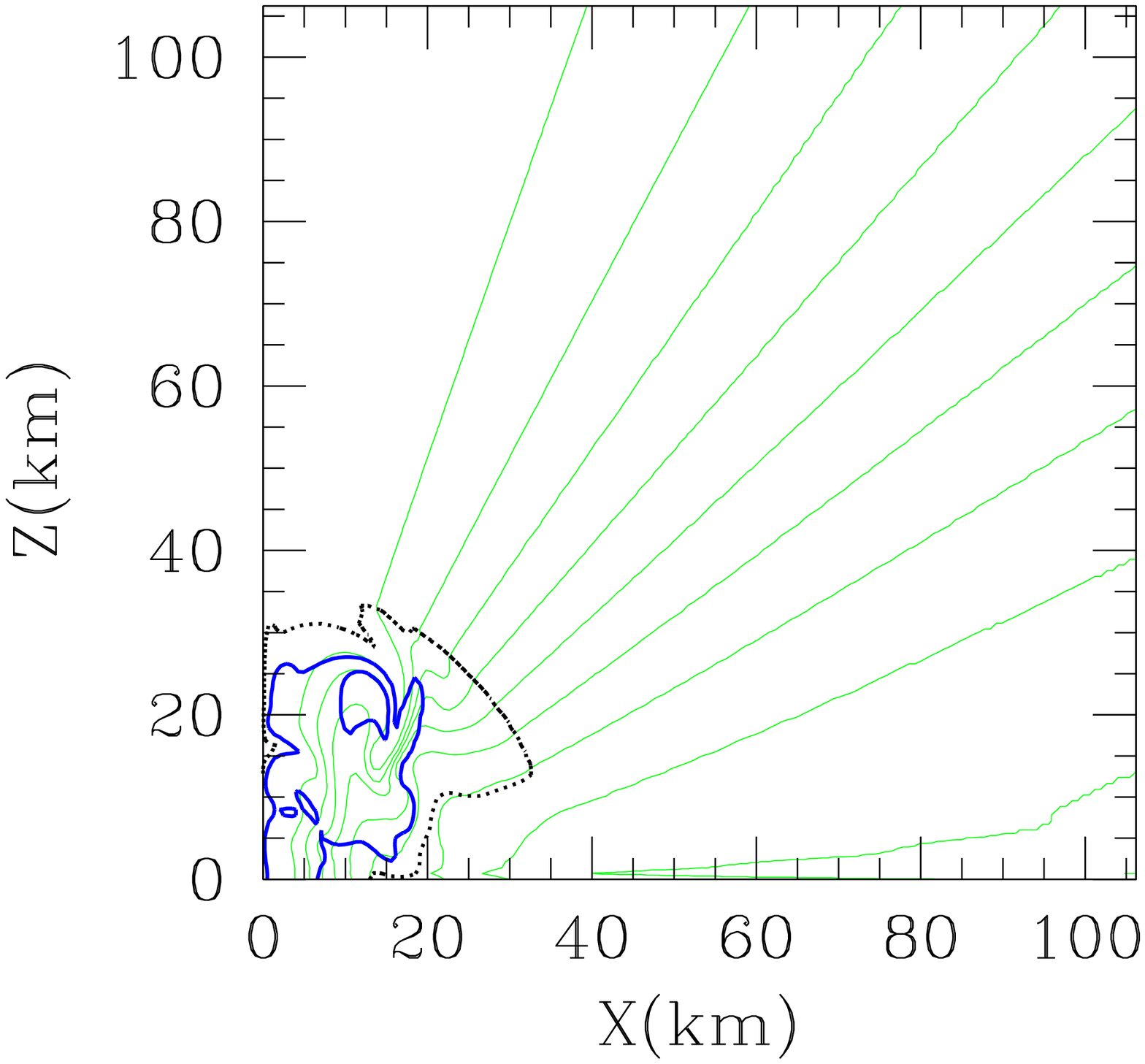}
\epsfxsize=2.15in
\leavevmode
\hspace{-1.45cm}\epsffile{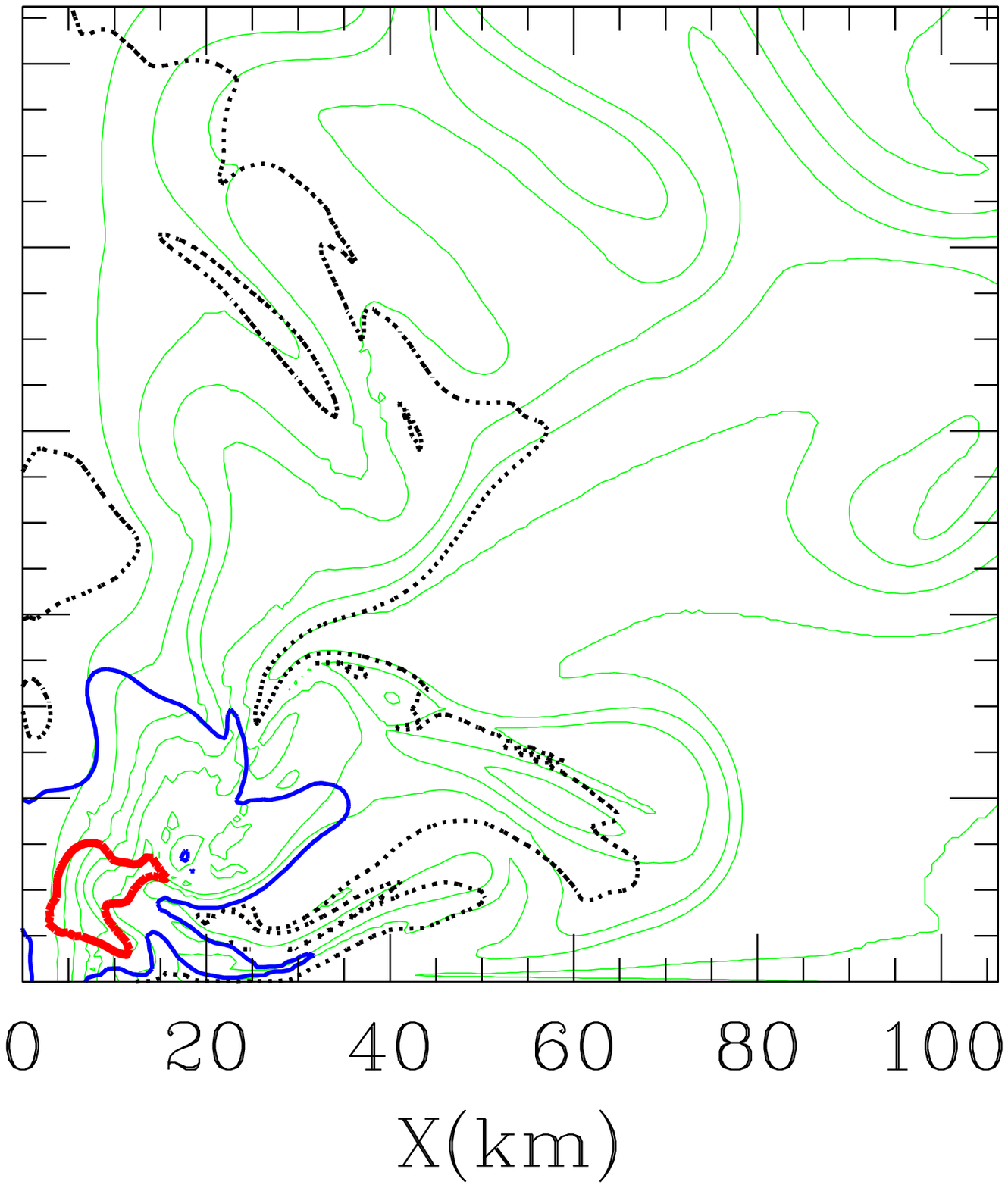}
\epsfxsize=2.15in
\leavevmode
\hspace{-1.45cm}\epsffile{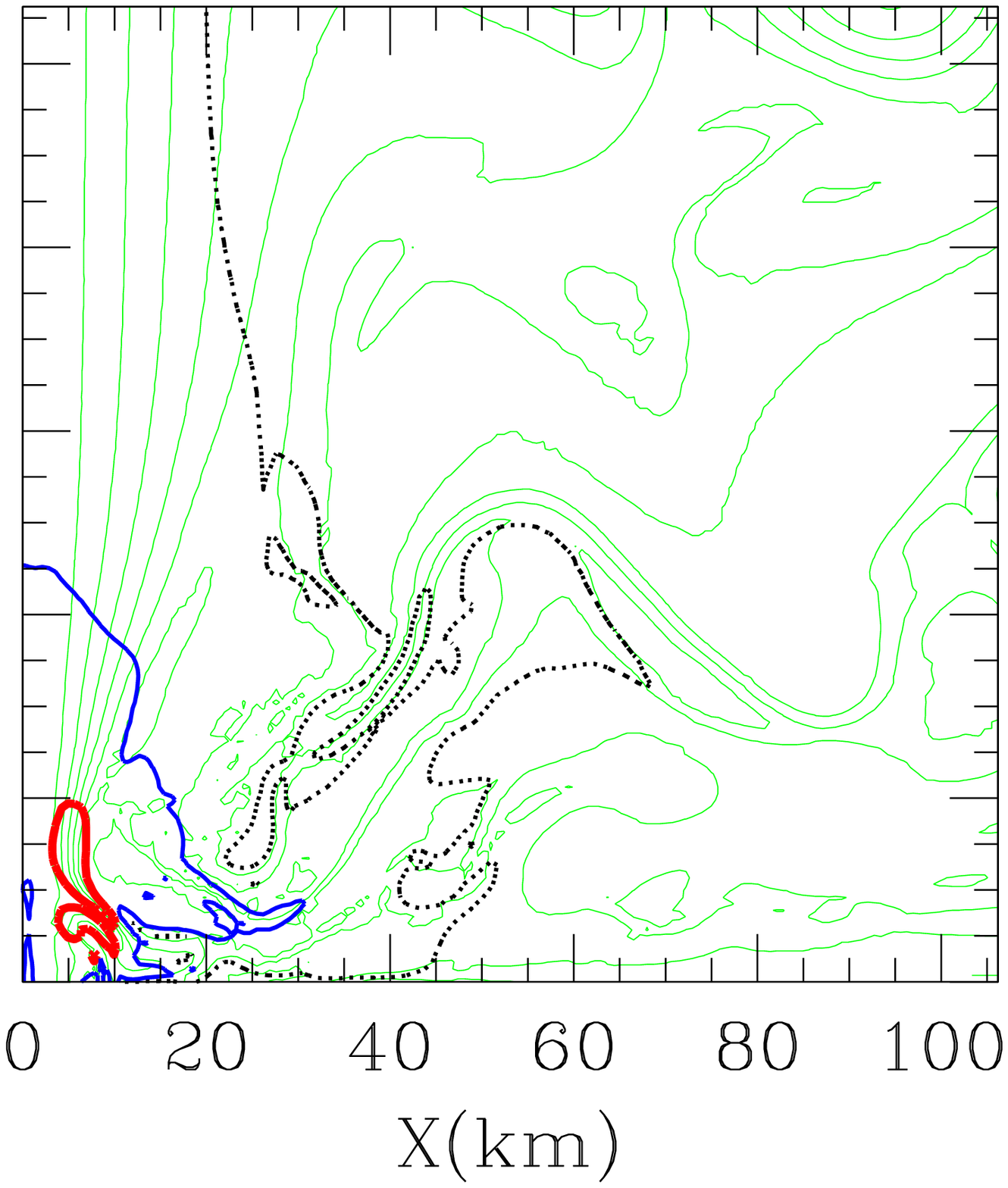}
\epsfxsize=2.15in
\leavevmode
\hspace{-1.45cm}\epsffile{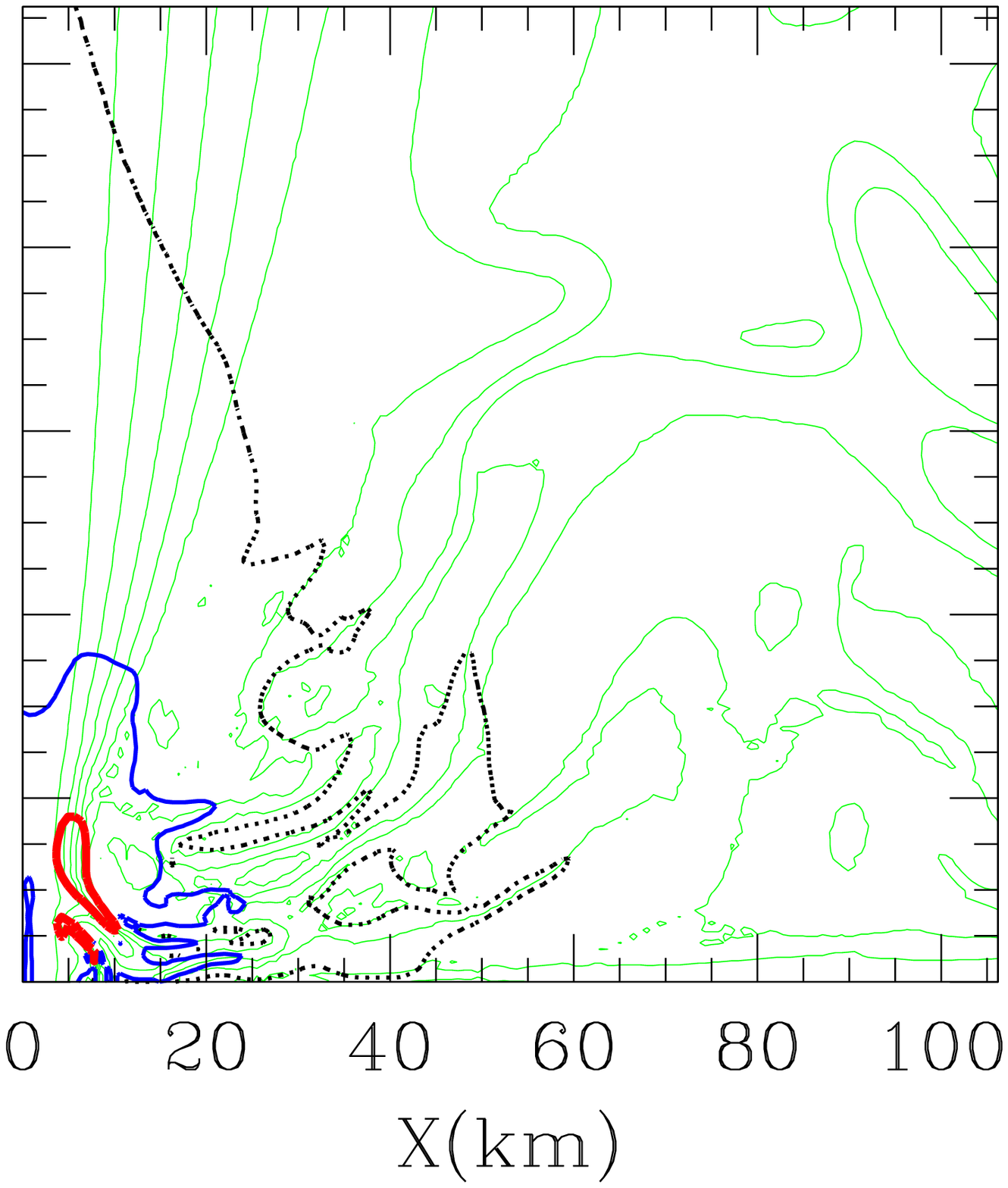}
\vspace{-5mm}
\caption{
Evolution of the density contour curves and velocity vectors for model A4.  
The contours and velocity vectors are drawn in the same manner as in
Fig.~\ref{FIG8}.
\label{FIG10}}
\end{center}
\end{figure*}

Model A4 has a different initial magnetic field profile from that of
models A1 and A2. Since the magnetic field is initially distributed
over a larger range of cylindrical radii for A4 (see Fig.~\ref{FIG1}),
field growth by compression is more efficient for A4.  Although the
maximum values of the magnetic field are initially identical for
models A1 and A4 (see Table II), those at bounce are larger for model
A4.  Model A4 also has a different field configuration after bounce
than A1 and A2.  In Fig.~\ref{FIG10}, we show snapshots of the density
contours, the velocity vectors, the magnetic field lines, and contour
curves of the magnetic pressure after the formation of the PNS for
model A4. By comparison between Fig.~\ref{FIG8} and Fig.~\ref{FIG10}
at $t \approx 116$ ms, it is found that the fraction of the
$z$-component of the magnetic field is larger for model A4 than for
model A2. This strong vertical field is advantageous for inducing the
MRI.  In addition, a stationary collimated field is formed more
quickly for A4 than for A2. This results in a stronger MHD outflow
(see Sec. \ref{sec:BB}).

\subsection{MHD outflow}\label{sec:BB}

\begin{figure*}[t]
\vspace{-4mm}
\begin{center}
\epsfxsize=2.15in
\leavevmode
\epsffile{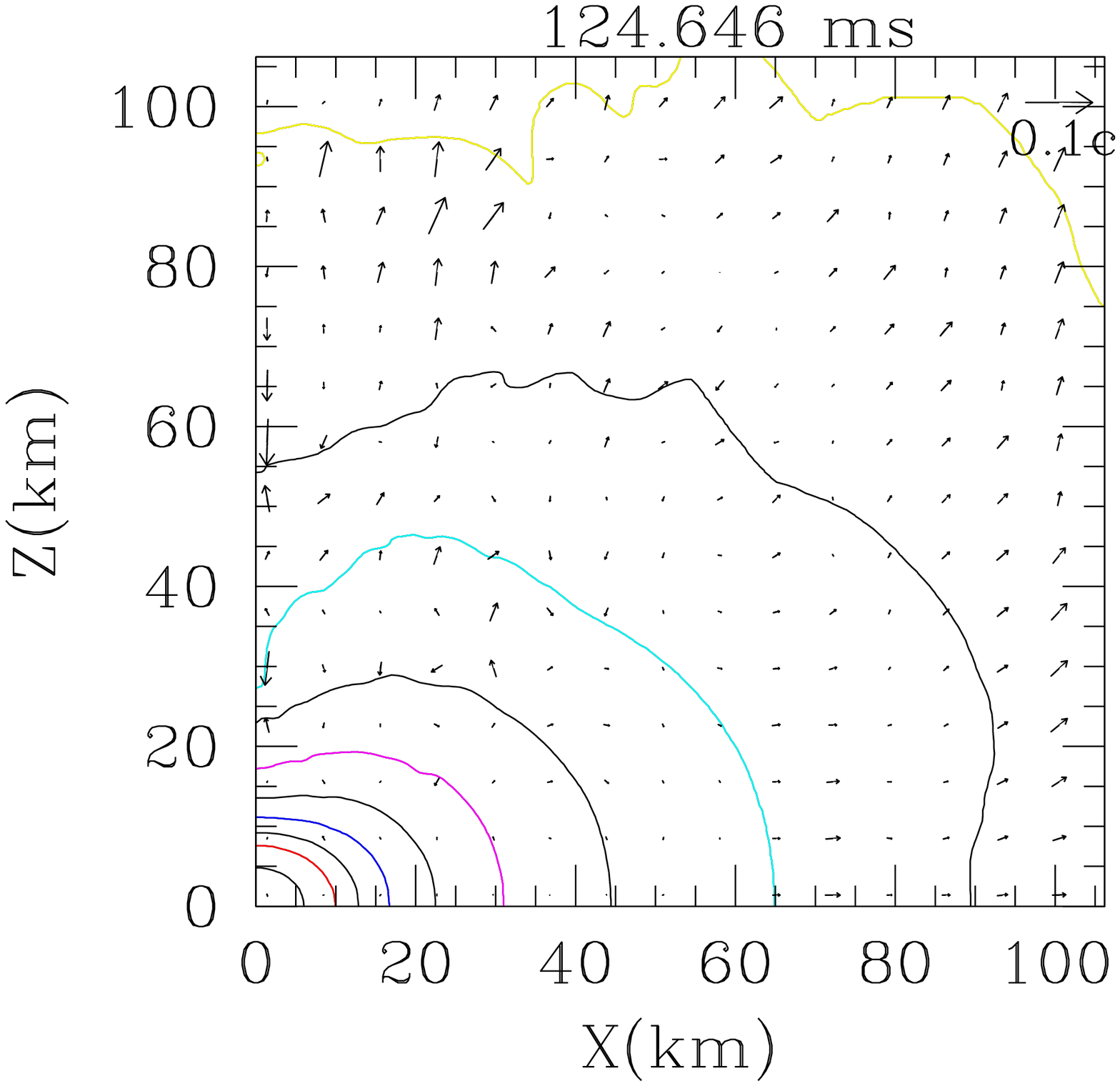}
\epsfxsize=2.15in
\leavevmode
\hspace{-1.45cm}\epsffile{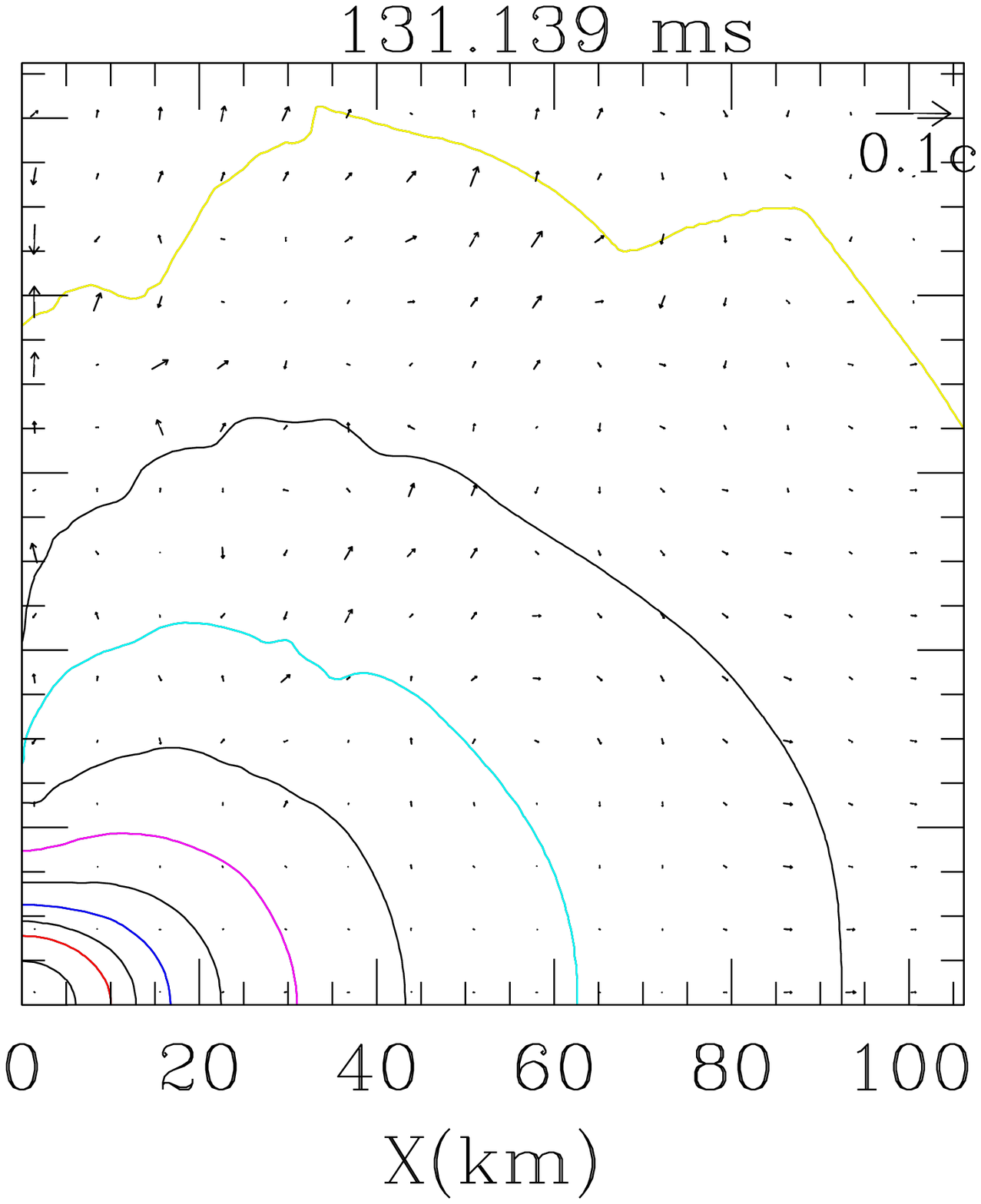}
\epsfxsize=2.15in
\leavevmode
\hspace{-1.45cm}\epsffile{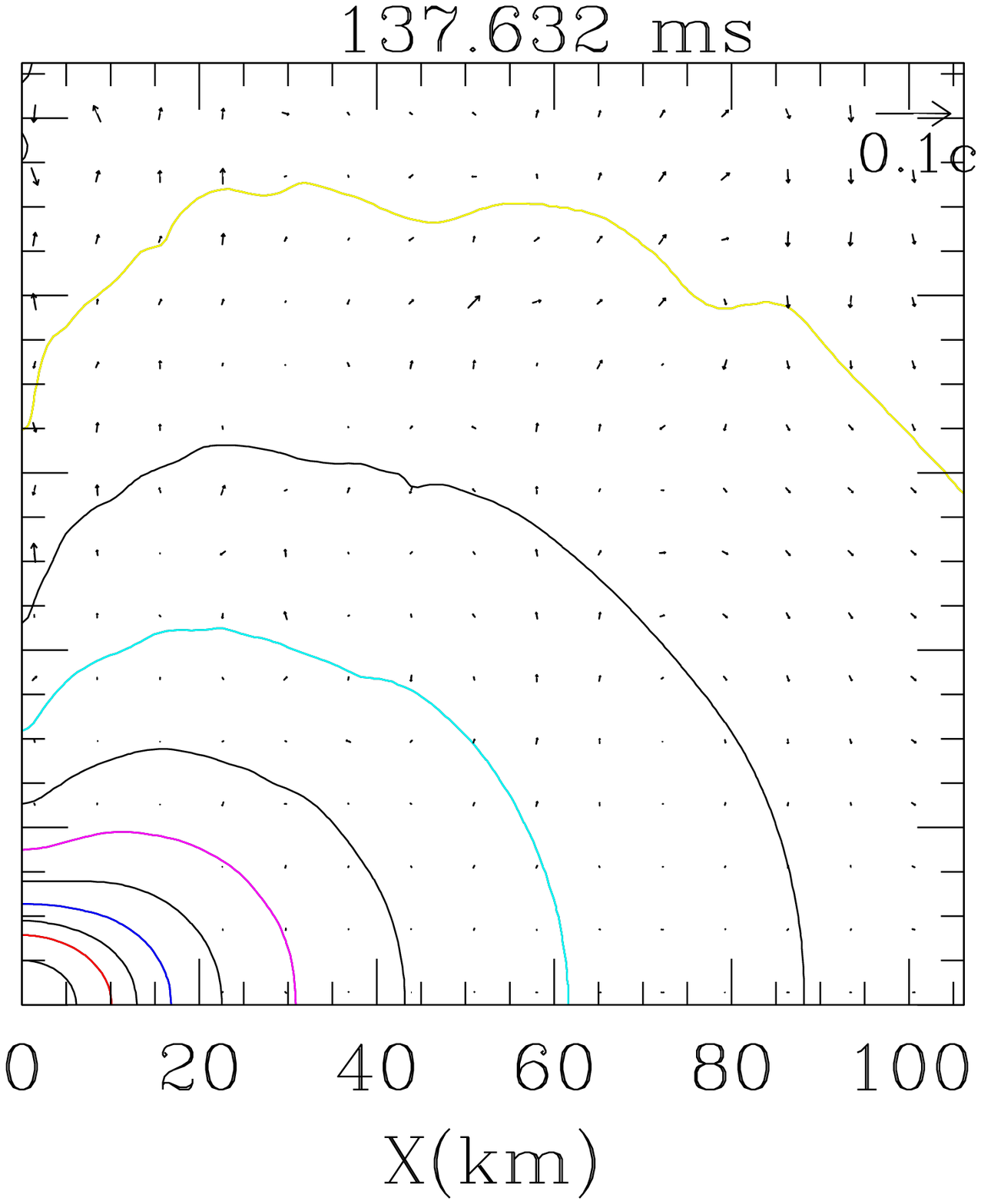}
\epsfxsize=2.15in
\leavevmode
\hspace{-1.45cm}\epsffile{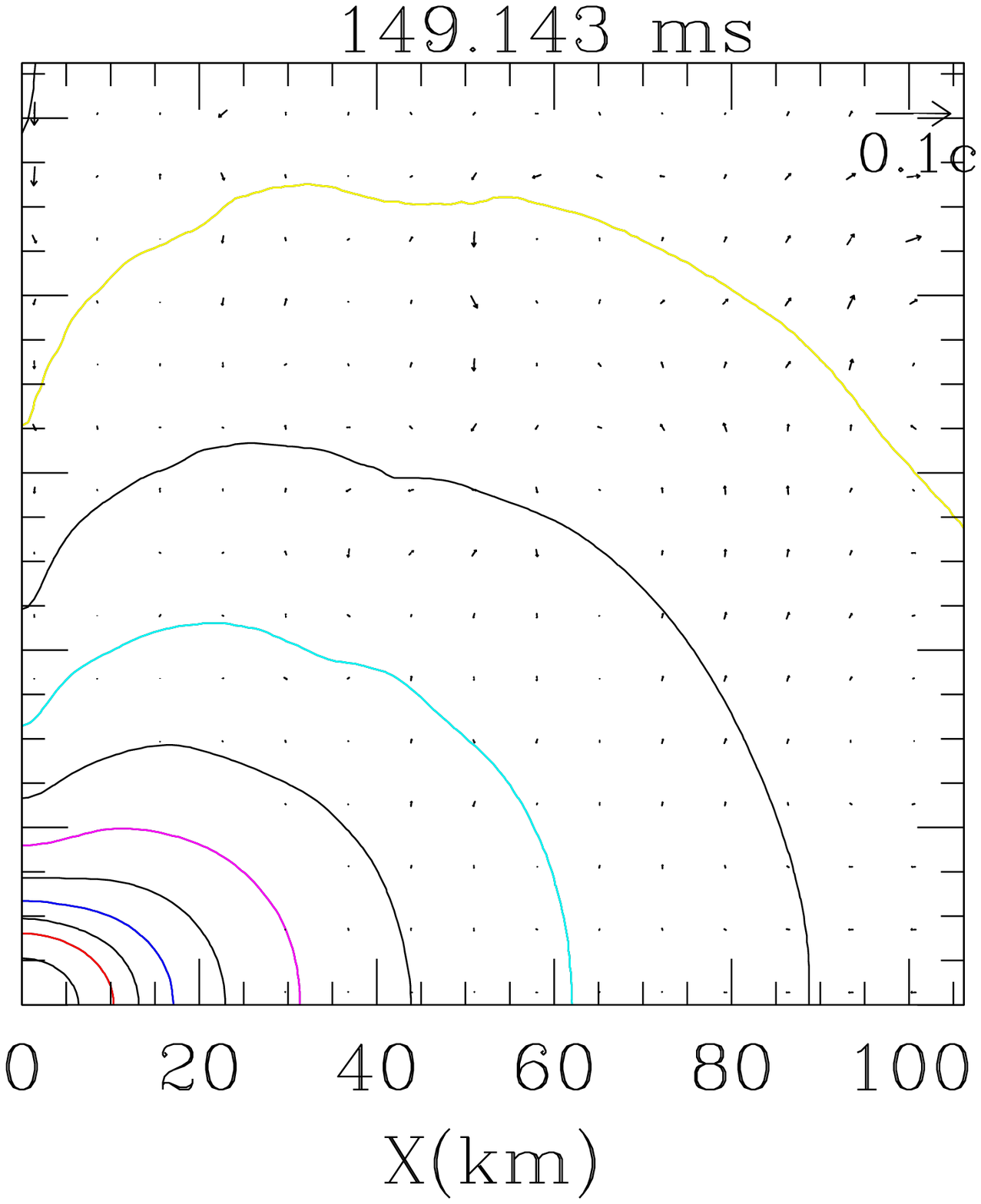}
\vspace{-5mm}
\caption{
Evolution of the density contours and velocity vectors for model A0.  
The contours and velocity vectors are drawn in the same manner as in
Fig.~\ref{FIG8}.
\label{FIG11}}
\end{center}
\end{figure*}

Both winding and the MRI increase the magnetic stress in the outer
regions of PNSs which have large degrees of differential
rotation. This stress induces an MHD outflow, particularly near the
rotation axis (see velocity vectors of Figs.~\ref{FIG8}
and~\ref{FIG10} for $t \agt 130$~ms). To prove that this is indeed due
to the magnetic stress, in Fig.~\ref{FIG11}, we show evolution of the
density contour curves and velocity vectors for model A0, which has no
magnetic field. In this model, the PNS relaxes to a stationary state
for $t \agt 130$~ms. Comparing Figs.~\ref{FIG8}, \ref{FIG10}, and
\ref{FIG11}, it is evident that the MHD outflow is driven by magnetic
effects.  One also sees that the density profile changes due to the
matter outflow (especially near the rotation axis) and that the
oblateness of the PNSs for models A2 and A4 is smaller than for model
A0 at $t \approx 150$ ms due to the angular momentum loss (compare the
density contour curves of $\rho=10^{12}~{\rm g/cm^3}$ for the three
models in the final snapshots).

We posit three possible sources for the MHD outflow: the
magneto-spring mechanism, turbulence induced by the MRI, and the
magneto-centrifugal mechanism. In the early phase, winding amplifies
the toroidal field and the magneto-spring mechanism works efficiently
to drive the MHD outflow. This outflow blows off matter preferentially
in the $z$-direction.  However, it is not sharply collimated; the
half-opening angle is $\sim 45$ degrees. Turbulent motion caused by
the MRI also induces an outflow.  In contrast to the magneto-spring
mechanism, however, the MRI-driven outflow is incoherent.

After the toroidal magnetic field saturates, a stationary, collimated,
helical magnetic field forms around the rotation axis. Matter then
flows out along the collimated field lines for $t \agt 150$~ms in case
A2 and for $t \agt 140$~ms in case A4, as shown in Figs.~\ref{FIG8}
and \ref{FIG10}.  This suggests that the magneto-centrifugal mechanism
(see Sec. \ref{sec:BP}) is operating.  This interpretation can be
inferred indirectly from two facts: (i) the MHD outflow is
continuously ejected from the PNS, even after the growth of the
toroidal field by winding saturates, and (ii) the outflow is absent in
the region very close to the rotation axis. This indicates that the
outflow is driven along the collimated magnetic field with an
appropriate inclination angle between the rotation axis and the
poloidal field lines, as is necessary for magneto-centrifugal
launching~\cite{BP,BP1}.

\begin{figure}[th]
\vspace{-2mm}
\begin{center}
\epsfxsize=3.2in
\leavevmode
\epsffile{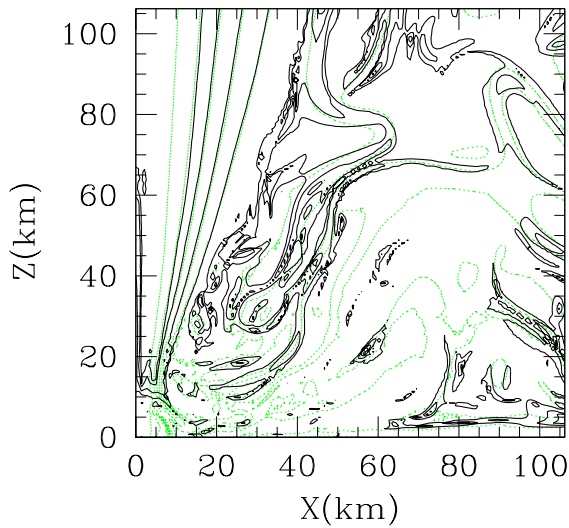}
\vspace{-6mm}
\caption{Contour of $k=\rho_* v^P/\cB^P$ (solid curves) and magnetic 
field lines (dotted curves) at $t=152.5$~ms for model A4.  The magnetic 
field lines are drawn in the same manner as in Fig.~\ref{FIG8}. The 
contours of $k$ are drawn for 
$k=6 \times 10^{4-0.4i}~{\rm g/(cm^2~s~G)}~(i=0$--4).  We note that these 
contours are drawn only for the values of $k$ found for material on the
collimated field lines. 
\label{FIG12}}
\end{center}
\end{figure}

Further evidence is found by confirming that the quantities 
$k \equiv \rho_* v^P/\cB^P$, which 
should be constant along the field lines in a stationary
state are actually constant~(see~\cite{Mestel,BP} and 
Appendix~B). 
Here $v^P$ and $\cB^P$ denote the poloidal components of $v^i$ and
$\cB^i$. Note that the definition of $k$ accounts for general 
relativistic corrections through $\rho_*$ and $\cB^P$. In Fig.~\ref{FIG12}, we 
show the contour curves for $k$ together with the magnetic field lines 
at $t=152.5$~ms for model A4. It is found that the two sets of curves 
overlap in the region of the collimated magnetic field lines, 
verifying that the structure of the magnetic field is almost stationary
in this region. On the other hand, the two sets of curves do
not overlap for the turbulent region far from the rotation axis.

\begin{figure*}[t]
\vspace{-4mm}
\begin{center}
\epsfxsize=3.1in
\leavevmode
\epsffile{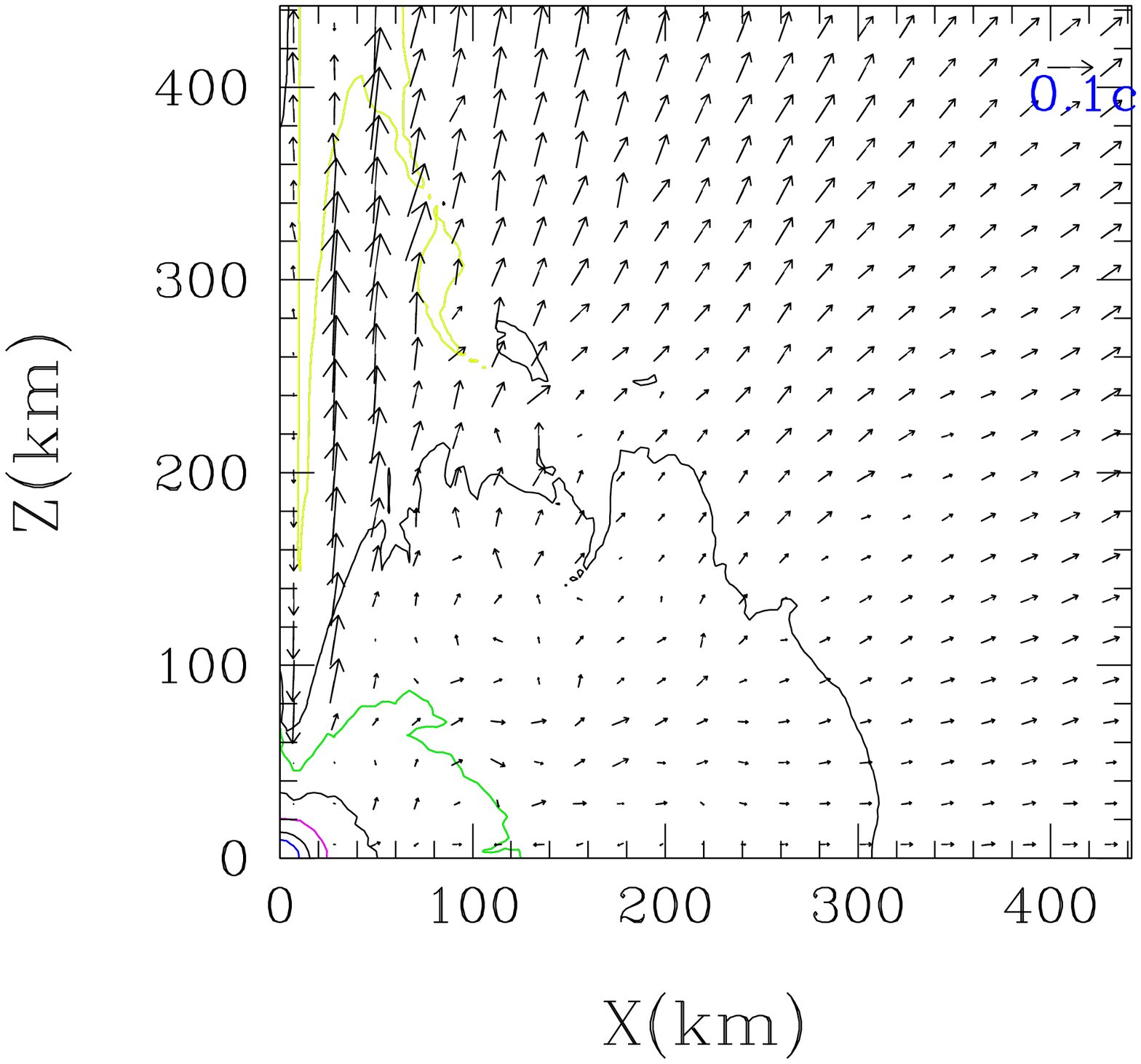}
\epsfxsize=3.1in
\leavevmode
~~~\epsffile{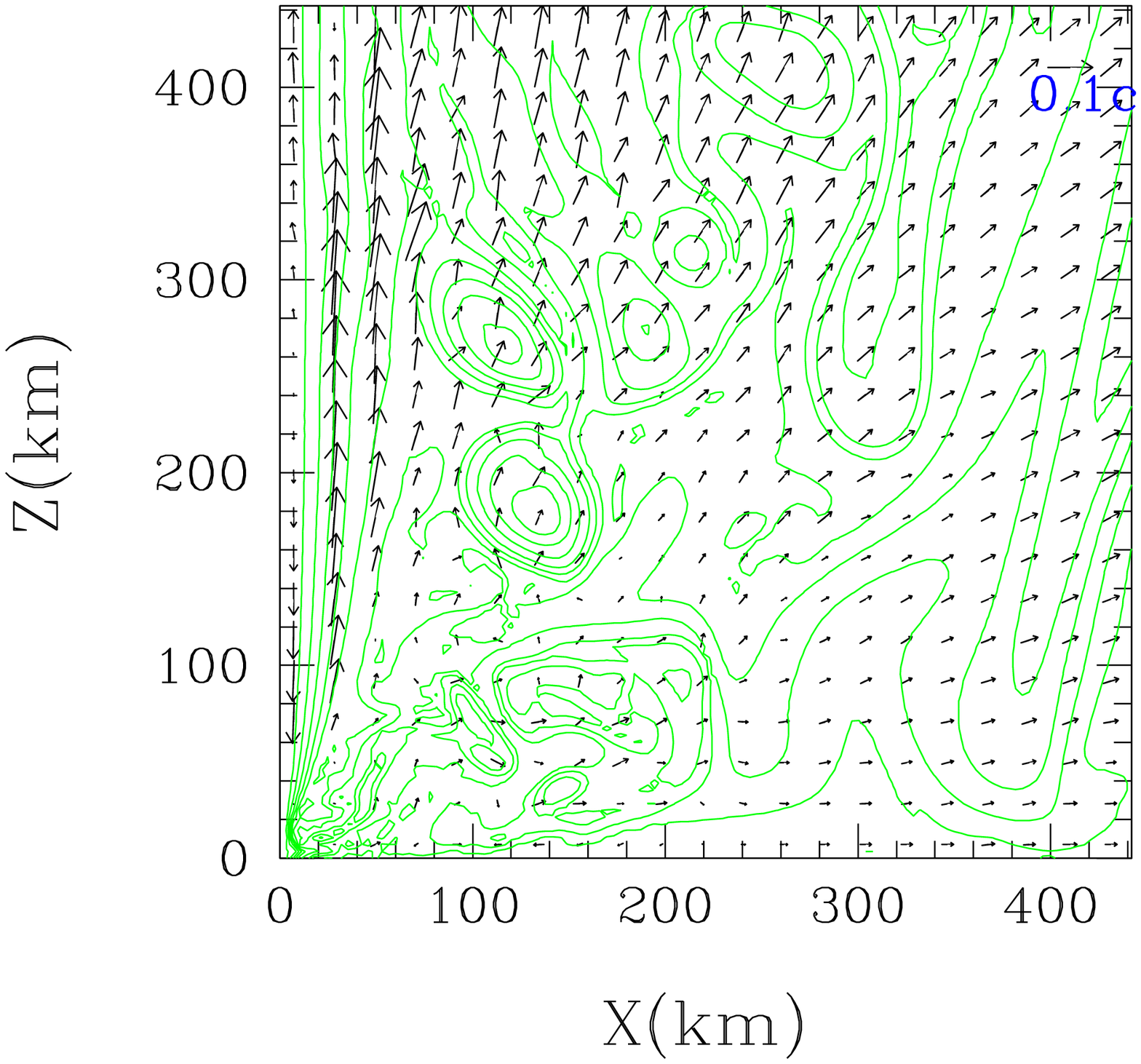}
\vspace{-5mm}
\caption{The density contours and velocity vectors (left) 
and the poloidal magnetic field lines and velocity vectors (right) 
at $t=152.5$ ms for model A4. 
The density contours are drawn for 
$\rho=10^{15-i}~{\rm g/cm^3}~(i=1$--7).
The velocity vectors and poloidal magnetic field lines are
drawn as in Fig.~\ref{FIG8}. 
\label{FIG13}}
\end{center}
\end{figure*}

To give a better overall picture of the outflow, we show in 
Fig.~\ref{FIG13} the density contours, velocity vectors, and 
poloidal magnetic field lines at $t=152.5$ ms for model A4. 
In this figure, a larger region than that shown in the last panel 
of Fig.~\ref{FIG10} is displayed.  For $r \agt 200$~km, a 
weakly-anisotropic outflow is found.  However, the velocity of the 
matter near the rotation axis is relatively large. For 
$r \alt 200$~km, on the other hand, the strong outflow 
is seen only along the collimated magnetic field in the vicinity of 
rotation axis. The greatly enhanced strength of the outflow along
the collimated field lines points to magneto-centrifugal
launching as a likely driver of the outflow.  Our results
suggest that jets may be launched during supernova explosions if 
the collimated magnetic fields found here are 
generic~\cite{WMW,AHM,MBA}.  For $r \alt 200$~km, a slower, incoherent 
flow pattern is also seen, reflecting the continuous driving of irregular 
matter motions by the MRI.

As mentioned in Sec.~\ref{sec:BP}, the magneto-centrifugal launching
mechanism requires a mechanism for ejecting material outward at the
base. This is particularly important if the opening angle of the
collimated field line is small \cite{BP,BP1}. In the present case, the
magnetic field is significantly perturbed in the vicinity of the
PNSs by the MRI.  Turbulence and/or irregular Poynting flux
generated by the MRI could serve to inject material into the 
magneto-centrifugal wind.

\begin{figure}[t]
\begin{center}
\epsfxsize=3.2in
\leavevmode
\epsffile{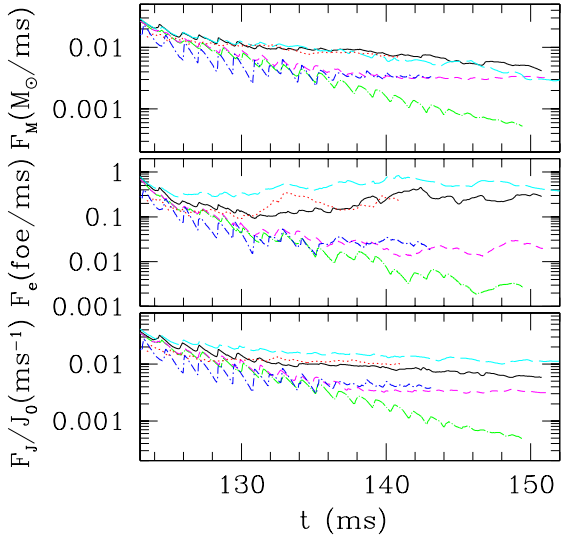}
\vspace{-5mm}
\caption{$F_M$, $F_e$, and $F_J$ at $r \approx 224$ km for models A0
(dot-long-dashed curves), A1 (dashed curves), A2 (solid curves), A4
(long-dashed curves), B (dotted curves), and C (dot-dashed curves).
The abbreviation 'foe' denotes $10^{51}$ ergs, while $J_0$ is the 
initial value of $J$.
\label{FIG14}}
\end{center}
\end{figure}

In addition to rest mass, energy and angular momentum are also 
carried away from the PNS by the outflows. 
In Fig.~\ref{FIG14} we show $F_M$, $F_e$, and $F_J$ evaluated at 
$r=L/4$ ($\approx 224$~km for models~A0--A4, $\approx 143$~km for 
model~B, and $\approx 161$~km for model~C) as a function of
time for models~A0--A4, B, and C. We also evaluated the fluxes at 
other radii and found that, aside from a time shift, the results 
depend weakly on the radius.

In the early ringdown phase in which shocks propagate outward, 
the loss of mass, energy, and angular momentum from the system is due
primarily to the matter outflow (as opposed to the Poynting outflow)
irrespective of the presence of magnetic field. Hence $F_M$,
$F_e$, and $F_J$ are roughly independent of the field strength
(compare results for model A0, A1, A2, and A4). In the absence of the
magnetic field (model A0; dot-dashed curves), however, the outflow rates
decay in an exponential manner, leaving a stationary PNS. 
On the other hand, in the presence of the magnetic field, the fluxes 
remain nonzero, and the PNSs continuously lose mass, 
energy, and angular momentum.  
The evolution of the fluxes for model~A1 show that the 
dominance of the MHD-driven outflow over the shock-driven outflow is 
delayed for $\sim 10$ ms after bounce, i.e.\ after the toroidal field 
becomes sufficiently strong. (Note that for model~A1, the growth of the 
toroidal field by winding saturates at the relatively late time of 
$\sim 30$ ms after bounce.  This is consistent with the fact that
the MHD-driven outflow becomes dominant later for this model than for 
A2--A4.)

The strength of the outflow depends on the initial magnetic field
strength (compare the results of models~A1 and A2). This is due to the
difference in the resulting ratio of the poloidal field strength to the
toroidal field strength $C_B$ for the PNS [see
Eq.~(\ref{dotE2})]. The toroidal fields are wound up to similar saturation
strengths in the various models regardless of the initial poloidal 
field strength (see Fig.~\ref{FIG7}). On the other hand, after bounce, 
the poloidal field is not uniformly amplified, though the MRI increases 
the strength locally.  As a result, the value of $C_B$ is smaller for model
A1 and consequently so are the values of $F_M$, $F_e$, and $F_J$.  

In addition to the magnitude of the initial seed magnetic field, the
strength of the outflow (and particularly of the magneto-centrifugal
outflow), also depends strongly on the initial field profile. The
relative size of the $z$-component of the magnetic field is larger for
model A4 than for model A2 (compare Figs.~\ref{FIG8} and \ref{FIG10}).
The larger $z$-component is favorable for inducing the MRI and the
resulting MHD outflow, although the magnetic energy of the PNSs for
models A2 and A4 are nearly identical (see Fig.~\ref{FIG6}).

Although the outflow rates depend on the initial magnetic field
profile and strength, the ratio of $F_J/F_M \sim 1.5$--$2.5J/M$ is
always larger than $J/M$ for all cases considered here.  Thus, the 
MHD outflow carries away material with a larger specific angular 
momentum than that of the PNSs, leading to spindown. 
Figure~\ref{FIG14} shows that the spindown time scale $J_0/F_J$ 
($J_0$ is the initial value of $J$) is $\sim $100--300 ms for models~A1, 
A2, and A4, and is approximately proportional to $C_B^{-1}$    
[see Eq.~(\ref{dotE2}) and note that the value of $C_B$ for A2 is 
about twice of that for A1]. Since $C_B \agt 0.1$ is indicated by 
our simulations, a large amount of the angular momentum of the PNSs 
is carried away within $\sim 1$~s after the onset of the MHD outflow.

\begin{figure}[th]
\begin{center}
\epsfxsize=3.2in
\leavevmode
\epsffile{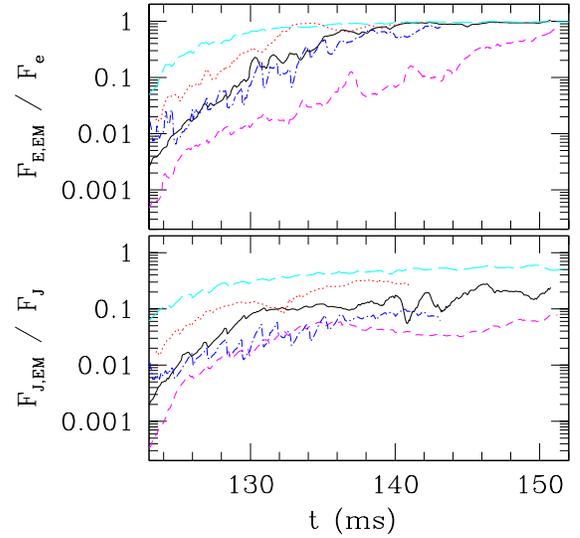}
\vspace{-5mm}
\caption{$F_{E,{\rm EM}}/F_e$ and $F_{J,{\rm EM}}/F_J$ for models A1
(dashed curves), A2 (solid curves), A4 (long-dashed curves),
B (dotted curves), and C (dot-dashed curves). 
\label{FIG15}}
\end{center}
\end{figure}

In Fig.~\ref{FIG15}, we show the ratios of $F_{E,{\rm EM}}/F_e$ and
$F_{J,{\rm EM}}/F_J$ as functions of time for models~A1, A2, A4, B, and C.
The plot indicates that the outflow is largely Poynting dominated.  The ratio 
of the EM flux to the total flux clearly depends on the initial magnetic 
field strength and profile. Comparing the results shown in Fig.~\ref{FIG15}, 
$F_{E,{\rm EM}}/F_e$ and $F_{J,{\rm EM}}/F_J$ are larger for larger values 
of $F_M$ and $F_J$ \cite{note3}.  

The large relative fraction of the Poynting flux at radii around 
$\sim 200$ km also implies that, during the outward
propagation of the magnetic energy, conversion to matter
kinetic energy is suppressed in our model. In realistic supernovae,
shock waves formed at bounce do not propagate outward unimpeded but rather
stall at a radius $\sim 150$--200 km, forming a standing shock
\cite{Wilson,Janka,Lieb,Burrow}. Thus, the MHD outflow driven from the
PNS will eventually hit the stalled shock, where the
Poynting energy flux as well as the kinetic energy of the matter
outflow may be converted to thermal energy. The total energy 
from the outflow during the 10 ms period following the relaxation 
of $F_M$ to a roughly constant value is $\agt 10^{50}$ ergs for all 
of the models.  If this energy were converted to thermal energy at 
the stalled shock, it could play a significant role in 
driving the supernova explosion, as pointed out in \cite{TQB,ABM,MBA}.

\begin{figure}[th]
\begin{center}
\epsfxsize=3.2in
\leavevmode
\epsffile{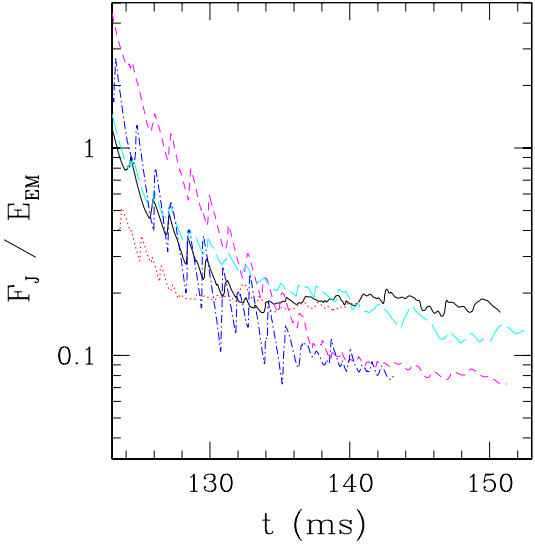}
\vspace{-5mm}
\caption{$F_{J}/E_{\rm EM}$ for models A1 (dashed curve), 
A2 (solid curve), A4 (long-dashed curve), B (dotted curve),
C (dot-dashed curve).
\label{FIG16}}
\end{center}
\end{figure}

In Fig.~\ref{FIG16}, the ratio of $F_J$ to $E_{\rm EM}$ is shown for
models~A1, A2, A4, B, and C.  By Eq. (\ref{dotJE}), this ratio should be 
of order $C_B$ if the MHD outflow theory \cite{BP} holds in the vicinity 
of the PNS.  Figure~\ref{FIG16} shows that, in the early
phase ($t \alt 130$ ms) in which matter outflow driven by shocks is
dominant, the value $F_J/E_{\rm EM}$ decreases steeply.  However, once
the MHD outflow dominates, the ratio relaxes to an approximately 
constant value of $\sim 0.1$--0.2 for all of the models. (Our simulations
indicate that  $C_B=|B^P|/|B^T|\sim 0.1$--1 for these models.)  We also find
that its value for model~A1 is about half as large as the values 
for models~A2 and A4, reflecting the smaller value of $C_B$ for model~A1. 
Thus, our results agree qualitatively with the MHD outflow theory.  
The MHD outflows found here are ultimately powered by
rotation. Thus, the outflow is expected to weaken as the PNSs spin 
down, although at least in the first few tens of milliseconds 
after the saturation of the field growth, the strength is
approximately constant as shown in Fig.~\ref{FIG16}.

\subsection{Outflow-induced spindown of the PNSs}\label{sec:BR}

Assuming that the PNSs spin down due to the MHD outflow, 
the loss rate of the angular momentum is described by Eq.~(\ref{dotJE}) 
with $\eta_*=O(1)$.  The spin angular momentum of the PNS
is approximately written by $J_*=I_* \Omega_*$ where $I_*$ is a moment
of inertia and $\Omega_*$ is an average angular velocity.  After the
saturation of the magnetic field growth, $E_{\rm EM}$ is approximately 
equal to $\zeta T_{\rm rot}$, where $\zeta$ is around 0.1--0.2
according to our results (see Fig.~\ref{FIG6}), and where $T_{\rm rot}$ 
is approximately $I_* \Omega_*^2/2$. Gathering these results, we have 
a relation for the evolution of the angular momentum of the PNS:
\beqn
{d (I_* \Omega_*) \over dt} \approx -{\zeta \eta_* \over 2}I_* \Omega_*^2
C_B.
\eeqn
Assuming that $I_*$ and $C_B$ are constant, we then have
\beqn
{d P_* \over dt}={d \over dt}
\biggl({2\pi \over \Omega_*}\biggr) \approx \pi \zeta \eta_*
C_B,
\label{period}
\eeqn
where $P_*=2\pi/\Omega_*$. 
Equation~(\ref{period}) suggests that the rotational period increases 
linearly with time as
\beqn
P_* &\approx& P_0 + \pi \zeta \eta_* C_B t \nonumber \\
&\approx& P_0 + 30
\biggl({\zeta \over 0.1}\biggr) 
\biggl({\eta_* \over 0.1}\biggr) 
\biggl({C_B \over 0.1}\biggr)
\biggl({t \over 10~{\rm s}}\biggr)~{\rm ms}, \label{pdot}
\eeqn
where $P_0$ is the initial value of $P_*$.
The value of $C_B$ during the stationary MHD outflow phase is somewhat
unclear for a real system. Our present result, however, indicates that 
$|B^P|\sim |B^T|$ near the rotation axis, giving $C_B \sim 0.1$--1.

\begin{figure}[th]
\vspace{-2mm}
\begin{center}
\epsfxsize=3.1in
\leavevmode
\epsffile{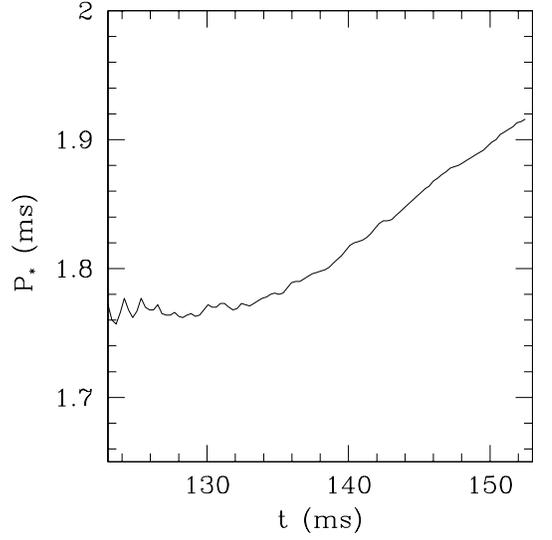}
\vspace{-5mm}
\caption{$P_*$ as a function of time for model~A4. 
\label{FIG17}}
\end{center}
\end{figure}

In Fig.~\ref{FIG17}, we show the evolution of $P_*$ for model~A4,
in which the PNS and magnetic field lines reach
a quasistationary state for $t \agt 140$ ms. To obtain $P_*$, we compute 
$\Omega_*$ from
\beqn
\Omega_*=\Big[\int_{\rho \geq 10^{-3}\rho_{\rm nuc}} \rho_* \Omega d^3x\Big]
\Big/\Big[\int_{\rho \geq 10^{-3}\rho_{\rm nuc}} \rho_* d^3x\Big]. 
\eeqn
The figure shows that the average spin period increases
approximately linearly in time for $t \agt 140$ ms. This agrees with
Eq.~(\ref{pdot}). For this case, we find $dP_*/dt \approx 0.013$. 
Thus the order of magnitude for the
spindown rate is in agreement with Eq.~(\ref{period}). 

The rapid spindown will continue until the rotation in the vicinity 
of the PNS becomes sufficiently uniform, i.e., until the magnetic field 
relaxes approximately to a stationary state. If the magnetic field is 
amplified until saturation with $\zeta \sim 0.1$ shortly after bounce, 
and if the  degree of differential rotation remains high for
$\sim 1000$~s after the saturation, the rotation will slow to a period 
longer than $\sim 1$~s. For poloidal magnetic field strengths 
$\agt 10^{15}$ G, the resulting neutron star will be a 
magnetar~\cite{AXP,TCQ}. However, we note that these results 
may be modified when the PNS cools, which occurs on a timescale 
($\sim 20$~s) which is much longer than that of our simulations.

We know that a class of young pulsars (Crab-like pulsars) has rotational 
periods of several 10~ms~\cite{Lyne}. The birth of such rapidly rotating 
young pulsars thus seems to require that the differential rotation in the 
vicinity of the PNS is quickly lost in order to avoid 
winding and the MRI, which lead to outflows and spindown.

\subsection{Gravitational waveforms}

\begin{figure}[th]
\vspace{-2mm}
\begin{center}
\epsfxsize=3.3in
\leavevmode
\epsffile{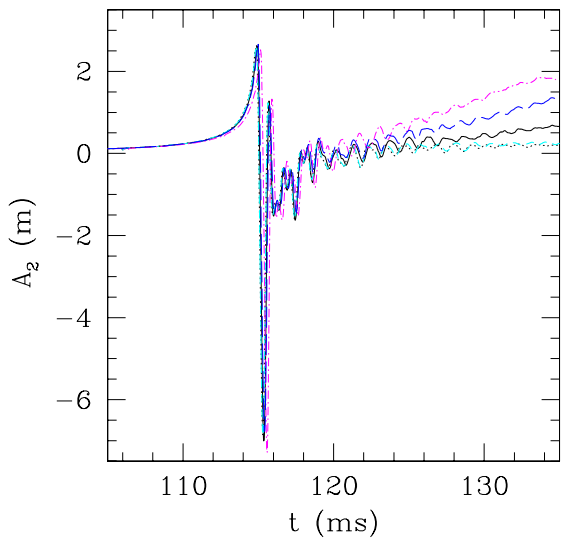}
\vspace{-7mm}
\caption{$A_2$ as a function of time for models A0 (dotted curve),
A1 (dashed curve), A2 (solid curve), A3 (dot-dashed curve),
and A4 (long-dashed curve).  
\label{FIG18}}
\end{center}
\end{figure}

In Fig.~\ref{FIG18}, we show gravitational waveforms for models~A0--A4. 
Gravitational waveforms are computed using the
quadrupole formula described in Sec.~\ref{GW}. According to the 
classification system of~\cite{Newton4,HD}, the waveforms for these
models are of type I, which is the most common type of waveform. The 
properties of type I waveforms may be summarized as follows: During 
the infall phase, a gravitational wave precursor is emitted due to 
the changing quadrupole moment as the collapsing core flattens. 
In the bounce phase, spiky burst
waves are emitted for a short time $\sim 1$~ms, and the
amplitude and the frequency of gravitational waves reaches a 
maximum. In the ring-down phase, gravitational waves associated with 
several oscillation modes of the PNS, for which the
frequencies are $\sim 1$ kHz, are emitted with amplitudes which are 
gradually damped due to shock dissipation at the outer edge of the 
PNS.

We find that the MHD outflow modifies the gravitational waveforms 
in the ringdown phase; the wave 
amplitude gradually increases with time during this phase, although 
the waveforms are otherwise unchanged. That this secular increase is 
caused by the MHD outflow may be understood through the following simple
analysis. Consider a matter outflow of mass $m$ and velocity $v$ in the
$z$ direction, and assume that the velocity changes slowly. The 
contribution to $A_2$ from this outflow comes from $\ddot
I_{zz}$, for which the correction is $\delta A_2 \approx 2 m v^2$. 
The MHD outflow is continuously ejected from the vicinity of the
PNS, and hence, the total mass of the outflow increases
with an approximately constant rate. We find that $A_2$ indeed increases
roughly linearly with time.  Furthermore, the magnitude is
approximately given by
\beqn
\delta A_2 \approx 3 \biggl({m \over 0.1M_{\odot}}\biggr)
\biggl({v \over 0.1c}\biggr)^2~{\rm m}, 
\eeqn
which is in approximate agreement with the numerical results
for the mass increase rate $dm/dt =0.1$--$5M_{\odot}$/s. 

The secular increase of the wave amplitude for model A4 is faster than
for model A2 although the magnetic energies of the PNSs are nearly
identical.  This is because the MHD outflow is more efficiently
driven for model A4 as shown in Sec.~\ref{sec:BB}. 

As discussed in Sec.~\ref{sec:BB}, the MHD outflow from a newly-formed 
PNS will likely hit a standing shock at a
radius $\sim 150$--200 km \cite{Wilson,Janka,Lieb,Burrow}. The value
of $m$ (and hence the linear growth of the amplitude) will then saturate 
in a few 10 ms.  Nevertheless, the amplitude $A_2$ could eventually 
reach a few meters, which is comparable to the amplitude of the spiky 
waves emitted at the bounce.  The outflow signal may thus be detectable 
for an event within our Galaxy by the ground-based laser 
interferometric detectors, 
since the expected amplitude is $\sim 10^{-20}$ [see Eq.~(\ref{hamplitude})].  
Detecting gravitational waves in the first few tens of ms of the supernova 
collapse event may thus provide information about the anisotropic outflow.

\subsection{Comparison of codes}

\begin{figure}[th]
\begin{center}
\epsfxsize=3.2in
\leavevmode
\epsffile{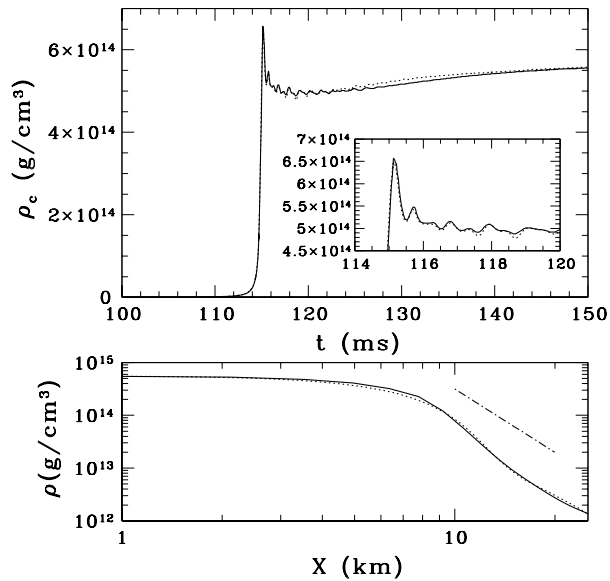}
\vspace{-2mm}
\caption{Results for model A2 obtained with the codes of~\cite{SS05} (solid 
lines) and of~\cite{DLSS} (dotted lines).
({\em top panel}) Evolution of the central density for model A2.  The
inset shows the bounce and immediate post-bounce behavior.
({\em lower panel}) Comparison of the density  profiles at $t=150.8$~ms. 
The dot-dashed line segment shows a slope of 
$\rho \propto \varpi^{-4}$.  The results from the code of~\cite{SS05}
correspond to the standard resolution.  For the results from the
code of~\cite{DLSS}, a fisheye grid was used for the PNS evolution 
phase (see Appendix~A for details).
\label{FIG5}
}
\end{center}
\end{figure}

\begin{figure}[th]
\begin{center}
\epsfxsize=3.2in
\leavevmode
\epsffile{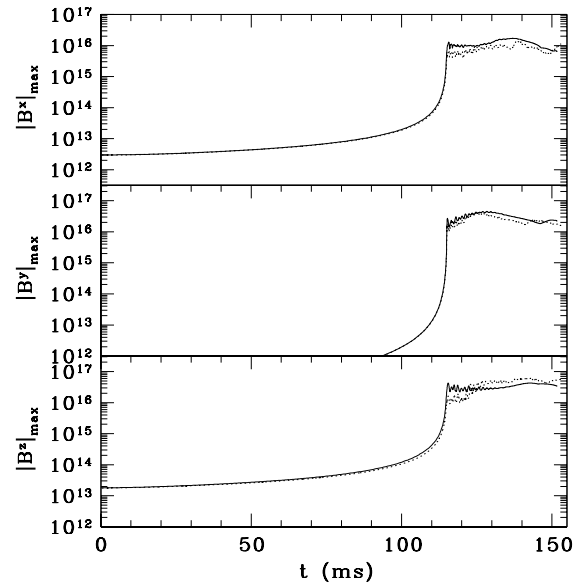}
\vspace{-2mm}
\caption{Evolution of the maximum values of $|B^i|$ for model A2.  
As in Fig.~\ref{FIG5}, results from the codes of~\cite{SS05} (solid 
lines) and of~\cite{DLSS} (dotted lines) are compared. 
\label{FIG19}
}
\end{center}
\end{figure}

\begin{figure}[th]
\begin{center}
\epsfxsize=3.2in
\leavevmode
\epsffile{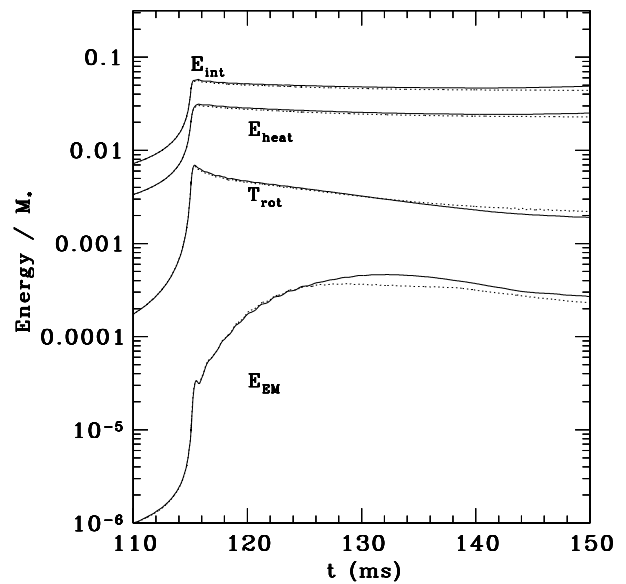}
\vspace{-2mm}
\caption{Individual components of the energy for model A2, 
defined as in Fig.~\ref{FIG6}.  
Results from the codes of~\cite{SS05} and~\cite{DLSS} are 
shown as solid and dotted lines, respectively.  
\label{FIG20}
}
\end{center}
\end{figure}

To demonstrate agreement between the codes of~\cite{DLSS} and
\cite{SS05}, we plot some representative quantities in 
Figs.~\ref{FIG5}--\ref{FIG20}. Figure~\ref{FIG5} compares the central density 
evolution and
late-time density profiles for model A2 from both codes.   The plot of central
density shows agreement on the maximum density at bounce, the slow increase in 
the central density caused by the spindown of the PNS, and the post-bounce
oscillations (see the inset of Fig.~\ref{FIG5}).  (Results from the code 
of~\cite{DLSS} were shifted forward in time by $0.35$ ms to align the core 
bounce.  This small difference in the coordinate time likely results from the 
slightly different lapse prescriptions used in the two simulations.)  The two 
codes also give the same principal features of the density profile at late 
times (e.g.\ $t=150.8$~ms), including the falloff behavior for the density 
outside the PNS.  The behavior of the magnetic 
field components, as compared in Fig.~\ref{FIG19}, also shows good agreement.  
Some differences arise after the PNS forms due to the fact that the magnetic 
field evolution is dominated by the turbulent effects of the MRI.  In this 
regime, the precise value of the maximum magnetic field component depends on 
the details of the turbulent motions; small differences in local quantities
can be amplified by turbulence.  Finally, we
show the evolution of the individual components of the energy for model
A2 (compare to Fig.~\ref{FIG6}), which also demonstrates good agreement
between the two codes.  We note that both the central 
density and magnetic field components grow by a 
factor of order $10^4$ during the evolution of model A2.  Given this large
dynamic range, the agreement shown here is rather noteworthy. 

\section{Summary and discussion}

We have presented general relativistic numerical simulations of
magnetized stellar core collapse to a PNS in axisymmetry and have
followed the subsequent evolution of the PNS by magnetorotational
effects. The following is a summary of the results:

\noindent
1. The magnetic field is amplified during the collapse and the
subsequent evolution of the PNS. During the infall phase, the poloidal
field grows due to the compression of the matter in which the field is
frozen. As the poloidal field strength grows, the amplification of the
toroidal field by winding is accelerated. After bounce, the
compression stops, and so does the rapid growth of the poloidal field.
However, the differential rotation in the vicinity of the PNS induces
further growth of the toroidal field through winding. The winding
continues until the magnetic field drains away the kinetic energy
stored in differential rotation (which is roughly 10--20\% of the
total rotational kinetic energy).  The ratio of the magnetic energy to
the rotational kinetic energy thus saturates at approximately the same
value regardless of the initial magnetic field profile and strength.
The poloidal magnetic field in the outer region of the PNS also grows
via the MRI. In contrast to magnetic winding, the MRI amplifies the
field strength in a turbulent manner.  The magnetic stress increases
until it becomes $\sim 10\%$ of the rotational kinetic energy density. 
After the magnetic field saturates, we find a stationary, collimated
helical magnetic field near the rotation axis.

\noindent
2. Because of the enhanced magnetic stress, an MHD outflow is launched
from the vicinity of the PNS, particularly near the rotation axis. Our
results suggest that three mechanisms drive the outflow. One is the
magneto-spring mechanism in which magnetic stress primarily due to the
toroidal field blows off matter along the rotation axis (but without
strong collimation), leading to the so-called `tower-like' field
structure. The second mechanism is MRI-induced turbulence, which leads
to a weaker and less coherent matter outflow than the magneto-spring
effect. The third mechanism is the Blandford-Payne magneto-centrifugal
mechanism, which plays a dominant role after the saturation of the
magnetic field growth and after the formation of a collimated helical
magnetic field near the rotation axis. In the present context, matter
is injected into the outflow by MRI turbulence and then flung out
along the helical magnetic field lines. By these three mechanisms, the
energy carried away by the MHD outflow in the first 10 ms after the
saturation of the toroidal magnetic field growth may be $\agt
10^{50}$~ergs for a PNS of rotation period $\sim 1$--2 ms and for a
typical ratio of the poloidal to toroidal field strength of $C_B\agt
0.1$.

\noindent
3: The MHD outflow carries away material with large specific angular
momentum, and hence, plays a crucial role in spinning down the
PNSs. We find that the angular momentum loss rate is $\sim 0.1$--$0.2
E_{\rm EM}C_B$ as long as differential rotation persists in the
vicinity of the PNS. Since $E_{\rm EM}$ is $\sim 10$--20\% of the
rotational kinetic energy $T_{\rm rot}$ after the field saturates, we
have $|\dot{J}|\sim 0.01 T_{\rm rot}C_B$. This implies that the
angular momentum (and rotational kinetic energy) may decrease rapidly
as show in Sec.~\ref{sec:BR}. If the differential rotation remains
strong enough in the first $\sim 1000$~s after the magnetic field
saturates, the spin period of the PNS increases to $\agt 1$ s.

In turbulent flows resulting from the MRI, small eddies would be
formed, dissipating kinetic energy into thermal energy and
transporting angular momentum outward.  The thermal energy generated
in this process has been suggested as a power source for a supernova
explosion~\cite{TQB}. However, we find that the role of the turbulence
is likely less important as a source of energy than the MHD
outflow. This may be partly due to our choice of the initial seed
magnetic field. In the present work, the poloidal magnetic field at
the time of PNS formation is high enough that the toroidal field is
rapidly amplified and drives a strong MHD outflow in a few tens of ms
after the bounce. For a weaker initial poloidal field, more time would
be required to amplify the toroidal field [see Eq. (\ref{eqn:dtBphi})]
and, hence, to produce a strong MHD outflow.  In contrast, the MRI
grows on the same timescale ($\sim 10\Omega^{-1}$) even for very weak
fields.  Since the wavelength of the fastest-growing MRI mode scales
with the field strength, very weak fields will result in turbulence
composed of small eddies, for which the typical size is much smaller
than the stellar radius.  This small scale turbulence converts kinetic
energy into thermal energy, as modeled in~\cite{TQB}. In this case,
the magneto-spring and magneto-centrifugal mechanisms may not play a
significant role and the collimated magnetic field may not
form. Unfortunately, simulations for such weak initial fields are
currently unable to capture this behavior, since the wavelength of the
MRI is much too short to be resolved. To determine whether such
processes indeed occur, a much more powerful supercomputer will be
required.

We do not observe the magnetic field amplification mechanism found 
in~\cite{ABM,MBA}.  In this process, an MHD outflow induced by 
the magneto-spring mechanism is first driven along the rotation axis. 
This outflow then triggers convective motions, leading to the
formation of large-scale eddies in the meridional plane which wind up 
the poloidal magnetic field lines. The strengthened poloidal field then
leads to further amplification of the toroidal field. In our simulations, 
however, we do not find noticeable convection. The plausible
reason for this discrepancy is that the simulations of~\cite{ABM,MBA} 
use a treatment of the equation of state and microphysics, which differs
from ours. In their simulations, neutrino cooling is taken into account,
and this cooling leads to negative entropy gradients and subsequent
convection near the neutrinosphere. In order to see this convective-type
MRI \cite{MRIrev}, we would have to take neutrino processes into account. 
This is an issue to be investigated in the future.

The simulations presented here were carried out assuming equatorial
plane symmetry, and the center of the PNS thus remains at the origin.
In the absence of this symmetry, the PNS may move, due to
back-reaction of MHD outflows. The lack of equatorial symmetry is
crucial in the explosion mechanism proposed by~\cite{OBDE06}, in which
acoustic waves from $l=1$ $g$-modes provide an important source of
energy.  In addition, the outflow may develop anisotropically, since
it is driven in part by inhomogeneous MRI-induced
turbulence. Anisotropic outflow could also arise if the magnetic field
profile is anisotropic in the supernova progenitor. Our numerical
results show that $\agt 0.1M_{\odot}$ of material will be ejected in
the first $\sim 100$~ms after the growth of the toroidal field
saturates. The typical outflow velocity is $\sim 0.1c$. If the
anisotropy in the direction of the ejected mass is $\agt 10 \%$, the
PNS will move with a velocity $\agt 10^{-3}c \sim 300$ km/s due to the
back-reaction.  Anisotropic MHD outflows may thus be able to explain
high-velocity pulsars \cite{HVP}. Simulations without equatorial plane
symmetry would be required to explore this possibility.

An additional limitation of our simulations is the assumption of
axisymmetry.  Nonaxisymmetric instabilities (such as bar modes and/or
one-armed spirals) may arise in the formation of PNSs and contribute
strongly to the gravitational wave signal~\cite{SKE,ootb05}.  In this
paper, we mainly consider the case in which the progenitor is rigidly
rotating and hence the resulting PNSs are weakly differentially
rotating with $T_{\rm rot}/|W| \lesssim 0.12$. 
This implies that the PNSs found in this paper would be
stable against nonaxisymmetric deformations. However, for more 
rapidly rotating PNSs, the possibility of nonaxisymmetric deformations 
have to
be taken into account. The behavior of the MRI is also expected to be
different in a full 3D calculation because of the effect of
nonaxisymmetric MRI induced by toroidal magnetic field 
\cite{NMRI}. Turbulence may arise and persist more readily in 3D due
to the lack of symmetry.  More specifically, according to the
axisymmetric anti-dynamo theorem~\cite{moffatt78}, sustained growth of
the magnetic field energy is not possible through axisymmetric
turbulence. Simulations in full 3D will eventually be necessary in
order to fully understand the role of the MRI in PNS evolution and jet
formation.  Given current computational resources, however, we
consider this a challenge for the future.

\acknowledgments

MS thanks T.K. Suzuki for helpful discussions. Numerical computations
were performed on the FACOM VPP5000 at ADAC at NAOJ, on the NEC SX8
at YITP in Kyoto University, and  on the NEC SX6 at
ISAS at JAXA, and at the NCSA at UIUC.  This work was supported in
part by Japanese Monbukagakusho Grants (Nos.\ 17030004 and 17540232)
and NSF Grants PHY-0205155 and PHY-0345151, NASA Grants NNG04GK54G and
NNG046N90H at UIUC.

\appendix

\section{Multiple transition fisheye coordinates}

The multiple transition fisheye coordinates~\cite{CLZ06}
$\bar{x}^i$ are related to the
original coordinates $x^i$ through the following transformation:
\beqn
  x^i &=& \frac{\bar{x}^i}{\bar{r}} r(\bar{r}) ,
\label{eq:fisheyet1}  \\
  r(\bar{r}) &=& a_n \bar{r} + \sum_{i=1}^n \kappa_i \ln
\frac{\cosh [(\bar{r}+\bar{r}_{0i})/s_i]}{\cosh [(\bar{r}-\bar{r}_{0i})/s_i]},
\label{eq:fisheyet2}  \\
 \kappa_i &=& \frac{(a_{i-1}-a_i)s_i}{2\tanh(\bar{r}_{0i}/s_i)} ,
\label{eq:fisheyet3}
\eeqn
where $r=\sqrt{x^2+y^2+z^2}$, $\bar{r} = \sqrt{\bar{x}^2 + \bar{y}^2
+ \bar{z}^2}$, $n$, $a_i$, $r_{0i}$ and $s_i$ are constant parameters.
We perform simulations on this fisheye grid with $0<\bar{x}<\bar{L}$,
and $0<\bar{z}<\bar{L}$. The Cartoon method is used to impose
axisymmetry as usual. We use a grid size of
$N \times 3 \times N$ with uniform grid spacing
$\bar{\Delta} = \bar{L}/N$. It follows from the
transformation~(\ref{eq:fisheyet1}) that this corresponds to a
resolution in the original grid of $\Delta \approx (dr/d\bar{r}) 
\bar{\Delta} \approx
a_i \bar{\Delta}$ in regions well separated from the transitions, 
i.e.\ where $\bar{r}_{0i}+s_i \ll \bar{r} \ll
\bar{r}_{0i+1}-s_{i+1}$. The ratio $\Delta/\bar{\Delta}$ smoothly 
changes from $a_i$ to $a_{i+1}$ in the transition region 
$\bar{r}_{0i+1}-s_{i+1} 
\lesssim \bar{r} \lesssim \bar{r}_{0i+1}+s_{i+1}$. 

Our implementation is as follows. We first perform simulations in the original 
coordinates $x^i$ with the regridding technique as discussed in 
Sec.~\ref{sec:simulations} until the value of $\Phi_c$ reaches 0.16. 
At this time, 
we interpolate the data to the fisheye coordinates $\bar{x}^i$ 
with the fisheye parameters $n=3$, $(a_0, a_1, a_2, a_3) = 
(0.1, 0.4, 0.8, 1.0)$. 
We choose the value of $\bar{L}$ so that the 
outer boundary is approximately the same as that of the original grid. 
We set $N=600$, and the values of $\bar{r}_{0i}$ so that 
$(r_{01}, r_{02}, r_{03}) \approx$ (100~km,~200~km,400~km), where 
$r_{0i} = r(\bar{r}_{0i})$. The transition width is set to be 
$s_i = 29 \bar{\Delta}$.
With this setting, the resulting 
resolution in the original grid becomes $\Delta \approx 0.7$~km for 
$r \lesssim 100$~km, $\Delta \approx 2.7$~km for 100~km $\lesssim r 
\lesssim 200$~km, $\Delta \approx 5.5$~km for 200~km $\lesssim r 
\lesssim 400$~km, and $\Delta \approx 6.8$~km for $r \gtrsim 400$~km. 
Therefore, using this technique, we can achieve high resolution 
in the central region with relatively few grid points. 

\section{Proof of $\ve{k=\rho_* v^P/\cB^P}$ constant along magnetic 
field lines}

In a stationary spacetime, $\partial_t \cB^i = 0 = \partial_t \rho_*$. 
It follows from the induction equation~(\ref{induction}) and the
continuity equation~(\ref{eq:continuity}) that 
\beqn
 \partial_j (v^j \cB^i - v^i \cB^j) &=& 0 , \label{eq:stat-ind} \\
 \partial_j (\rho_* v^j) &=& 0.
\eeqn
In an axisymmetric spacetime, the continuity equation becomes 
\beq
  \partial_P ( \varpi \rho_* v^P) = 0, \label{eq:stat-con}
\eeq
where $P$ denotes the poloidal components ($\varpi$ and $z$).
We introduce a one-form 
\beq
  \omega_i = \epsilon_{ijk} v^j B^k = [ijk] v^j \cB^k,
\eeq
where $\epsilon_{\alpha \beta \gamma}=n^{\mu}
\epsilon_{\mu \alpha \beta \gamma}$ is the 3-dimensional 
Levi-Civita tensor, and $[ijk]$ is the permutation symbol. 
It is easy to 
show that Eq.~(\ref{eq:stat-ind}) is equivalent to $D_{[i} \omega_{j]}=0$
(i.e.\ $\omega_i$ is a closed one-form), where $D_i$ denotes the 
covariant derivative associated with the three-metric $\gamma_{ij}$. 
Hence $\omega_i$ can be written as 
\beq
 \omega_i = D_i f = \partial_i f , \label{eq:omegai}
\eeq
where $f$ is a scalar function. We assume that Eq.~(\ref{eq:omegai}) 
holds globally. It follows that 
$\omega_{\varphi}= \varpi (v^z \cB^{\varpi}-v^{\varpi} \cB^z)=0$ in
an axisymmetric and stationary spacetime. This implies that 
$v^{\varpi}/\cB^{\varpi} = v^z/\cB^z$, i.e.\ 
$v^P = \mu \cB^P$, where $\mu$ is a scalar function. Substituting 
$v^P = \mu \cB^P$ into Eq.~(\ref{eq:stat-con}) we have 
$\partial_P (\varpi \rho_* \mu \cB^P) =0$. Using the no-monopole constraint 
$\partial_P (\varpi \cB^P)=0$, we obtain $B^P \partial_P k = 0 = 
B^j \partial_j k$, where $k\equiv \rho_* \mu = \rho_* v^P/\cB^P$. Hence 
$k$ is constant along magnetic field lines in an axisymmetric, 
stationary spacetime.


\begin{thebibliography}{99}
\bibitem{OAM0} In the work of~\cite{YS,SKY,OAM}, a logarithmic grid
spacing is adopted for the radial coordinate and a sufficient
resolution was guaranteed only in the central region. However, the MRI
is active not at the central region, but near the surface of
the PNS. The wavelength of the fastest-growing mode of the
MRI is proportional to the strength of the poloidal magnetic field
(see Sec.~\ref{sec:MRI}). The simulation of~\cite{OAM} can resolve it
in the strongest magnetic strength.  However, in this case, other 
MHD effects dominate and it is difficult to identify the role
of the MRI as mentioned in~\cite{OAM}. On the other hand, for the
weaker magnetic field strength, they could not resolve the MRI because
of insufficient grid resolution.
\bibitem{note1} In a real supernova, the shock will stall at a radius of $\sim
150$--200 km forming a standing shock \cite{Wilson,Janka,Lieb,Burrow}. 
In this paper, we do not pay attention to the issue for the propagation of the
shock and longterm dynamics of supernova explosion.
\bibitem{note3} Note that the fact $F_{\rm E,EM}/F_e \approx 1$ does
not imply that approximately all the energy are carried away by the
Poynting flux since $F_{e}$ is composed of the bulk kinetic, thermal,
rotational kinetic, electromagnetic, and gravitational potential
energy. The last one is negative and decreases the value of
$F_e$. Thus, $F_{\rm E,EM}/F_e \approx 1$ implies that absolute value
of the gravitational potential energy is nearly equal to the sum of
the bulk kinetic, thermal, and rotational kinetic energy.
\end{thebibliography}
\end{document}